\documentclass{article}
\usepackage[utf8x]{}
\usepackage[top=1.5cm,bottom=1.5cm,left=1.5cm,right=1.5cm]{geometry}
\usepackage{graphicx}
\graphicspath{ {./images/} }
\usepackage{amsmath}
\usepackage{slashed}
\usepackage{cancel}
\usepackage{tikz}
\usetikzlibrary{patterns}

\usepackage{hyperref}

\title{Tests of the Charge Convexity Conjecture in Caswell-Banks-Zaks Theory}
\author{Ofer Aharony, Yacov-Nir Breitstein}
\date{May 15, 2023}

\numberwithin{equation}{section}
\tikzset{every picture/.style={line width=0.75pt}} 

\tikzset{
pattern size/.store in=\mcSize, 
pattern size = 5pt,
pattern thickness/.store in=\mcThickness, 
pattern thickness = 0.3pt,
pattern radius/.store in=\mcRadius, 
pattern radius = 1pt}
\makeatletter
\pgfutil@ifundefined{pgf@pattern@name@_klt208kxf}{
\pgfdeclarepatternformonly[\mcThickness,\mcSize]{_klt208kxf}
{\pgfqpoint{0pt}{0pt}}
{\pgfpoint{\mcSize+\mcThickness}{\mcSize+\mcThickness}}
{\pgfpoint{\mcSize}{\mcSize}}
{
\pgfsetcolor{\tikz@pattern@color}
\pgfsetlinewidth{\mcThickness}
\pgfpathmoveto{\pgfqpoint{0pt}{0pt}}
\pgfpathlineto{\pgfpoint{\mcSize+\mcThickness}{\mcSize+\mcThickness}}
\pgfusepath{stroke}
}}
\makeatother

\tikzset{
pattern size/.store in=\mcSize, 
pattern size = 5pt,
pattern thickness/.store in=\mcThickness, 
pattern thickness = 0.3pt,
pattern radius/.store in=\mcRadius, 
pattern radius = 1pt}
\makeatletter
\pgfutil@ifundefined{pgf@pattern@name@_49wcz6le7}{
\pgfdeclarepatternformonly[\mcThickness,\mcSize]{_49wcz6le7}
{\pgfqpoint{0pt}{0pt}}
{\pgfpoint{\mcSize+\mcThickness}{\mcSize+\mcThickness}}
{\pgfpoint{\mcSize}{\mcSize}}
{
\pgfsetcolor{\tikz@pattern@color}
\pgfsetlinewidth{\mcThickness}
\pgfpathmoveto{\pgfqpoint{0pt}{0pt}}
\pgfpathlineto{\pgfpoint{\mcSize+\mcThickness}{\mcSize+\mcThickness}}
\pgfusepath{stroke}
}}
\makeatother

\begin{document}

\global\long\def\swap{\leftrightarrow}%

\global\long\def\goto{\rightarrow}%

\global\long\def\ergo{\Rightarrow}%

\begin{titlepage}

\maketitle

\centering
\textit{Department of Particle Physics and Astrophysics, Weizmann Institute of Science,}

\textit{Rehovot 7610001, Israel}

\vspace{15pt}

e-mail: ofer.aharony@weizmann.ac.il, yacov-nir.breitstein@weizmann.ac.il

\vspace{40pt}

\begin{abstract}

\large

The Charge Convexity Conjecture (CCC) states that in a unitary conformal field theory in $d\geq 3$ dimensions with a global symmetry, the minimal dimension of operators in certain representations of the symmetry, as a function of the charge $q$ of the representation (or a generalized notion of it), should be convex. More precisely, this was conjectured to be true when $q$ is restricted to positive integer multiples of some integer $q_0$.
The CCC was tested on a number of examples, most of which are in $d<4$ dimensions, and its version in which $q_0$ is taken to be the charge of the lowest-dimension positively-charged operator was shown to hold in all of them.

In this paper we test the conjecture in a non-trivial example of a $d=4$ theory, which is the family of Caswell-Banks-Zaks IR fixed points of $SU(N_c)$ gauge theory coupled to $N_f$ massless fermions and $N_s$ massless scalars.
In these theories, the lowest-dimension gauge-invariant operators that transform non-trivially under the global symmetry are mesons. These may consist of two scalars, two fermions or one of each.

We find that the CCC holds in all applicable cases, providing significant new evidence for its validity, and suggesting a stronger version for non-simple global symmetry groups.  

\end{abstract}

\end{titlepage}

\section{Introduction and summary}

Motivated by a possible generalization of the weak gravity conjecture (\cite{paper4}, see also \cite{paper5}) to asymptotically-anti-de-Sitter space-times, it was conjectured in \cite{paper7} that the dimensions of operators in unitary conformal field theories (CFTs) in $d \geq 3$ space-time dimensions with global symmetries obey a {\it Charge Convexity Conjecture} (CCC). In the simplest case where the global symmetry is $U(1)$, we can denote by $\Delta(q)$ the dimension of the lowest-dimension operator with some positive integer charge $q$. The conjecture then states that for some positive integer $q_0$ and for any positive integers $q_1,q_2$, the inequality
\begin{equation} \label{eq:CCC}
\Delta((q_1+q_2) q_0) \geq \Delta(q_1 q_0) + \Delta(q_2 q_0)
\end{equation}
is obeyed. The conjecture in \cite{paper7} included the further condition that $q_0$ should not be parameterically large in any parameters of the CFT, and the conjecture was shown to hold in many examples (generically the conjecture becomes stronger as $q_0$ is taken to be smaller). Most of the tests of the conjecture were in some perturbative expansion (small coupling, large $N$, etc.), and in this context it is enough to show that $\Delta(2q_0) > 2 \Delta(q_0)$ for the conjecture to hold (when the perturbative expansion is valid). A natural generalization of the conjecture to more complicated global symmetries is that there should be a representation $r_0$ of the global symmetries, such that if $\Delta(n)$ is the dimension of the lowest-dimension operator in the $n$'th symmetric power of the representation $r_0$, then for any positive integers $n_1,n_2$, $\Delta(n_1+n_2) \geq \Delta(n_1)+\Delta(n_2)$. Further discussion of the conjecture and its motivations may be found in \cite{paper7} (see also \cite{paper18}-\cite{paper21}).
There, the non-abelian generalization was focused to a representation $r_0$ of a simple component of the symmetry group. One can consider an expansion of this to any product of simple components (and $U(1)$'s), where the represenation $r_0$ should have non-trivial weights in each of the components included in the product. 

A counter-example to the original phrasing of the conjecture was found in \cite{paper17}, which constructed a $d=3$ example where $q_0$ must be chosen to be exponentially large in some parameters of the CFT. However, slightly weaker versions of the conjecture are still consistent with all known examples. The simplest version which is not ruled out states that $q_0$ should be the charge of the lowest-dimension positively-charged operator; this is consistent with all known examples. For a general global symmetry, if we look at a specific product of simple factors of the global symmetry, then the corresponding generalization would be that $r_0$ should be the representation of the lowest-dimension operator that is charged under all of these simple factors. As far as we know, there are no counter-examples to this version of the conjecture.

The perturbative tests of the conjecture that were performed up to now were mostly in $d<4$ dimensions, because it is much easier to construct and to analyze CFTs in lower dimensions. In $d=4$ the simplest examples of non-trivial unitary CFTs are low-energy fixed points of asymptotically-free gauge theories, and the computations of anomalous dimensions in these theories are more complicated than the ones analyzed in \cite{paper7}. In this paper we test the conjecture in a simple example of this class - the Caswell-Banks-Zaks (\cite{paper8}-\cite{paper11}) fixed points of $SU(N_c)$ gauge theory coupled to $N_f$ massless fermions and $N_s$ massless scalars in the fundamental representation. As we review below, when the number of fermions and scalars is close to the asymptotic freedom bound (and when the number of scalars is small enough) these theories flow to IR fixed points which are weakly coupled (for large $N_c$) and can be analyzed in perturbation theory (in the gauge coupling, and in $\phi^4$-type couplings when scalars are present). The lowest-dimension gauge-invariant operators charged under the global symmetry in these theories are mesons, and the next-lowest charged operators can be thought of as products of mesons. In perturbation theory it is enough to check that the dimension of a two-meson operator (in a symmetric product of the global symmetry representations) is larger than twice the dimension of single-meson operators for the conjecture \eqref{eq:CCC} to hold. For mesons made from two scalars, it was argued already in \cite{paper7} that the conjecture holds, and we review this below. In this paper we test the conjecture for the other mesons of the theory, made from two fermions or from a fermion and a scalar. These operators carry charges under 3 possible subgroups of the global symmetry, and we find that in all 3 cases, the conjecture (in its version of the previous paragraph) holds (in perturbation theory). This provides significant new evidence for the validity of the conjecture, including the expansion to products of simple components discussed above.

There are many possible generalizations of our analysis to other perturbative fixed points in $d=4$. One can consider other gauge groups, or additional matter fields. In particular one can add additional fields in the adjoint representation, as in supersymmetric gauge theories.

A version of this paper with less detailed computations than the ones presented below has been published in \cite{paperSelf}.

\section{Caswell-Banks-Zaks fixed points in 4 dimensions} \label{Banks-Zaks}

4-dimensional $SU(N_c)$ gauge theories with $N_f$ flavors of massless Dirac fermions and $N_s$ flavors of massless scalars in the fundamental representation, have an infrared-stable fixed point for certain ratios of $N_f,N_s, N_c$ \cite{paper8}-\cite{paper11}. The Lagrangian of the theory can be explicitly written as in \cite{paper11}:
 
\begin{equation} \label{eq:5}
    \mathcal{L} = -\frac{1}{4} F_{\mu \nu}^A F^{A \mu \nu} + \text{Tr}_f \left( \bar{\psi} i \slashed{D} \psi \right) + \text{Tr}_s \left( D_{\mu} \phi^{\dagger} D^{\mu} \phi \right) -\tilde{h} \text{Tr}_s \left( \phi^{\dagger} \phi \phi^{\dagger} \phi \right) -\tilde{f} \left( \text{Tr}_s \left( \phi^{\dagger} \phi \right) \right)^2,
\end{equation}
where the traces $ \text{Tr}_f,\text{Tr}_s$ are over the flavor indices of the fermions and scalars, respectively. The scalars are viewed here as an $N_c \times N_s$ matrix, and the fermions as an $N_c \times N_f$ Dirac fermion-valued matrix.
The global symmetry of the fermions is $SU(N_f)_L \times SU(N_f)_R \times U(1)_{B}$, and that of the scalars is $SU(N_s) \times U(1)_{B'}$. A gauge-invariant operator that transforms non-trivially under the $U(1)_{B,B'}$ symmetries will be either a baryon, a mixed meson ($\phi^{\dagger}\psi$ or $\bar{\psi}\phi$), or a product of one or more of these with a $U(1)_{B,B'}$ neutral operator.
In the large $N$ approximation, baryons are non-trivial and their $U(1)$ charges will diverge, so they are irrelevant to the discussion of the CCC. In the case of mixed mesons, their $U(1)$ charges are proportional to their charge under the respective global $SU(N)$ symmetry, so we do not need to consider these symmetries separately.

The scaling limit in which the fixed point can be proven to occur is $N_c \rightarrow \infty$, with 
 
\begin{equation}
    x_s = \frac{N_s}{N_c},\qquad x_f = \frac{N_f}{N_c},\qquad \lambda = \frac{N_c g^2}{16\pi^2},\qquad h = \frac{N_c \Tilde{h}}{16\pi^2},\qquad f = \frac{N_c N_s \tilde{f}}{16\pi^2}
\end{equation}
held fixed. The perturbative expansion is in the 't Hooft couplings, $\lambda,h,f$, which are considered to be of the same parametric order.
The $\beta$ function of the 't Hooft gauge coupling $\lambda$ is, to two loop order:
\begin{equation}
    \beta_{\lambda} = -\frac{22-x_s-4x_f}{3} \lambda^2+ b_1 \lambda^3,
\end{equation}
where $b_1 = b_1(x_s,x_f,N_c)$ is of parametric order $1$ in the large $N_c$ limit.
The parameter regime compatible with reliable weakly coupled fixed points is when the theory has a number of flavors slightly below the asymptotic freedom bound:
\begin{equation}\label{eq:2}
    \varepsilon \equiv \frac{1}{75} \left( 22-x_s-4x_f \right) \ll 1 \indent \& \indent \varepsilon>0.
\end{equation}
Higher order corrections are then smaller than the two loop term, and can safely be neglected. The non-trivial fixed point for the gauge coupling $\lambda$ to this order is given by:
\begin{equation}
    \lambda^* = \frac{\varepsilon}{1+x_s/50}+O(\varepsilon^2).
\end{equation}

It is shown in \cite{paper22},\cite{paper11} that there are fixed points for the quartic scalar couplings with real coupling constants, obtained by solving for the vanishing scalar $\beta$ functions $\beta_h,\beta_f$ to one loop order (and fine-tuning the scalar masses to zero).
At leading order in the couplings, the four non-trivial fixed points of the quartic scalar interactions, $(h^*_+,f^*_{\pm +}),(h^*_-,f^*_{\pm -})$, are given by:
\begin{equation}
    h^*_{\pm} = \lambda^* \frac{3\pm \sqrt{6-2x_s}}{4(1+x_s)},\qquad
    f^*_{\pm +} = \lambda^* \left( -\frac{\sqrt{6-3x_s}}{4} \pm A_+ \right), \qquad
    f^*_{\pm -} = \lambda^* \left( \frac{\sqrt{6-3x_s}}{4} \pm A_- \right),
\end{equation}
with
\begin{equation}
    A_\pm = \frac{3\sqrt{2-(13\pm 6\sqrt{6-3x_s})x_s+x_s^2-2x_s^3}}{4\sqrt{3}(1+x_s)}.
\end{equation}
The range of parameters where any of these four fixed points obtain real values is approximately:
\begin{equation}\label{eq:3}
    x_s = \frac{N_s}{N_c} \leq 0.84.
\end{equation}
The theory at any fixed point in this regime is referred to as the Caswell-Banks-Zaks theory. The CCC was verified in \cite{paper7} for operators made of scalars. Note that the bounds (\ref{eq:2}),(\ref{eq:3}) don't allow a CFT with only scalar matter, but do allow one with only fermionic matter.
For the remainder of this paper, we denote $N=N_c$ for simplicity.

\subsection{Global symmetry representations}
The global symmetry group of the fermions is $SU(N_f)_L \times SU(N_f)_R \times U(1)_B$, and that of the scalars is $SU(N_s)\times U(1)_{B'}$. As explained above, there is no need to discuss the $U(1)$ charges separately.
The non-abelian components are discussed below.

\subsubsection{Scalar mesons}

The (anti)scalar fields are in a trivial representation of the $SU(N_f)$ symmetry components, and in a (anti)fundamental representation of the $SU(N_s)$ component. The scalar meson $\phi^* \phi$ therefore transforms in the representation $\bf{\bar{N_s}} \times \bf{N_s} = (\bf{N_s^2-1})+\bf{1}$.
Only the former of the two resulting irreducible representations (irreps.) is relevant to the CCC, and it will be our focus.

\subsubsection{Fermion mesons}

It is convenient to separate the Dirac fermions into Weyl fermions:
\begin{equation}
    \psi = \underbrace{\frac{1}{2}(1+\gamma^5)\psi}_{\equiv \psi_R} + \underbrace{\frac{1}{2}(1-\gamma^5)\psi}_{\equiv \psi_L}, \indent
    \Bar{\psi} = \underbrace{\frac{1}{2}\Bar{\psi}(1+\gamma^5)}_{=\Bar{\psi}_L} + \underbrace{\frac{1}{2}\Bar{\psi}(1-\gamma^5)}_{=\Bar{\psi}_R}.
\end{equation}
In this notation, the Weyl fermions belong to the following group representations: 
\begin{equation}
    \psi_L \sim \left( \bf{N_f},\bf{1} \right), \indent
    \psi_R \sim \left( \bf{1},\bf{N_f} \right), \indent
    \Bar{\psi}_L \sim \left( \bf{\Bar{N_f}},\bf{1} \right), \indent
    \Bar{\psi}_R \sim  \left( \bf{1},\bf{\Bar{N_f}} \right).
\end{equation}
Here the left-hand side of the parentheses refers to the representation of $SU(N_f)_L$ and the right-hand side to $SU(N_f)_R$.

In 4 dimensions all the degrees of freedom quadratic in the fermions can be described by mesons in the different Lorentz representations:
\begin{equation}
    \left\{ 1,\gamma^5,\gamma^{\mu}, \gamma^5\gamma^{\mu}, \gamma^{\mu \nu} \right\},
\end{equation}
contracted between $\bar{\psi}$ and $\psi$, where $\mu,\nu$ are space-time indices, $\gamma^5 \equiv i\gamma^0 \gamma^1 \gamma^2 \gamma^3 = \text{diag}(-1,-1,1,1)$ (the last expression in a particular basis that separates $\psi_L,\psi_R$),  and $\gamma^{\mu\nu} \equiv\frac{1}{2} \left[ \gamma^{\mu},\gamma^{\nu} \right]$.
To see the representations of the different mesons, we can contract the fermions while keeping in mind that each $\gamma^{\mu}$ interchanges left- and right-handed Weyl fermions, while $\gamma^5$ does not - so the pseudoscalar and psuedovector are in the same representations as the scalar and vector, respectively.

For the scalar representation:
\begin{equation} \label{eq:reps1}
    \Bar{\psi} \psi = \begin{pmatrix}
        \Bar{\psi}_R \\
        \Bar{\psi}_L
    \end{pmatrix}
    \begin{pmatrix}
        \psi_L & \psi_R
    \end{pmatrix}
    = \Bar{\psi}_R \psi_L + \Bar{\psi}_L \psi_R
    \indent \Rightarrow \indent
    \Bar{\psi} \psi, \Bar{\psi} \gamma^5 \psi \in \left( \bf{N_f}, \bf{\Bar{N_f}} \right) + \left( \bf{\Bar{N_f}},\bf{N_f} \right).
\end{equation}
For the vector representation:
\begin{equation} \label{eq:reps2}
    \Bar{\psi} \gamma^{\mu} \psi = \begin{pmatrix}
        \Bar{\psi}_R \\
        \Bar{\psi}_L
    \end{pmatrix}
    \gamma^{\mu}
    \begin{pmatrix}
        \psi_L & \psi_R
    \end{pmatrix}
    \sim \Bar{\psi}_R \cdots \psi_R + \Bar{\psi}_L \cdots \psi_L
    \indent \Rightarrow 
    \Bar{\psi} \gamma^{\mu} \psi, \Bar{\psi} \gamma^5 \gamma^{\mu} \psi \in \left( \bf{1}, \bf{N_f^2-1} \right) + \left( \bf{N_f^2-1},\bf{1} \right) + 2\cdot (\bf{1},\bf{1}).
\end{equation}

The trivial representation of vectors is irrelevant to the CCC, 
and so will be ignored from this point forward. The tensors have $[\gamma^{\mu}, \gamma^{\nu}]$, so there are 2 interchanges - which restore the original contraction between the Weyl fermions. Therefore, they are in the same flavor representation as the scalar and pseudoscalar: $\Bar{\psi} \gamma^{\mu\nu} \psi \in \left( \bf{N_f}, \bf{\Bar{N_f}} \right) + \left( \bf{\Bar{N_f}},\bf{N_f} \right)$. It can be seen from here, that the only mixing of multi-meson operators in symmetric products of the flavor representations in (\ref{eq:reps1}),(\ref{eq:reps2}) will be between the 2-scalar and the 2-tensor mesons (where the space-time indices of the 2-tensor mesons will be contracted to give a Lorentz scalar). It is useful to further distinguish the representations using the chiral projections mentioned above, to divide the different mesons into their irreducible representations:
\begin{align}
    & \bar{\psi}(1-\gamma^5)\psi,\bar{\psi}(1-\gamma^5)\gamma^{\mu\nu}\psi \in \left( \bf{N_f}, \bf{\bar{N_f}} \right); \indent 
    \bar{\psi}(1+\gamma^5)\psi,\bar{\psi}(1+\gamma^5)\gamma^{\mu\nu}\psi \in \left( \bf{\bar{N_f}},\bf{N_f} \right); \nonumber\\
    & \bar{\psi}(1-\gamma^5)\gamma^{\mu}\psi \in \left( \bf{1},\bf{N_f^2-1} \right); \indent
    \bar{\psi}(1+\gamma^5)\gamma^{\mu}\psi \in \left( \bf{N_f^2-1},\bf{1} \right).
\end{align}
The operators in different representations, and their symmetric products with themselves, do not mix with each other. Our calculations treat the left- and right-handed projections simultaneously. 

\subsubsection{Mixed mesons}

Similarly to the previous cases, the mixed mesons transform in the following representations 
of $SU(N_s),SU(N_f)_L,SU(N_f)_R$:

\begin{equation}
    \bar{\psi} \phi \in (\bf{N_s},\bf{\bar{N_f}},\bf{1})+(\bf{N_s},\bf{1},\bf{\bar{N_f}}); \indent
    \phi^* \psi \in (\bf{\bar{N_s}},\bf{N_f},\bf{1})+(\bf{\bar{N_s}},\bf{1},\bf{N_f}).
\end{equation}

The conjecture as stated in \cite{paper7} only requires one operator in a representation with weights of order one of each simple subgroup.
In the case of fermionic mesons, both the representation of mesons contracted as scalars and tensors, and that of the mesons contracted as vectors, have such non-trivial components in both $SU(N_f)$'s. The chirally projected mixed meson has non-trivial weights in $SU(N_s)$ as well as either one of the $SU(N_f)$'s. Thus, verifying the conjecture in either of the fermionic representations in addition to the scalar one, or even only in the mixed meson case, will suffice for verifying it in the theory.
However, we show below that it holds for all of the above representations, giving evidence for a stronger version of the conjecture.

\section{Anomalous dimensions of single mesons} \label{section 4.1}

\subsection{Fermionic mesons} \label{single fermionic mesons}
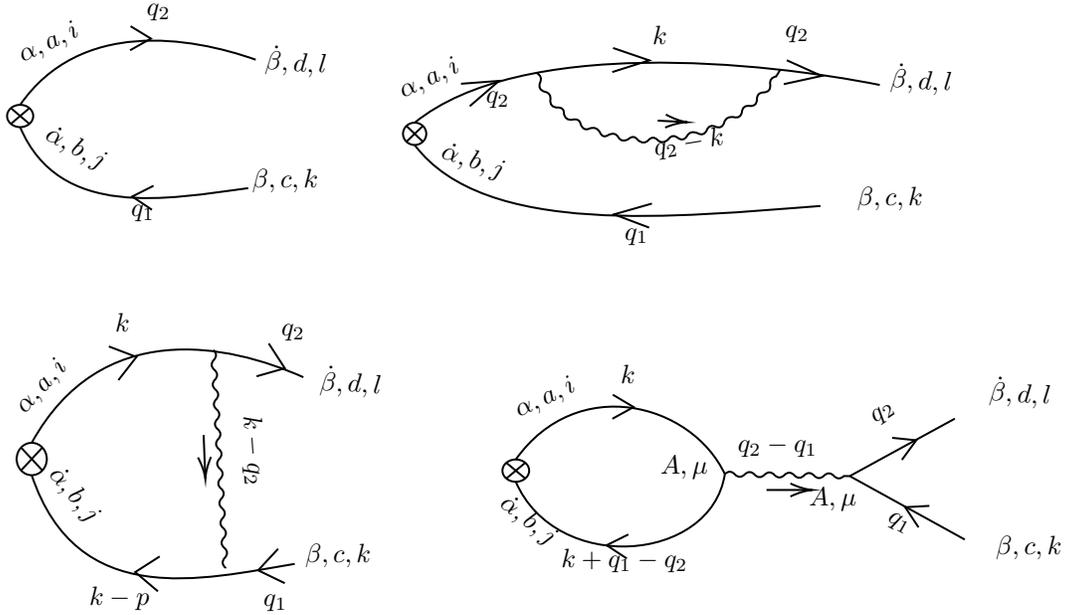
\begin{figure}[ht]
\centering

\begin{tikzpicture}[x=0.75pt,y=0.75pt,yscale=-1,xscale=1]

\draw   (5,60.86) .. controls (5,57.76) and (7.95,55.24) .. (11.59,55.24) .. controls (15.23,55.24) and (18.19,57.76) .. (18.19,60.86) .. controls (18.19,63.97) and (15.23,66.49) .. (11.59,66.49) .. controls (7.95,66.49) and (5,63.97) .. (5,60.86) -- cycle ; \draw   (6.93,56.89) -- (16.26,64.84) ; \draw   (16.26,56.89) -- (6.93,64.84) ;
\draw    (11.59,55.24) .. controls (41.86,11.75) and (97.43,20.76) .. (130.4,32.46) ;
\draw    (11.59,66.49) .. controls (31.5,113.49) and (86.13,102.68) .. (126.63,97.28) ;
\draw   (67.28,15.54) -- (77.53,21.79) -- (67.05,27.69) ;
\draw   (78.06,108.2) -- (68.12,101.52) -- (78.86,96.07) ;
\draw   (204.69,69.86) .. controls (204.69,66.76) and (207.32,64.24) .. (210.56,64.24) .. controls (213.8,64.24) and (216.42,66.76) .. (216.42,69.86) .. controls (216.42,72.97) and (213.8,75.49) .. (210.56,75.49) .. controls (207.32,75.49) and (204.69,72.97) .. (204.69,69.86) -- cycle ; \draw   (206.41,65.89) -- (214.71,73.84) ; \draw   (214.71,65.89) -- (206.41,73.84) ;
\draw    (210.56,64.24) .. controls (264.4,20.75) and (386.48,33.5) .. (445.13,45.2) ;
\draw    (210.56,75.49) .. controls (245.97,122.49) and (343.16,111.68) .. (415.22,106.28) ;
\draw   (310.63,26.54) -- (328.86,32.79) -- (310.22,38.69) ;
\draw   (328.81,117.2) -- (311.12,110.52) -- (330.22,105.07) ;
\draw    (272.13,39.2) .. controls (274.21,40.31) and (274.76,41.94) .. (273.79,44.07) .. controls (273.02,46.36) and (273.79,47.87) .. (276.09,48.6) .. controls (278.38,49.07) and (279.34,50.46) .. (278.99,52.77) .. controls (278.86,55.19) and (280,56.46) .. (282.43,56.58) .. controls (284.8,56.5) and (285.94,57.51) .. (285.85,59.61) .. controls (286.28,62.03) and (287.73,63.06) .. (290.18,62.71) .. controls (292.56,62.2) and (293.94,63) .. (294.31,65.11) .. controls (295.26,67.44) and (296.73,68.14) .. (298.72,67.21) .. controls (301.06,66.36) and (302.84,67.03) .. (304.05,69.23) .. controls (304.96,71.26) and (306.36,71.68) .. (308.24,70.47) .. controls (310.51,69.31) and (312.2,69.69) .. (313.29,71.62) .. controls (314.5,73.53) and (316.23,73.81) .. (318.49,72.45) .. controls (320.16,70.96) and (321.68,71.11) .. (323.04,72.89) .. controls (325.03,74.66) and (326.83,74.73) .. (328.43,73.1) .. controls (329.95,71.42) and (331.51,71.39) .. (333.1,73) .. controls (334.79,74.57) and (336.61,74.42) .. (338.57,72.56) .. controls (339.9,70.71) and (341.47,70.49) .. (343.26,71.9) .. controls (345.14,73.25) and (346.69,72.95) .. (347.92,70.98) .. controls (349.05,68.99) and (350.85,68.52) .. (353.31,69.57) .. controls (355.36,70.67) and (356.87,70.17) .. (357.86,68.06) .. controls (358.73,65.94) and (360.22,65.35) .. (362.31,66.28) .. controls (364.48,67.12) and (365.93,66.43) .. (366.66,64.22) .. controls (367.27,62) and (368.91,61.08) .. (371.58,61.45) .. controls (373.86,61.97) and (375.21,61.08) .. (375.62,58.77) .. controls (375.91,56.48) and (377.21,55.49) .. (379.5,55.8) .. controls (381.85,55.98) and (383.07,54.89) .. (383.18,52.53) .. controls (383.16,50.2) and (384.32,49.01) .. (386.65,48.96) .. controls (389.02,48.77) and (390.1,47.48) .. (389.88,45.09) .. controls (389.54,42.76) and (390.53,41.37) .. (392.86,40.92) -- (395.13,37.2) ;
\draw    (333.13,63.2) -- (346,63.89) ;
\draw [shift={(348,64)}, rotate = 183.08] [color={rgb, 255:red, 0; green, 0; blue, 0 }  ][line width=0.75]    (10.93,-3.29) .. controls (6.95,-1.4) and (3.31,-0.3) .. (0,0) .. controls (3.31,0.3) and (6.95,1.4) .. (10.93,3.29)   ;
\draw   (233.93,44.6) -- (253.14,43.15) -- (238.35,55.92) ;
\draw   (398.24,32.74) -- (415.81,40.64) -- (396.71,44.8) ;
\draw   (9.83,234) .. controls (9.83,229.48) and (13.23,225.82) .. (17.43,225.82) .. controls (21.63,225.82) and (25.03,229.48) .. (25.03,234) .. controls (25.03,238.51) and (21.63,242.17) .. (17.43,242.17) .. controls (13.23,242.17) and (9.83,238.51) .. (9.83,234) -- cycle ; \draw   (12.05,228.21) -- (22.81,239.78) ; \draw   (22.81,228.21) -- (12.05,239.78) ;
\draw    (17.43,225.82) .. controls (52.33,162.63) and (116.42,175.71) .. (154.44,192.72) ;
\draw    (17.43,242.17) .. controls (40.39,310.47) and (103.39,294.77) .. (150.09,286.92) ;
\draw   (56.32,177.18) -- (70.25,182.5) -- (61.14,194.17) ;
\draw   (81.08,300.79) -- (69.62,291.08) -- (81.99,283.16) ;
\draw    (109,179) .. controls (110.76,180.57) and (110.85,182.23) .. (109.28,183.99) .. controls (107.71,185.75) and (107.8,187.41) .. (109.56,188.98) .. controls (111.31,190.56) and (111.4,192.23) .. (109.83,193.98) .. controls (108.26,195.74) and (108.35,197.4) .. (110.11,198.97) .. controls (111.87,200.54) and (111.96,202.2) .. (110.39,203.96) .. controls (108.82,205.72) and (108.91,207.38) .. (110.67,208.95) .. controls (112.42,210.53) and (112.51,212.2) .. (110.94,213.95) .. controls (109.37,215.71) and (109.46,217.37) .. (111.22,218.94) .. controls (112.98,220.51) and (113.07,222.17) .. (111.5,223.93) .. controls (109.93,225.69) and (110.02,227.35) .. (111.78,228.92) .. controls (113.54,230.49) and (113.63,232.16) .. (112.06,233.92) .. controls (110.49,235.67) and (110.58,237.34) .. (112.33,238.91) .. controls (114.09,240.48) and (114.18,242.14) .. (112.61,243.9) .. controls (111.04,245.66) and (111.13,247.32) .. (112.89,248.89) .. controls (114.65,250.46) and (114.74,252.12) .. (113.17,253.88) .. controls (111.6,255.64) and (111.69,257.31) .. (113.45,258.88) .. controls (115.2,260.45) and (115.29,262.12) .. (113.72,263.87) .. controls (112.15,265.63) and (112.24,267.29) .. (114,268.86) .. controls (115.76,270.43) and (115.85,272.09) .. (114.28,273.85) .. controls (112.71,275.61) and (112.8,277.28) .. (114.56,278.85) .. controls (116.31,280.42) and (116.4,282.09) .. (114.83,283.84) .. controls (113.26,285.6) and (113.35,287.26) .. (115.11,288.83) -- (115.13,289.2) -- (115.13,289.2) ;
\draw   (136.98,175.69) -- (145.08,188.2) -- (130.83,192.25) ;
\draw   (145.04,296.67) -- (131.62,290.18) -- (141.69,279.32) ;
\draw    (104.13,221.87) -- (104.92,242) ;
\draw [shift={(105,244)}, rotate = 267.76] [color={rgb, 255:red, 0; green, 0; blue, 0 }  ][line width=0.75]    (10.93,-3.29) .. controls (6.95,-1.4) and (3.31,-0.3) .. (0,0) .. controls (3.31,0.3) and (6.95,1.4) .. (10.93,3.29)   ;
\draw   (254.83,239.86) .. controls (254.83,236.76) and (257.85,234.24) .. (261.57,234.24) .. controls (265.29,234.24) and (268.31,236.76) .. (268.31,239.86) .. controls (268.31,242.97) and (265.29,245.49) .. (261.57,245.49) .. controls (257.85,245.49) and (254.83,242.97) .. (254.83,239.86) -- cycle ; \draw   (256.8,235.89) -- (266.34,243.84) ; \draw   (266.34,235.89) -- (256.8,243.84) ;
\draw    (261.57,234.24) .. controls (292.52,190.75) and (350,205.63) .. (367,241.63) ;
\draw    (261.57,245.49) .. controls (281.92,292.49) and (360,286.63) .. (367,241.63) ;
\draw   (310.51,201.54) -- (320.99,207.79) -- (310.28,213.69) ;
\draw   (317.53,284.2) -- (307.37,277.52) -- (318.34,272.07) ;
\draw    (367,241.63) .. controls (368.69,239.99) and (370.36,240.02) .. (372,241.71) .. controls (373.64,243.4) and (375.31,243.43) .. (377,241.79) .. controls (378.69,240.15) and (380.36,240.18) .. (382,241.87) .. controls (383.64,243.56) and (385.31,243.59) .. (387,241.95) .. controls (388.69,240.31) and (390.36,240.34) .. (392,242.03) .. controls (393.64,243.72) and (395.31,243.75) .. (397,242.11) .. controls (398.69,240.47) and (400.36,240.5) .. (402,242.19) .. controls (403.63,243.88) and (405.3,243.91) .. (406.99,242.27) .. controls (408.68,240.63) and (410.35,240.66) .. (411.99,242.35) .. controls (413.63,244.04) and (415.3,244.07) .. (416.99,242.43) .. controls (418.68,240.79) and (420.35,240.82) .. (421.99,242.51) .. controls (423.63,244.2) and (425.3,244.23) .. (426.99,242.59) -- (430,242.63) -- (430,242.63) ;
\draw    (430,242.63) -- (483,213.63) ;
\draw    (430,242.63) -- (488,274.63) ;
\draw   (451.39,223.3) -- (463.53,224.55) -- (456.31,234.42) ;
\draw   (464.69,268.5) -- (457.95,258.33) -- (470.16,257.65) ;
\draw    (388,249.63) -- (409,249.63) ;
\draw [shift={(411,249.63)}, rotate = 180] [color={rgb, 255:red, 0; green, 0; blue, 0 }  ][line width=0.75]    (10.93,-3.29) .. controls (6.95,-1.4) and (3.31,-0.3) .. (0,0) .. controls (3.31,0.3) and (6.95,1.4) .. (10.93,3.29)   ;

\draw (74,3) node [anchor=north west][inner sep=0.75pt]   [align=left] {$\displaystyle q_{2}$};
\draw (66,104) node [anchor=north west][inner sep=0.75pt]   [align=left] {$\displaystyle q_{1}$};
\draw (141.83,164.21) node [anchor=north west][inner sep=0.75pt]   [align=left] {$\displaystyle q_{2}$};
\draw (132.83,300.99) node [anchor=north west][inner sep=0.75pt]   [align=left] {$\displaystyle q_{1}$};
\draw (5.45,205.4) node [anchor=north west][inner sep=0.75pt]  [rotate=-312.53] [align=left] {$\displaystyle \alpha ,a,i$};
\draw (161.84,185.21) node [anchor=north west][inner sep=0.75pt]  [rotate=-3.11] [align=left] {$\displaystyle \dot{\beta } ,d,l$};
\draw (33.25,233.79) node [anchor=north west][inner sep=0.75pt]  [rotate=-50.52] [align=left] {$\displaystyle \dot{\alpha } ,b,j$};
\draw (153.89,275.19) node [anchor=north west][inner sep=0.75pt]   [align=left] {$\displaystyle \beta ,c,k$};
\draw (45,297) node [anchor=north west][inner sep=0.75pt]   [align=left] {$\displaystyle k-p$};
\draw (58,159) node [anchor=north west][inner sep=0.75pt]   [align=left] {$\displaystyle k$};
\draw (134.88,209.84) node [anchor=north west][inner sep=0.75pt]  [rotate=-85.66] [align=left] {$\displaystyle k-q_{2}$};
\draw (437.66,210.39) node [anchor=north west][inner sep=0.75pt]  [rotate=-325.93] [align=left] {$\displaystyle q_{2}$};
\draw (451.08,258.06) node [anchor=north west][inner sep=0.75pt]  [rotate=-29.28] [align=left] {$\displaystyle q_{1}$};
\draw (257.64,204.06) node [anchor=north west][inner sep=0.75pt]  [rotate=-334.8] [align=left] {$\displaystyle \alpha ,a,i$};
\draw (499.47,190.38) node [anchor=north west][inner sep=0.75pt]  [rotate=-3.11] [align=left] {$\displaystyle \dot{\beta } ,d,l$};
\draw (258.63,248.08) node [anchor=north west][inner sep=0.75pt]  [rotate=-38.51] [align=left] {$\displaystyle \dot{\alpha } ,b,j$};
\draw (502.24,271.25) node [anchor=north west][inner sep=0.75pt]   [align=left] {$\displaystyle \beta ,c,k$};
\draw (7.26,25.06) node [anchor=north west][inner sep=0.75pt]  [rotate=-334.8] [align=left] {$\displaystyle \alpha ,a,i$};
\draw (134.31,23.38) node [anchor=north west][inner sep=0.75pt]  [rotate=-3.11] [align=left] {$\displaystyle \dot{\beta } ,d,l$};
\draw (27.55,62.21) node [anchor=north west][inner sep=0.75pt]  [rotate=-28.22] [align=left] {$\displaystyle \dot{\alpha } ,b,j$};
\draw (127.2,86.25) node [anchor=north west][inner sep=0.75pt]   [align=left] {$\displaystyle \beta ,c,k$};
\draw (199.1,43.06) node [anchor=north west][inner sep=0.75pt]  [rotate=-334.8] [align=left] {$\displaystyle \alpha ,a,i$};
\draw (449.57,32.38) node [anchor=north west][inner sep=0.75pt]  [rotate=-3.11] [align=left] {$\displaystyle \dot{\beta } ,d,l$};
\draw (225.51,71.21) node [anchor=north west][inner sep=0.75pt]  [rotate=-18.22] [align=left] {$\displaystyle \dot{\alpha } ,b,j$};
\draw (432.2,95.25) node [anchor=north west][inner sep=0.75pt]   [align=left] {$\displaystyle \beta ,c,k$};
\draw (329,14) node [anchor=north west][inner sep=0.75pt]   [align=left] {$\displaystyle k$};
\draw (328.75,70.37) node [anchor=north west][inner sep=0.75pt]  [rotate=-350.89] [align=left] {$\displaystyle q_{2} -k$};
\draw (245,47) node [anchor=north west][inner sep=0.75pt]   [align=left] {$\displaystyle q_{2}$};
\draw (396,14) node [anchor=north west][inner sep=0.75pt]   [align=left] {$\displaystyle q_{2}$};
\draw (315,116) node [anchor=north west][inner sep=0.75pt]   [align=left] {$\displaystyle q_{1}$};
\draw (313,185) node [anchor=north west][inner sep=0.75pt]   [align=left] {$\displaystyle k$};
\draw (372,221) node [anchor=north west][inner sep=0.75pt]   [align=left] {$\displaystyle q_{2} -q_{1}$};
\draw (283,278) node [anchor=north west][inner sep=0.75pt]   [align=left] {$\displaystyle k+q_{1} -q_{2}$};
\draw (333,231) node [anchor=north west][inner sep=0.75pt]   [align=left] {$\displaystyle A,\mu $};
\draw (408,247) node [anchor=north west][inner sep=0.75pt]   [align=left] {$\displaystyle A,\mu $};

\end{tikzpicture}

\caption{\label{fig:diagram4} Tree-level and one-loop Feynman diagrams for the correlation function of a single meson with a fermion-antifermion pair.}

\end{figure}

We begin by computing the generic-index 
correlation function:
\begin{equation}
    \left< (\psi_{\alpha ai}\Bar{\psi}_{\Dot{\alpha} bj})(p) \psi_{\beta ck}(q_1) \Bar{\psi}_{\Dot{\beta}dl}(-q_2) \right>.
\end{equation}
Here $a,b,c,d$ are color indices (and the first two will be contracted inside the meson to give a gauge-invariant object), $i,j,k,l$ flavor indices in the appropriate representation of $SU(N_f)\times SU(N_f)_R$ (the specific representation depends on the type of spinor structure contraction; all possibilities will be addressed in what follows, so this is kept generic), and $\alpha, \Dot{\alpha}, \beta, \Dot{\beta}$ are fermionic indices. The relevant diagrams up to one loop order are shown in Figure \ref{fig:diagram4}. We denote them by $I$ through $IV$ respectively, row by row and left to right within rows. Note that these and other diagrams we compute in this work are not gauge-invariant. We compute them in the Feynman gauge, and after contracting the color indices of the mesons, our results for the meson anomalous dimensions will be gauge-invariant, so they will not depend on this gauge choice. We will use dimensional regulariation with $d=4-2\epsilon$.

\subsubsection{Diagram computation}

The tree-level diagram is equal (for our normalization of the composite operator) to:
\begin{equation}
    I = \frac{i \slashed{q_2}_{\alpha \Dot{\beta}}}{q_2^2} \frac{i \slashed{q_1}_{\beta \Dot{\alpha}}}{q_1^2} \delta_{ad} \delta_{il} \delta_{bc} \delta_{jk} =
    - \frac{\slashed{q_2}_{\alpha \Dot{\beta}}}{q_2^2} \frac{\slashed{q_1}_{\beta \Dot{\alpha}}}{q_1^2} \delta_{ad} \delta_{il} \delta_{bc} \delta_{jk}.
\end{equation}

For the propagator correction diagram, we can first compute the propagator corrections to leading loop order. The relevant Feynman diagrams are shown in Figure \ref{fig:propagator1}.
Here $A$ is an adjoint color representation index, $\mu$ is a (Minkowski) spacetime index, and the others as before.
The tree level inverse propagator equals $-i \slashed{p}_{\alpha \Dot{\alpha}} \delta_{ab} \delta_{ij}$.
Let us compute the one loop diagram:

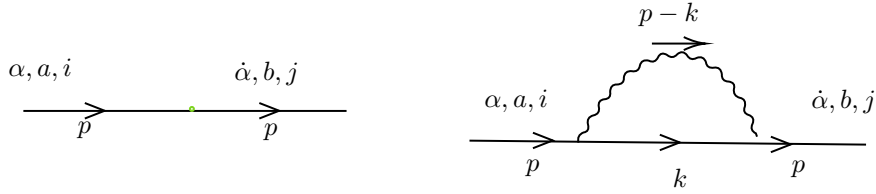
\begin{figure}[h]
\centering

\begin{tikzpicture}[x=0.75pt,y=0.75pt,yscale=-1,xscale=1]

\draw    (33,70.17) -- (196,70.17) ;
\draw  [color={rgb, 255:red, 126; green, 211; blue, 33 }  ,draw opacity=1 ] (117,69.17) .. controls (117,68.61) and (117.45,68.17) .. (118,68.17) .. controls (118.55,68.17) and (119,68.61) .. (119,69.17) .. controls (119,69.72) and (118.55,70.17) .. (118,70.17) .. controls (117.45,70.17) and (117,69.72) .. (117,69.17) -- cycle ;
\draw   (64,66.17) -- (74,70.08) -- (64,74) ;
\draw   (152,66.17) -- (162,70.08) -- (152,74) ;
\draw    (258,85.17) -- (458,87.17) ;
\draw   (290,81.17) -- (300,85.08) -- (290,89) ;
\draw   (410,83.17) -- (420,87.08) -- (410,91) ;
\draw    (313,86.17) .. controls (312.35,83.68) and (313.21,81.97) .. (315.56,81.03) .. controls (317.61,80.73) and (318.34,79.38) .. (317.76,76.99) .. controls (317.22,74.6) and (318.12,73.08) .. (320.46,72.43) .. controls (322.77,71.9) and (323.7,70.49) .. (323.23,68.19) .. controls (322.82,65.88) and (323.76,64.58) .. (326.06,64.27) .. controls (328.32,64.08) and (329.28,62.87) .. (328.94,60.66) .. controls (329.01,58.05) and (330.15,56.77) .. (332.37,56.84) .. controls (334.84,56.71) and (336,55.58) .. (335.84,53.45) .. controls (336.11,51.01) and (337.46,49.89) .. (339.87,50.1) .. controls (342.16,50.5) and (343.52,49.57) .. (343.93,47.31) .. controls (344.49,45.06) and (346.02,44.23) .. (348.52,44.83) .. controls (350.5,45.8) and (352.03,45.2) .. (353.12,43.05) .. controls (354.09,41.06) and (355.62,40.7) .. (357.71,41.97) .. controls (359.52,43.44) and (361.2,43.31) .. (362.75,41.6) .. controls (364.58,40.03) and (366.23,40.19) .. (367.72,42.08) .. controls (368.9,44.06) and (370.52,44.51) .. (372.57,43.42) .. controls (374.86,42.57) and (376.42,43.31) .. (377.27,45.62) .. controls (377.58,47.77) and (378.95,48.67) .. (381.36,48.33) .. controls (383.62,48.02) and (384.93,49.15) .. (385.28,51.74) .. controls (385.18,54.03) and (386.29,55.23) .. (388.61,55.34) .. controls (391,55.67) and (391.93,56.87) .. (391.39,58.94) .. controls (391.01,61.35) and (392.01,62.89) .. (394.39,63.56) .. controls (396.56,64) and (397.39,65.5) .. (396.86,68.05) .. controls (396.03,70.14) and (396.7,71.53) .. (398.85,72.23) .. controls (401.02,73.06) and (401.65,74.56) .. (400.72,76.72) .. controls (399.74,78.85) and (400.32,80.45) .. (402.45,81.5) -- (403,83.17) ;
\draw   (354,82.17) -- (364,86.08) -- (354,90) ;
\draw    (350,36.17) -- (375,36.17) ;
\draw [shift={(377,36.17)}, rotate = 180] [color={rgb, 255:red, 0; green, 0; blue, 0 }  ][line width=0.75]    (10.93,-3.29) .. controls (6.95,-1.4) and (3.31,-0.3) .. (0,0) .. controls (3.31,0.3) and (6.95,1.4) .. (10.93,3.29)   ;

\draw (138,44) node [anchor=north west][inner sep=0.75pt]   [align=left] {$\displaystyle \dot{\alpha } ,b,j$};
\draw (24,42) node [anchor=north west][inner sep=0.75pt]   [align=left] {$\displaystyle \alpha ,a,i$};
\draw (429,58) node [anchor=north west][inner sep=0.75pt]   [align=left] {$\displaystyle \dot{\alpha } ,b,j$};
\draw (264,58) node [anchor=north west][inner sep=0.75pt]   [align=left] {$\displaystyle \alpha ,a,i$};
\draw (153,75) node [anchor=north west][inner sep=0.75pt]   [align=left] {$\displaystyle p$};
\draw (59,74) node [anchor=north west][inner sep=0.75pt]   [align=left] {$\displaystyle p$};
\draw (285,93) node [anchor=north west][inner sep=0.75pt]   [align=left] {$\displaystyle p$};
\draw (419,94) node [anchor=north west][inner sep=0.75pt]   [align=left] {$\displaystyle p$};
\draw (359,98) node [anchor=north west][inner sep=0.75pt]   [align=left] {$\displaystyle k$};
\draw (342,14) node [anchor=north west][inner sep=0.75pt]   [align=left] {$\displaystyle p-k$};

\end{tikzpicture}

\caption{\label{fig:propagator1} Feynman diagrams for the fermion propagator, at tree level and one loop level.}
\end{figure}

\begin{equation}
    \delta D_{\alpha \Dot{\alpha}}(p) = (ig)^2 \delta_{ij} t^A_{ca}t^A_{bc} \int \frac{d^d k}{(2\pi)^d} \gamma^{\mu} \frac{i \slashed{k}}{k^2} \gamma_{\mu} \frac{-i}{(p-k)^2}.
\end{equation}
We have $t^A_{bc} t^A_{ca}=(t^A t^A)_{ba}=C_F\delta_{ab} = \frac{N^2-1}{2N}\delta_{ab}$, and to leading order in $\epsilon$ $\gamma^{\mu} \slashed{k} \gamma_{\mu} = -2\slashed{k}$, so the divergent term (as $\epsilon \to 0$) is:

\begin{align} \label{eq:fermion propagator}
    & \delta D_{\alpha \Dot{\alpha}}(p) = \frac{2g^2 C_F}{(2\pi)^4} \delta_{ij} \delta_{ab} \int d^d k \frac{\slashed{k}}{k^2 (p-k)^2} =
    \frac{2g^2 C_F}{(2\pi)^4} \delta_{ij} \delta_{ab} \int_0^1 dx \int d^d k \frac{\slashed{k}}{(k^2-2xp\cdot k+xp^2)^2} = \nonumber \\
    & \underset{k\rightarrow k-xp}{=} \frac{2g^2 C_F}{(2\pi)^4} \delta_{ij} \delta_{ab} \int_0^1 dx \int d^d k \frac{x \slashed{p}}{(k^2+ \underbrace{x(1-x)p^2}_{-\Delta})^2} \underset{k \rightarrow k_E}{=} 
    \frac{2ig^2 C_F}{(2\pi)^4} \delta_{ij} \delta_{ab} \int_0^1 dx x\slashed{p} \int \frac{d^d k_E}{(k_E^2+\Delta)^2} = \nonumber \\
    & = \frac{2ig^2 C_F}{(4\pi)^2} \delta_{ij} \delta_{ab} \int_0^1 dx x\slashed{p} \Gamma(\epsilon) \underbrace{\Delta^{-\epsilon}}_{\approx \left( p_E^2 \right)^{-\epsilon}} =
    \frac{ig^2 C_F}{(4\pi)^2 \epsilon} \delta_{ij} \delta_{ab} \left( p_E^2 \right)^{-\epsilon} \slashed{p}.
\end{align}

Here and throughout we use the subscript $E$ to denote a Euclidean momentum. From this we can extract the fermion renormalization function, once we introduce a renormalization scale $M$. We do this by separating the dimensionful part into $\left( p_E^2 \right)^{-\epsilon} = \left( \frac{p_E^2}{M^2} \right)^{-\epsilon} M^{-2\epsilon} \approx M^{-2\epsilon}$. Thus:
\begin{equation} \label{fermion_anom_dim}
    Z_{\psi} = 1- \frac{g^2 C_F}{(4\pi)^2 \epsilon} M^{-2\epsilon} + O(g^4),
\end{equation}
in agreement with the result in \cite{book1} for our case. 
We insert this result into diagram II and get:
\begin{equation}
    II = \frac{i \slashed{q_2}_{\alpha \gamma}}{q_2^2} \delta D_{\gamma \Dot{\gamma}}(q_2) \delta_{bc} \delta_{jk} \frac{i \slashed{q_2}_{\Dot{\gamma} \Dot{\beta}}}{q_2^2} \frac{i \slashed{q_1}_{\beta \Dot{\alpha}}}{q_1^2}=
    \frac{g^2 C_F}{(4\pi)^2 \epsilon} M^{-2\epsilon} \frac{\slashed{q_2}_{\alpha \Dot{\beta}}}{q_2^2} \frac{\slashed{q_1}_{\beta \Dot{\alpha}}}{q_1^2} \delta_{il} \delta_{jk} \delta_{bc} \delta_{ad} ,
\end{equation}
with another equal contribution coming from the complementary diagram with a one-loop correction to the antifermion propagator.
Diagram III, with the gluon exchange between fermions, gives by similar manipulations:
\begin{equation}
    III = -\frac{g^2}{4(4\pi)^2 \epsilon q_1^2 q_2^2} \delta_{il}\delta_{jk} t^A_{da}t^A_{bc} (\gamma^{\nu} \gamma^{\mu} \slashed{q_2})_{\alpha \Dot{\beta}}(\slashed{q_1} \gamma_{\mu} \gamma_{\nu})_{\beta \Dot{\alpha}} M^{-2\epsilon}
\end{equation}
Diagram IV contains a vertex factor of $t^A_{ba}$. Upon contraction in the color indices we'll get $t^A_{aa} = 0$ as the gauge group is $SU(N)$, and so this diagram will vanish.  

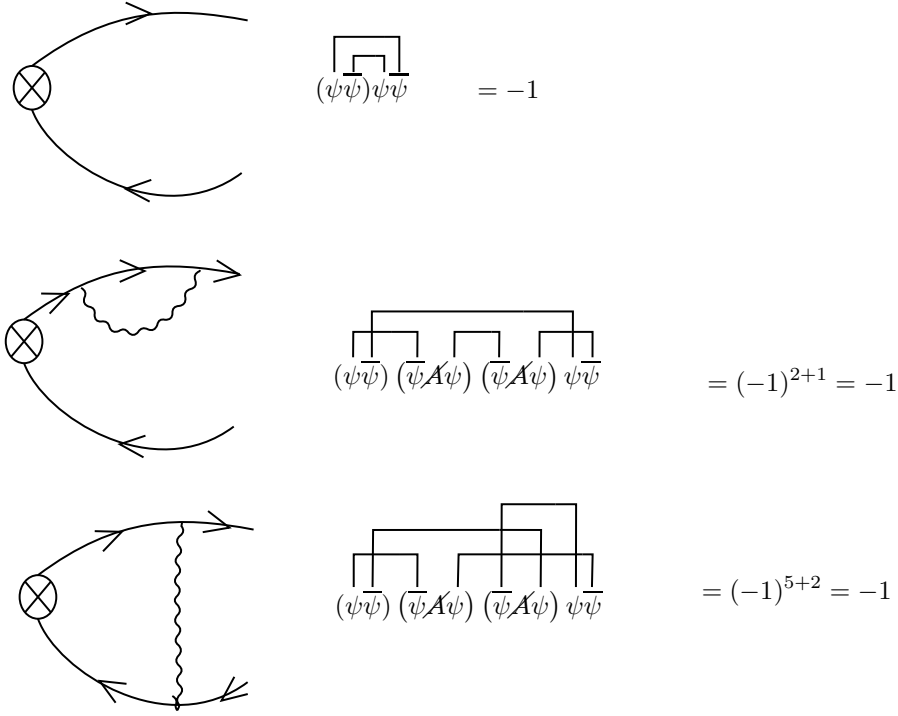
\begin{figure}[h]
\centering
\begin{tikzpicture}[x=0.75pt,y=0.75pt,yscale=-1,xscale=1]

\draw   (59.5,109) .. controls (59.5,102.92) and (63.64,98) .. (68.75,98) .. controls (73.86,98) and (78,102.92) .. (78,109) .. controls (78,115.08) and (73.86,120) .. (68.75,120) .. controls (63.64,120) and (59.5,115.08) .. (59.5,109) -- cycle ; \draw   (62.21,101.22) -- (75.29,116.78) ; \draw   (75.29,101.22) -- (62.21,116.78) ;
\draw    (68.75,98) .. controls (108.75,68) and (140.5,68) .. (177.5,75) ;
\draw    (68.75,120) .. controls (78.5,147) and (134.5,182) .. (174.5,152) ;
\draw   (221.21,101.1) -- (221.21,83.29) -- (246.75,83.29) ;
\draw   (246.75,83.29) -- (254,83.29) -- (254,101.1) ;
\draw   (230.91,101.08) -- (230.91,92.9) -- (242.12,92.9) ;
\draw   (242.12,92.9) -- (246.4,92.9) -- (246.39,101.1) ;
\draw   (55.5,237) .. controls (55.5,230.92) and (59.64,226) .. (64.75,226) .. controls (69.86,226) and (74,230.92) .. (74,237) .. controls (74,243.08) and (69.86,248) .. (64.75,248) .. controls (59.64,248) and (55.5,243.08) .. (55.5,237) -- cycle ; \draw   (58.21,229.22) -- (71.29,244.78) ; \draw   (71.29,229.22) -- (58.21,244.78) ;
\draw    (64.75,226) .. controls (104.75,196) and (136.5,196) .. (173.5,203) ;
\draw    (64.75,248) .. controls (74.5,275) and (130.5,310) .. (170.5,280) ;

\draw    (93.5,210) .. controls (95.65,211.25) and (96.22,212.86) .. (95.2,214.84) .. controls (94.45,217.19) and (95.23,218.74) .. (97.53,219.49) .. controls (99.78,219.92) and (100.71,221.24) .. (100.3,223.45) .. controls (100.31,225.96) and (101.45,227.13) .. (103.72,226.97) .. controls (106.01,226.64) and (107.38,227.59) .. (107.83,229.82) .. controls (108.64,232.07) and (110.24,232.72) .. (112.62,231.78) .. controls (114.45,230.48) and (116.03,230.74) .. (117.36,232.55) .. controls (119.07,234.22) and (120.71,234.13) .. (122.27,232.28) .. controls (123.5,230.31) and (125.17,229.87) .. (127.27,230.95) .. controls (129.41,231.85) and (130.96,231.11) .. (131.92,228.73) .. controls (132.39,226.46) and (133.68,225.58) .. (135.79,226.1) .. controls (138.28,226.19) and (139.55,225.08) .. (139.58,222.77) .. controls (139.47,220.48) and (140.58,219.27) .. (142.91,219.13) .. controls (145.32,218.77) and (146.38,217.36) .. (146.1,214.89) .. controls (145.51,212.75) and (146.42,211.3) .. (148.82,210.55) .. controls (151.07,209.94) and (151.83,208.52) .. (151.11,206.28) .. controls (150.32,204.07) and (151.03,202.51) .. (153.24,201.61) -- (153.5,201) ;
\draw   (230.69,245.04) -- (230.78,232.21) -- (249.85,232.21) ;
\draw   (249.85,232.21) -- (263.16,232.21) -- (263.07,245.04) ;

\draw   (240.15,245.04) -- (240.15,221.9) -- (316.71,221.9) ;
\draw   (316.71,221.9) -- (341.55,221.9) -- (341.49,245.04) ;

\draw   (281.73,245.04) -- (281.85,232.21) -- (300.53,232.21) ;
\draw   (300.53,232.21) -- (304.45,232.21) -- (304.32,245.04) ;

\draw   (324.7,245.04) -- (324.74,232.21) -- (340.65,232.21) ;
\draw   (340.65,232.21) -- (351.46,232.21) -- (351.43,245.04) ;

\draw   (62.5,366) .. controls (62.5,359.92) and (66.64,355) .. (71.75,355) .. controls (76.86,355) and (81,359.92) .. (81,366) .. controls (81,372.08) and (76.86,377) .. (71.75,377) .. controls (66.64,377) and (62.5,372.08) .. (62.5,366) -- cycle ; \draw   (65.21,358.22) -- (78.29,373.78) ; \draw   (78.29,358.22) -- (65.21,373.78) ;
\draw    (71.75,355) .. controls (111.75,325) and (143.5,325) .. (180.5,332) ;
\draw    (71.75,377) .. controls (81.5,404) and (137.5,439) .. (177.5,409) ;

\draw    (139.5,416) .. controls (141.59,417.15) and (142.13,418.78) .. (141.11,420.9) .. controls (141.64,423.85) and (142.3,423.72) .. (143.09,420.53) .. controls (141.51,418.8) and (141.56,417.1) .. (143.24,415.45) .. controls (144.89,413.74) and (144.87,412.11) .. (143.17,410.57) .. controls (141.46,408.93) and (141.41,407.18) .. (143.02,405.31) .. controls (144.63,403.72) and (144.58,402.16) .. (142.86,400.62) .. controls (141.13,399.02) and (141.07,397.34) .. (142.68,395.57) .. controls (144.29,393.74) and (144.23,391.96) .. (142.5,390.21) .. controls (140.79,388.88) and (140.74,387.24) .. (142.35,385.31) .. controls (143.96,383.36) and (143.92,381.69) .. (142.21,380.28) .. controls (140.52,378.85) and (140.48,377.14) .. (142.11,375.14) .. controls (143.75,373.63) and (143.73,371.9) .. (142.04,369.93) .. controls (140.36,368.44) and (140.35,366.95) .. (142.02,365.45) .. controls (143.69,363.47) and (143.7,361.73) .. (142.05,360.23) .. controls (140.4,358.72) and (140.43,357) .. (142.14,355.07) .. controls (143.85,353.66) and (143.9,351.97) .. (142.3,350) .. controls (140.7,348.49) and (140.79,346.85) .. (142.56,345.06) .. controls (144.31,343.8) and (144.43,342.21) .. (142.9,340.29) .. controls (141.41,338.37) and (141.58,336.64) .. (143.43,335.11) .. controls (145.29,333.7) and (145.51,332.08) .. (144.1,330.27) -- (144.5,328) ;
\draw   (231,361) -- (231,344.5) -- (253.85,344.52) ;
\draw   (253.85,344.52) -- (263.16,344.5) -- (263.07,361) ;

\draw   (283.5,361) -- (283.89,344.5) -- (346.46,344.5) ;
\draw   (346.46,344.5) -- (351.5,344.5) -- (351.15,361) ;

\draw   (240.5,361) -- (240.78,332.2) -- (304.48,332.16) ;
\draw   (304.48,332.16) -- (325.5,332.2) -- (325.29,361.01) ;

\draw   (305.5,361) -- (305.62,319.2) -- (332.85,319.22) ;
\draw   (332.85,319.22) -- (343.16,319.2) -- (343.07,361) ;

\draw   (116,66) -- (128.5,71.5) -- (116,77) ;
\draw   (124.98,295.54) -- (113.03,288.94) -- (125.97,284.59) ;
\draw   (161.92,196.11) -- (173.43,203.46) -- (160.23,206.98) ;
\draw   (113,196) -- (125.5,201.5) -- (113,207) ;
\draw   (73.35,212.86) -- (87.01,213.07) -- (77.63,223) ;
\draw   (128.07,166.46) -- (116.02,160.03) -- (128.9,155.49) ;
\draw   (177.56,415) -- (164.32,418.34) -- (170.8,406.32) ;
\draw   (110.92,419.39) -- (102.83,408.38) -- (116.41,409.85) ;
\draw   (157.53,322.64) -- (168.31,331.02) -- (154.85,333.31) ;
\draw   (100.38,332.79) -- (114.03,333.12) -- (104.56,342.97) ;

\draw (210,100) node [anchor=north west][inner sep=0.75pt]   [align=left] {$\displaystyle ( \psi \overline{\psi }) \psi \overline{\psi }$};
\draw (292,104) node [anchor=north west][inner sep=0.75pt]   [align=left] {$\displaystyle =-1$};
\draw (219,245) node [anchor=north west][inner sep=0.75pt]   [align=left] {$\displaystyle ( \psi \overline{\psi })\left(\overline{\psi }\cancel{A} \psi \right)\left(\overline{\psi }\cancel{A} \psi \right) \psi \overline{\psi }$};
\draw (220,361) node [anchor=north west][inner sep=0.75pt]   [align=left] {$\displaystyle ( \psi \overline{\psi })\left(\overline{\psi }\cancel{A} \psi \right)\left(\overline{\psi }\cancel{A} \psi \right) \psi \overline{\psi }$};
\draw (408,249) node [anchor=north west][inner sep=0.75pt]   [align=left] {$\displaystyle =( -1)^{2+1} =-1$};
\draw (405,354) node [anchor=north west][inner sep=0.75pt]   [align=left] {$\displaystyle =( -1)^{5+2} =-1$};

\end{tikzpicture}

\caption{\label{fig:fermion_contractions1} Fermion operator contractions for the Feynman diagrams of a single meson correlation function with a fermion-antifermion pair.}
\end{figure}

Next, we take a moment to consider possible sign changes from the fermion contractions. The different contractions are shown schematically in Figure \ref{fig:fermion_contractions1}. The $\slashed{A}$ terms do not play a part and should simply help to illustrate where the different $\psi$ terms come from. The spinor indices are dealt with separately. The calculation of the signs is according to the parity of the number of swaps of fermionic operators necessary to reach canonical form - which is the number of contraction line intersections in Figure \ref{fig:fermion_contractions1}, plus the number of contractions with $\bar{\psi}$ preceding $\psi$. 
We see that all diagrams get a factor of $-1$ relative to the convention that sets $Z_{\psi} = 1+O(g^2)$. This has no effect here, but we take it into account for good measure as it will be significant in the bi-meson case.

The next step is color index contractions. Since we are interested in mesons which are gauge singlets, we contract the color indices by $\delta_{ab}$ in all diagrams.
For diagrams $I,II$ we simply get:
\begin{align}
    & I \rightarrow \delta_{ab} (I) = \underbrace{\delta_{ab} \delta_{ad} \delta_{bc}}_{=\delta_{cd}} \delta_{il} \delta_{jk} \frac{\slashed{q_2}_{\alpha \Dot{\beta}}}{q_2^2} \frac{\slashed{q_1}_{\beta \Dot{\alpha}}}{q_1^2} =
    \delta_{cd} \delta_{il} \delta_{jk} \frac{\slashed{q_2}_{\alpha \Dot{\beta}}}{q_2^2} \frac{\slashed{q_1}_{\beta \Dot{\alpha}}}{q_1^2} \\
    & II \rightarrow \delta_{ab} (II) = -\delta_{ab} \delta_{bc} \delta_{ad} \frac{g^2 C_F M^{-2\epsilon}}{(4\pi)^2 \epsilon} \frac{\slashed{q_2}_{\alpha \Dot{\beta}}}{q_2^2} \frac{\slashed{q_1}_{\beta \Dot{\alpha}}}{q_1^2} \delta_{il} \delta_{jk}  =
    -\delta_{cd} \frac{g^2 C_F M^{-2\epsilon}}{(4\pi)^2 \epsilon} \frac{\slashed{q_2}_{\alpha \Dot{\beta}}}{q_2^2} \frac{\slashed{q_1}_{\beta \Dot{\alpha}}}{q_1^2} \delta_{il} \delta_{jk}.
\end{align}
For diagram $III$, we contract the color factor separately:

\begin{align}
    & \delta_{ab} (III) \propto \delta_{ab} t^A_{da} t^A_{bc} =
    (t^A t^A)_{dc} = C_F \delta_{dc} \label{eq:color contraction}\\
    & \Rightarrow III \rightarrow \delta_{ab} (III) = \delta_{cd} \frac{g^2 C_F M^{-2\epsilon}}{4(4\pi)^2 \epsilon} \delta_{il}\delta_{jk} (\gamma^{\nu} \gamma^{\mu} \slashed{q_2})_{\alpha \Dot{\beta}}(\slashed{q_1} \gamma_{\mu} \gamma_{\nu})_{\beta \Dot{\alpha}}.
\end{align}

\subsubsection{Contraction with different spinor structures}

We begin with the chiral scalars. To get the correlation functions for the scalar mesons $\Bar{\psi} (1 \pm \gamma^5) \psi$, we need to contract the spinor indices with $\left( 
1\pm \gamma^5 \right)_{\Dot{\alpha} \alpha}$:

\begin{itemize}
    \item For diagrams $I,II$:
    \begin{align}
        & \slashed{q_2}_{\alpha \Dot{\beta}} \slashed{q_1}_{\beta \Dot{\alpha}} \left( 1\pm \gamma^5 \right)_{\Dot{\alpha} \alpha} = \left[ \slashed{q_1} \left( 1\pm \gamma^5 \right) \slashed{q_2} \right]_{\beta \Dot{\beta}} \\
        \Rightarrow & I_s \equiv \left( 1\pm \gamma^5 \right)_{\Dot{\alpha} \alpha} \cdot (I) = \frac{\left[ \slashed{q_1} \left( 1\pm \gamma^5 \right) \slashed{q_2} \right]_{\beta \Dot{\beta}}}{q_1^2 q_2^2} \delta_{il} \delta_{jk} \delta_{cd}\\
        & II_s \equiv \left( 1\pm \gamma^5 \right)_{\Dot{\alpha} \alpha} \cdot (II) =
        -\frac{g^2 C_F M^{-2\epsilon}}{(4\pi)^2 \epsilon} \frac{\left[ \slashed{q_1} \left( 1\pm \gamma^5 \right) \slashed{q_2} \right]_{\beta \Dot{\beta}}}{q_1^2 q_2^2} \delta_{il} \delta_{jk} \delta_{cd}.
    \end{align}

    \item For diagram $III$:
    \begin{align}
        & \left( 1\pm \gamma^5 \right)_{\Dot{\alpha} \alpha} (\gamma^{\nu} \gamma^{\mu} \slashed{q_2})_{\alpha \Dot{\beta}}(\slashed{q_1} \gamma_{\mu} \gamma_{\nu})_{\beta \Dot{\alpha}} =
        \left[ \slashed{q_1} \gamma_{\mu} \gamma_{\nu} \left( 1\pm \gamma^5 \right) \gamma^{\nu} \gamma^{\mu} \slashed{q_2} \right]_{\beta \Dot{\beta}} =
        \delta_{\mu}^{\mu} \delta_{\nu}^{\nu} \left[ \slashed{q_1} \left( 1\pm \gamma^5 \right) \slashed{q_2}\right]_{\beta \Dot{\beta}} = 16 \left[ \slashed{q_1} \left( 1\pm \gamma^5 \right) \slashed{q_2} \right]_{\beta \Dot{\beta}} \nonumber\\
        \Rightarrow & III_s \equiv \left( 1\pm \gamma^5 \right)_{\Dot{\alpha} \alpha} \cdot (III) =
        \frac{4g^2 C_F M^{-2\epsilon}}{(4\pi)^2 \epsilon} \frac{\left[ \slashed{q_1} \left( 1\pm \gamma^5 \right) \slashed{q_2} \right]_{\beta \Dot{\beta}}}{q_1^2 q_2^2} \delta_{il}\delta_{jk} \delta_{cd}.
    \end{align}
    
\end{itemize}

The full one-loop correlation function is:
\begin{equation}
    \left< (\psi_{\alpha ai}\Bar{\psi}_{\Dot{\alpha} bj})(p) \psi_{\beta ck}(q_1) \Bar{\psi}_{\Dot{\beta}dl}(q_2) \right> (1\pm \gamma^5)_{\Dot{\alpha} \alpha} = I_s + 2\cdot II_s +III_s =
    \left( 1+\frac{2g^2 C_F M^{-2\epsilon}}{(4\pi)^2 \epsilon} \right) \frac{\left[ \slashed{q_1} \left( 1\pm \gamma^5 \right) \slashed{q_2} \right]_{\beta \Dot{\beta}}}{q_1^2 q_2^2} \delta_{il}\delta_{jk}.
\end{equation}

It can be expressed as a renormalized correlation function multiplied by operator renormalization functions as follows:
\begin{align}
    & \psi = \sqrt{Z_{\psi}} [\psi]; \indent \Bar{\psi} = \sqrt{Z_{\psi}} [\Bar{\psi}];
    \indent (\Bar{\psi} \psi) = Z_s [\Bar{\psi} \psi] \\
    & \left< (\psi_{\alpha ai}\Bar{\psi}_{\Dot{\alpha} bj})(p) \psi_{\beta ck}(q_1) \Bar{\psi}_{\Dot{\beta}dl}(q_2) \right> (1\pm \gamma^5)_{\Dot{\alpha} \alpha} =
    Z_s Z_{\psi} \left< \left[ (\psi_{\alpha ai}\Bar{\psi}_{\Dot{\alpha} bj}) \right] (p) \left[ \psi_{\beta ck} \right] (q_1) \left[ \Bar{\psi}_{\Dot{\beta}dl} \right] (q_2) \right> (1\pm \gamma^5)_{\Dot{\alpha} \alpha}.
\end{align}
We also denote: 
\begin{equation}
    Z_{tot}^s \equiv Z_s Z_{\psi} = Z_{tot,0}^s + \delta Z_{tot}^s; \indent 
    Z_s = Z_s^0 + \delta Z_s; \indent Z_{\psi} = 1+\delta Z_{\psi},
\end{equation}
and obtain:
\begin{align}
    & Z_{tot,0}^s + \delta Z_{tot}^s = \left( Z_s^0 + \delta Z_s \right) \left( 1+\delta Z_{\psi} \right) \\
    & \Rightarrow Z_s^0 = Z_{tot,0}^s; \indent
    \delta Z_s = \delta Z_{tot}^s - Z_s^0 \delta Z_{\psi} = 
    \delta Z_{tot}^s - Z_{tot,0}^s \delta Z_{\psi}.
\end{align}
At scale $M$ this is, using (\ref{fermion_anom_dim}):
\begin{align}
    & Z_{tot,0}^s = 1; \indent
    \delta Z_{tot}^s = \frac{2g^2 C_F M^{-2\epsilon}}{(4\pi)^2 \epsilon} \\
    & \Rightarrow Z_s^0 = 1; \indent
    \delta Z_s = \frac{2g^2 C_F M^{-2\epsilon}}{(4\pi)^2 \epsilon} + \frac{g^2 C_F M^{-2\epsilon}}{(4\pi)^2 \epsilon} =
    \frac{3g^2 C_F M^{-2\epsilon}}{(4\pi)^2 N \epsilon}.
\end{align}

Finally we can compute the anomalous dimension:
\begin{equation}
    \gamma_s = \frac{M}{Z_s} \frac{\partial Z_s}{\partial M} = 
    -2\epsilon M \cdot \left( \frac{3g^2 C_F M^{-1-2\epsilon}}{(4\pi)^2 \epsilon} \right) = 
    -\frac{6g^2 C_F M^{-2\epsilon}}{(4\pi)^2}
    \overset{\epsilon \rightarrow 0}{\rightarrow} -\frac{6g^2 C_F}{(4\pi)^2},
\end{equation}
in agreement with \cite{paper16}.

For the chiral vector mesons, we contract with $\left[ \left(1 \pm \gamma^5\right) \gamma^{\rho}\right]_{\Dot{\alpha} \alpha}$, with $\rho$ a spacetime index.
\begin{itemize}
    \item For diagrams $I,II$:
    \begin{equation}
        \slashed{q_2}_{\alpha \Dot{\beta}} \slashed{q_1}_{\beta \Dot{\alpha}} \left[ \left(1 \pm \gamma^5\right) \gamma^{\rho}\right]_{\Dot{\alpha} \alpha} =
        \left[ \slashed{q_1} \left(1 \pm \gamma^5\right) \gamma^{\rho} \slashed{q_2} \right]_{\beta \Dot{\beta}} \Rightarrow \left\{ \begin{matrix}
        I_v = \frac{\left[ \slashed{q_1} \left(1 \pm \gamma^5\right) \gamma^{\rho} \slashed{q_2} \right]_{\beta \Dot{\beta}}}{q_1^2 q_2^2} \delta_{il} \delta_{jk} \delta_{cd}\\
        II_v = -\frac{g^2 C_F M^{-2\epsilon}}{(4\pi)^2 \epsilon} \frac{\left[ \slashed{q_1} \left(1 \pm \gamma^5\right) \gamma^{\rho} \slashed{q_2} \right]_{\beta \Dot{\beta}}}{q_1^2 q_2^2} \delta_{il} \delta_{jk} \delta_{cd} \end{matrix} \right. ,
    \end{equation}
    \item For diagram $III$:
    \begin{align}
        & \left( \gamma^{\nu} \gamma^{\mu} \slashed{q_2} \right)_{\alpha \Dot{\beta}} \left( \slashed{q_1} \gamma_{\mu} \gamma_{\nu} \right)_{\beta \Dot{\alpha}} \left[ \left(1 \pm \gamma^5\right) \gamma^{\rho}\right]_{\Dot{\alpha} \alpha} =
        \left[ \slashed{q_1} \gamma_{\mu} \gamma_{\nu} \left(1 \pm \gamma^5\right) \gamma^{\rho} \gamma^{\nu} \gamma^{\mu} \slashed{q_2} \right]_{\beta \Dot{\beta}} = \nonumber \\
        & = \left[ \slashed{q_1} \left(1 \pm \gamma^5\right) \gamma_{\mu} \gamma_{\nu} \gamma^{\rho} \gamma^{\nu} \gamma^{\mu} \slashed{q_2} \right]_{\beta \Dot{\beta}} 
        = -2\left[ \slashed{q_1} \left(1 \pm \gamma^5\right) \gamma_{\mu} \gamma^{\rho} \gamma^{\mu} \slashed{q_2} \right]_{\beta \Dot{\beta}} 
        = 4 \left[ \slashed{q_1} \left(1 \pm \gamma^5\right) \gamma^{\rho} \slashed{q_2} \right]_{\beta \Dot{\beta}} \nonumber \\
        \Rightarrow & III_v = \frac{g^2 C_F M^{-2\epsilon}}{(4\pi)^2 \epsilon} \frac{\left[ \slashed{q_1} (1\pm \gamma^5) \gamma^{\rho} \slashed{q_2} \right]_{\beta \Dot{\beta}}}{q_1^2 q_2^2} \delta_{il} \delta_{jk} \delta_{cd},
    \end{align}
\end{itemize}
and the renormalization functions are:
\begin{equation}
    Z_{tot}^v = 1 - \frac{g^2 C_F M^{-2\epsilon}}{(4\pi)^2 \epsilon} \indent
    \Rightarrow \indent \delta Z_v = -\frac{g^2 C_F M^{-2\epsilon}}{(4\pi)^2 \epsilon} +\frac{g^2 C_F M^{-2\epsilon}}{(4\pi)^2 \epsilon} = 0
\end{equation}
as expected from a Noether current associated with a conserved symmetry.
The anomalous dimension also vanishes accordingly:
\begin{equation}
    \gamma_v = 0.
\end{equation}
 
For the tensor mesons, we contract with $(\gamma^{\rho \sigma})_{\Dot{\alpha} \alpha} \equiv \frac{1}{2} \left[ \gamma^{\rho}, \gamma^{\sigma} \right]_{\Dot{\alpha} \alpha}$.
\begin{itemize}
    \item  For diagrams $I,II$:
    \begin{equation}
        \slashed{q_2}_{\alpha \Dot{\beta}} \slashed{q_1}_{\beta \Dot{\alpha}} (\gamma^{\rho \sigma})_{\Dot{\alpha} \alpha} =
        \left( \slashed{q_1} \gamma^{\rho \sigma} \slashed{q_2} \right)_{\beta \Dot{\beta}} \Rightarrow \left\{ \begin{matrix}
        I_t = \frac{\left( \slashed{q_1} \gamma^{\rho \sigma} \slashed{q_2} \right)_{\beta \Dot{\beta}}}{q_1^2 q_2^2} \delta_{il} \delta_{jk} \delta_{cd} \\
        II_t = -\frac{g^2 C_F M^{-2\epsilon}}{(4\pi)^2 \epsilon} \frac{\left( \slashed{q_1} \gamma^{\rho \sigma} \slashed{q_2} \right)_{\beta \Dot{\beta}}}{q_1^2 q_2^2} \delta_{il} \delta_{jk} \delta_{cd} \end{matrix} \right. ,
    \end{equation}
    \item For diagram $III$:
    \begin{equation}
        \left( \gamma^{\nu} \gamma^{\mu} \slashed{q_2} \right)_{\alpha \Dot{\beta}} \left( \slashed{q_1} \gamma_{\mu} \gamma_{\nu} \right)_{\beta \Dot{\alpha}} (\gamma^{\rho \sigma})_{\Dot{\alpha} \alpha} = 
        \frac{1}{2} \left( \slashed{q_1} \gamma_{\mu} \underbrace{\gamma_{\nu} \left[ \gamma^{\rho}, \gamma^{\sigma} \right] \gamma^{\nu}}_{=4\eta^{[\rho \sigma]} =0} \gamma^{\mu} \slashed{q_2} \right)_{\beta \Dot{\beta}} = 0
        \Rightarrow III_t = 0,
    \end{equation}

\end{itemize}
and the renormalization functions and anomalous dimension are accordingly:
\begin{align}
    & Z_{tot}^{t} = 1-\frac{2g^2 C_F M^{-2\epsilon}}{(4\pi)^2 \epsilon} \indent \Rightarrow
    Z_t = 1- \frac{g^2 C_F M^{-2\epsilon}}{(4\pi)^2 \epsilon}
    \indent \Rightarrow \gamma_t  = \frac{2g^2 C_F}{(4\pi)^2},
\end{align}
agreeing with \cite{paper16} as well.

\subsection{Mixed mesons}

\subsubsection{Diagrams}

We consider the mixed meson $\phi^{*}\psi$, and the correlation function
\begin{equation}
\left\langle \left(\phi_{ia'}^{*}\psi_{jb'\alpha}\right)(p)\phi_{i'a}(p_{1})\overline{\psi}_{j'b\dot{\alpha}}(-p_{2})\right\rangle \delta_{a'b'}.
\end{equation}
Again we assume from the start that the flavor indices give a non-vanishing correlation function: $i=i',j=j'$.
We have 4 contributing diagrams up to one-loop order, shown in Figure \ref{fig:single mixed meson diagrams}. They are the tree-level, scalar/fermion propagator
correction, and the gluon exchange. There are also scalar propagator
corrections from the $\phi^{4}$ interactions and from the "seagull" vertex, but they are independent
of the momenta and therefore do not contribute to the field renormalizations
and to the anomalous dimensions.
The Feynman rule for the scalar gauge vertex is given in Figure \ref{fig:Feynman rule scalar QCD}. 

The tree level diagram is equal to:
\begin{equation}
\label{eq:mixed meson tree level}
\frac{i}{p_{1}^{2}}\frac{i\cancel{p_{2}}_{\alpha\dot{\alpha}}}{p_{2}^{2}}\delta_{ab}=-\frac{\cancel{p_{2}}_{\alpha\dot{\alpha}}}{p_{1}^{2}p_{2}^{2}}\delta_{ab}.
\end{equation}
The diagram with a fermion propagator correction $\delta D_{\beta\dot{\beta}}(p_{2})$ is
\begin{align}
 & -\frac{\cancel{p_{2}}_{\alpha\beta}}{p_{1}^{2}p_{2}^{2}}\delta D_{\beta\dot{\beta}}(p_{2})\frac{i\cancel{p_{2}}_{\dot{\beta}\dot{\alpha}}}{p_{2}^{2}}=
 -\frac{\cancel{p_{2}}_{\alpha\beta}}{p_{1}^{2}p_{2}^{2}}\frac{ig^{2}C_F}{\left(4\pi\right)^{2}\epsilon}\left(p_{2E}^{2}\right)^{-\epsilon}\cancel{p_{2}}_{\beta\dot{\beta}}\frac{i\cancel{p_{2}}_{\dot{\beta}\dot{\alpha}}}{p_{2}^{2}}
 \approx\frac{g^{2}C_{F}}{\left(4\pi\right)^{2}\epsilon}\frac{\left(p_{E}^{2}\right)^{-\epsilon}}{p_{1}^{2}p_{2}^{2}}\cancel{p_{2}}_{\alpha\dot{\alpha}}\delta_{ab};
\end{align}
(color factors are trivial and only included in the end).

\begin{figure} 
\centering

\begin{tikzpicture}[x=0.75pt,y=0.75pt,yscale=-1,xscale=1]

\draw   (18,106.75) .. controls (18,103.3) and (21.13,100.5) .. (25,100.5) .. controls (28.87,100.5) and (32,103.3) .. (32,106.75) .. controls (32,110.2) and (28.87,113) .. (25,113) .. controls (21.13,113) and (18,110.2) .. (18,106.75) -- cycle ; \draw   (20.05,102.33) -- (29.95,111.17) ; \draw   (29.95,102.33) -- (20.05,111.17) ;
\draw  [dash pattern={on 4.5pt off 4.5pt}]  (25,100.5) .. controls (33.87,80.5) and (64.87,72.5) .. (102.87,78.5) ;
\draw [color={rgb, 255:red, 0; green, 0; blue, 0 }  ,draw opacity=1 ]   (25,113) .. controls (38.87,133.75) and (49.87,138.5) .. (107.87,139.5) ;
\draw   (61.72,85.12) -- (49.23,81.84) -- (57.82,72.2) ;
\draw   (46.48,124.52) -- (55.89,133.37) -- (43.75,137.74) ;

\draw   (149,107.75) .. controls (149,104.3) and (152.13,101.5) .. (156,101.5) .. controls (159.87,101.5) and (163,104.3) .. (163,107.75) .. controls (163,111.2) and (159.87,114) .. (156,114) .. controls (152.13,114) and (149,111.2) .. (149,107.75) -- cycle ; \draw   (151.05,103.33) -- (160.95,112.17) ; \draw   (160.95,103.33) -- (151.05,112.17) ;
\draw  [dash pattern={on 4.5pt off 4.5pt}]  (156,101.5) .. controls (164.87,81.5) and (195.87,73.5) .. (233.87,79.5) ;
\draw [color={rgb, 255:red, 0; green, 0; blue, 0 }  ,draw opacity=1 ]   (156,114) .. controls (169.87,134.75) and (180.87,139.5) .. (238.87,140.5) ;
\draw   (192.72,86.12) -- (180.23,82.84) -- (188.82,73.2) ;
\draw   (190.48,130.52) -- (199.89,139.37) -- (187.75,143.74) ;
\draw    (177.87,132.83) .. controls (177.88,130.25) and (179.04,128.95) .. (181.35,128.92) .. controls (183.72,128.94) and (184.94,127.81) .. (185.03,125.52) .. controls (185.26,123.23) and (186.63,122.27) .. (189.14,122.62) .. controls (191.19,123.41) and (192.68,122.73) .. (193.61,120.56) .. controls (194.9,118.5) and (196.48,118.2) .. (198.35,119.65) .. controls (199.83,121.36) and (201.47,121.55) .. (203.27,120.21) .. controls (205.5,119.26) and (207.02,119.94) .. (207.81,122.26) .. controls (208.16,124.57) and (209.47,125.65) .. (211.76,125.48) .. controls (214.2,125.71) and (215.27,127) .. (214.96,129.35) .. controls (214.47,131.63) and (215.32,133.01) .. (217.52,133.49) .. controls (219.84,134.4) and (220.58,135.98) .. (219.75,138.21) -- (219.87,138.5) ;
\draw   (260,109.75) .. controls (260,106.3) and (263.13,103.5) .. (267,103.5) .. controls (270.87,103.5) and (274,106.3) .. (274,109.75) .. controls (274,113.2) and (270.87,116) .. (267,116) .. controls (263.13,116) and (260,113.2) .. (260,109.75) -- cycle ; \draw   (262.05,105.33) -- (271.95,114.17) ; \draw   (271.95,105.33) -- (262.05,114.17) ;
\draw  [dash pattern={on 4.5pt off 4.5pt}]  (267,103.5) .. controls (275.87,83.5) and (306.87,75.5) .. (344.87,81.5) ;
\draw [color={rgb, 255:red, 0; green, 0; blue, 0 }  ,draw opacity=1 ]   (267,116) .. controls (280.87,136.75) and (291.87,141.5) .. (349.87,142.5) ;
\draw   (310.72,86.12) -- (298.23,82.84) -- (306.82,73.2) ;
\draw   (288.48,127.52) -- (297.89,136.37) -- (285.75,140.74) ;
\draw    (287.87,85.83) .. controls (290.18,86.12) and (291.26,87.42) .. (291.11,89.71) .. controls (291.38,92.18) and (292.67,93.21) .. (294.97,92.8) .. controls (297.23,92.13) and (298.76,92.84) .. (299.57,94.93) .. controls (301,96.94) and (302.67,97.17) .. (304.59,95.62) .. controls (305.84,93.82) and (307.46,93.51) .. (309.44,94.69) .. controls (311.77,95.45) and (313.22,94.66) .. (313.78,92.33) .. controls (313.95,90) and (315.18,88.83) .. (317.49,88.81) .. controls (319.84,88.49) and (320.82,87.12) .. (320.41,84.7) .. controls (319.72,82.49) and (320.45,81.08) .. (322.59,80.46) -- (322.87,79.83) ;
\draw   (373,109.75) .. controls (373,106.3) and (376.13,103.5) .. (380,103.5) .. controls (383.87,103.5) and (387,106.3) .. (387,109.75) .. controls (387,113.2) and (383.87,116) .. (380,116) .. controls (376.13,116) and (373,113.2) .. (373,109.75) -- cycle ; \draw   (375.05,105.33) -- (384.95,114.17) ; \draw   (384.95,105.33) -- (375.05,114.17) ;
\draw  [dash pattern={on 4.5pt off 4.5pt}]  (380,103.5) .. controls (388.87,83.5) and (419.87,75.5) .. (457.87,81.5) ;
\draw [color={rgb, 255:red, 0; green, 0; blue, 0 }  ,draw opacity=1 ]   (380,116) .. controls (393.87,136.75) and (404.87,141.5) .. (462.87,142.5) ;
\draw   (416.72,88.12) -- (404.23,84.84) -- (412.82,75.2) ;
\draw   (401.48,127.52) -- (410.89,136.37) -- (398.75,140.74) ;
\draw    (431,79) .. controls (432.71,80.62) and (432.76,82.29) .. (431.15,84) .. controls (429.54,85.71) and (429.59,87.38) .. (431.3,89) .. controls (433.02,90.61) and (433.07,92.27) .. (431.46,93.99) .. controls (429.85,95.7) and (429.9,97.37) .. (431.61,98.99) .. controls (433.32,100.61) and (433.37,102.28) .. (431.76,103.99) .. controls (430.15,105.7) and (430.2,107.37) .. (431.91,108.99) .. controls (433.62,110.6) and (433.67,112.27) .. (432.06,113.98) .. controls (430.45,115.69) and (430.5,117.36) .. (432.21,118.98) .. controls (433.93,120.59) and (433.98,122.26) .. (432.37,123.98) .. controls (430.76,125.69) and (430.81,127.36) .. (432.52,128.98) .. controls (434.23,130.59) and (434.28,132.26) .. (432.67,133.97) .. controls (431.06,135.68) and (431.11,137.35) .. (432.82,138.97) -- (432.87,140.5) -- (432.87,140.5) ;

\draw   (221.48,131.52) -- (230.89,140.37) -- (218.75,144.74) ;
\draw   (164.19,117.64) -- (170.24,129.04) -- (157.33,129.27) ;
\draw   (342.73,87.86) -- (333.09,79.28) -- (345.1,74.57) ;
\draw   (286.09,93.67) -- (273.38,95.89) -- (277.15,83.55) ;
\draw   (453.73,87.86) -- (444.09,79.28) -- (456.1,74.57) ;
\draw   (443.44,135.16) -- (453.99,142.6) -- (442.58,148.64) ;

\draw (26.87,84) node  [font=\footnotesize,rotate=-326.77] [align=left] {\begin{minipage}[lt]{21.76pt}\setlength\topsep{0pt}
$\displaystyle a'$
\end{minipage}};
\draw (24.36,126.7) node  [font=\footnotesize,rotate=-32.16] [align=left] {\begin{minipage}[lt]{21.76pt}\setlength\topsep{0pt}
$\displaystyle b',\alpha $
\end{minipage}};
\draw (119.87,79) node  [font=\footnotesize,rotate=-6.63] [align=left] {\begin{minipage}[lt]{21.76pt}\setlength\topsep{0pt}
$\displaystyle a$
\end{minipage}};
\draw (122.6,139.59) node  [font=\footnotesize,rotate=-0.33] [align=left] {\begin{minipage}[lt]{21.76pt}\setlength\topsep{0pt}
$\displaystyle b,\dot{\alpha }$
\end{minipage}};

\end{tikzpicture}

\caption{\label{fig:single mixed meson diagrams}
 Single mixed meson diagrams. Only diagrams relevant to the anomalous dimension up to one-loop are presented.}

\end{figure}
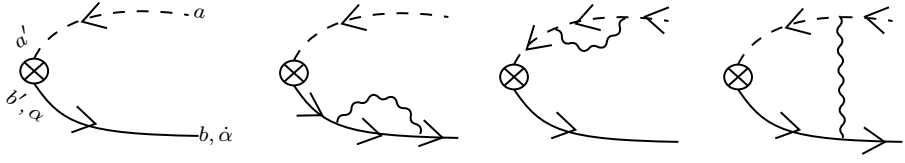

\begin{figure}
    \centering
    \begin{tikzpicture}[x=0.75pt,y=0.75pt,yscale=-1,xscale=1]
    
    \draw  [dash pattern={on 4.5pt off 4.5pt}]  (33,85) -- (170,86.17) ;
    \draw    (101.5,85.58) .. controls (99.76,83.99) and (99.69,82.33) .. (101.28,80.59) .. controls (102.87,78.85) and (102.8,77.18) .. (101.06,75.59) .. controls (99.33,74) and (99.26,72.33) .. (100.85,70.6) .. controls (102.44,68.86) and (102.37,67.19) .. (100.63,65.6) .. controls (98.89,64.01) and (98.82,62.35) .. (100.41,60.61) .. controls (102,58.87) and (101.93,57.2) .. (100.19,55.61) -- (100,51.17) -- (100,51.17) ;
    \draw   (69,81) -- (79,85.58) -- (69,90.17) ;
    \draw   (142,81) -- (152,85.58) -- (142,90.17) ;
    
    \draw (43,92) node [anchor=north west][inner sep=0.75pt]   [align=left] {$\displaystyle a,i,p$};
    \draw (122,93) node [anchor=north west][inner sep=0.75pt]   [align=left] {$\displaystyle b,j,p'$};
    \draw (63,40) node [anchor=north west][inner sep=0.75pt]   [align=left] {$\displaystyle A,\mu $};
    \draw (183,68) node [anchor=north west][inner sep=0.75pt]   [align=left] {$\displaystyle =ig( p+p')_{\mu }\left( t^{A}\right)_{ba}$};

    \end{tikzpicture}
    \caption{\label{fig:Feynman rule scalar QCD} Feynman rule for scalar QCD, to leading order in the gauge coupling.}    
\end{figure}

The scalar inverse propagator correction is at tree level $-ip^{2}=ip_{E}^{2}$.
At one-loop level, it is:
\begin{equation}
\frac{2ig^{2}C_{F}}{\left(4\pi\right)^{2}\epsilon}\delta_{ab}\left(p_{E}^{2}\right)^{1-\epsilon},
\end{equation}
and the scalar renormalization function is 
\begin{equation}
Z_{\phi}=1+\frac{2g^{2}C_{F}}{\left(4\pi\right)^{2}\epsilon}\left(p_{E}^{2}\right)^{-\epsilon}.
\end{equation}
The value of the corresponding diagram will be that of the tree level multiplied
by the ratio of the propagator corrections, giving:
\begin{equation}
-\frac{\cancel{p_{2}}_{\alpha\dot{\alpha}}}{p_{1}^{2}p_{2}^{2}}\cdot\frac{2g^{2}C_{F}}{\left(4\pi\right)^{2}\epsilon}\delta_{ab}\left(p_{E}^{2}\right)^{-\epsilon}.
\end{equation}

The gluon exchange diagram is equal to:
\begin{equation}
-\frac{g^{2}}{\left(4\pi\right)^{2}p_{1}^{2}p_{2}^{2}\epsilon}\left(t^{A}\right)_{bb'}\left(t^{A}\right)_{a'a}\cancel{p_{2}}_{\alpha\dot{\alpha}}\left(p_{E}^{2}\right)^{-\epsilon}.
\end{equation}

In the tree level diagrams and the ones with propagator corrections,
the color contraction is trivial and amounts to $\delta_{ab}$ (and
was implicit in the diagram calculations). In the diagram with the
gluon exchange it gives $C_{F}\delta_{ab}$, similarly to (\ref{eq:color contraction}).
Since there is only one pair of fermionic operators, and they are
ordered in the standard way, there are no sign changes from their
anticommutation relations.

\subsubsection{Resummation and renormalization function}

All the amplitudes have the same spinor form $\cancel{p_{2}}_{\alpha\dot{\alpha}}$,
and so there is no need to project onto different representations. Each diagram contributes once, so the total (relevant
part of the) correlation function is:
\begin{equation}
 \left\langle \left(\phi_{ia'}^{*}\psi_{jb'\alpha}\right)(p)\phi_{i'a}(p_{1})\overline{\psi}_{j'b\dot{\alpha}}(-p_{2})\right\rangle \delta_{a'b'}
=-\frac{\cancel{p_{2}}_{\alpha\dot{\alpha}}}{p_{1}^{2}p_{2}^{2}}\delta_{ab}\left(1+\frac{2g^{2}C_{F}}{\left(4\pi\right)^{2}\epsilon}\left(p_{E}^{2}\right)^{-\epsilon}\right),
\end{equation}
and so the total renormalization function, after setting a renormalization
scale $M$, is
\begin{equation}
Z_{\phi^{*}\psi}^{tot}=1+\frac{2g^{2}C_{F}}{\left(4\pi\right)^{2}\epsilon}M^{-2\epsilon}.
\end{equation}
We want to extract the wavefunction renormalization of the
mixed meson operator; so we decompose $Z_{\phi^{*}\psi}^{tot}$ as
$Z_{\phi^{*}\psi}^{tot}=Z_{\phi^{*}\psi}Z_{\psi}^{1/2}Z_{\phi}^{1/2}$
and get:
\begin{equation}
\delta Z_{\phi^{*}\psi}=\delta Z_{\phi^{*}\psi}^{tot}-\frac{1}{2}\delta Z_{\psi}-\frac{1}{2}\delta Z_{\phi}
=\frac{3g^{2}C_{F}}{2\left(4\pi\right)^{2}\epsilon}M^{-2\epsilon},
\end{equation}
and the anomalous dimension of the mixed meson is thus
\begin{equation}
\gamma_{\phi^{*}\psi}=\frac{\partial\log Z_{\phi^{*}\psi}}{\partial\log M}=\left(-2\epsilon\right)\cdot\left(\frac{3g^{2}C_{F}}{2\left(4\pi\right)^{2}\epsilon}\right)=-\frac{3g^{2}C_{F}}{\left(4\pi\right)^{2}}.
\end{equation}

\section{Anomalous dimensions of double mesons}

In this section, we keep contributions up to the leading order in $\frac{1}{N}$ contributing to the difference between the two sides of (\ref{eq:CCC}). 

\subsection{Scalar double mesons}

This subsection is a review of the discussion in \cite{paper7}, with some elaboration regarding symmetry group indices.
The Lagrangian is (\ref{eq:5}), and as our basic operator, we choose scalar mesons of the type $\phi^* \phi$, which is in the adjoint representation of $SU(N_s)$.
One can separate the loop level diagrams into ones involving only interactions within each meson (intra-meson interaction diagrams), and ones that involve interactions between the mesons (inter-meson interaction diagrams). The form of the $1/N$ expansion in the 't Hooft limit (in which meson operators factorize at large $N$ \cite{paper3}) implies that the inter-meson diagrams start contributing at order $1/N$ compared to the intra-meson diagrams (this is true for all types of mesons that we discuss).
In this case (unlike the fermion mesons
we discuss later) there is no mixing between operators - and therefore no need to diagonalize the renormalization matrix. Then the intra-meson diagrams will simply cancel out when computing the difference $\gamma_{(\phi^* \phi)^n}-n\gamma_{\phi^* \phi}$, and we only need to consider corrections from inter-meson diagrams. 
At one loop order, there are two types of these diagrams - gluon exchanges and $\phi^4$ interactions.
We consider in the following the same correlator as in \cite{paper7}, but begin with flavor indices explicit and generic - namely:
\begin{equation}
    \left< (\phi^*_{i a'} \phi_{b' j} \phi^*_{k c'} \phi_{d' l})(p){\phi^*_{i' a}(-p_1) \phi_{b j'}(q_1) \phi^*_{k' c}(-p_2) \phi_{d l'}(q_2)} \right> \delta_{a' b'} \delta_{c' d'},
\end{equation}
 with $a,b,c,d$ color indices and $i,j,k,l$ flavor indices. 
The tree-level amplitude is immediately found to be $\frac{1}{p_1^2 p_2^2 q_1^2 q_2^2}$ times a sum of products of $\delta$ symbols.

\paragraph{Gluon exchange diagrams} 

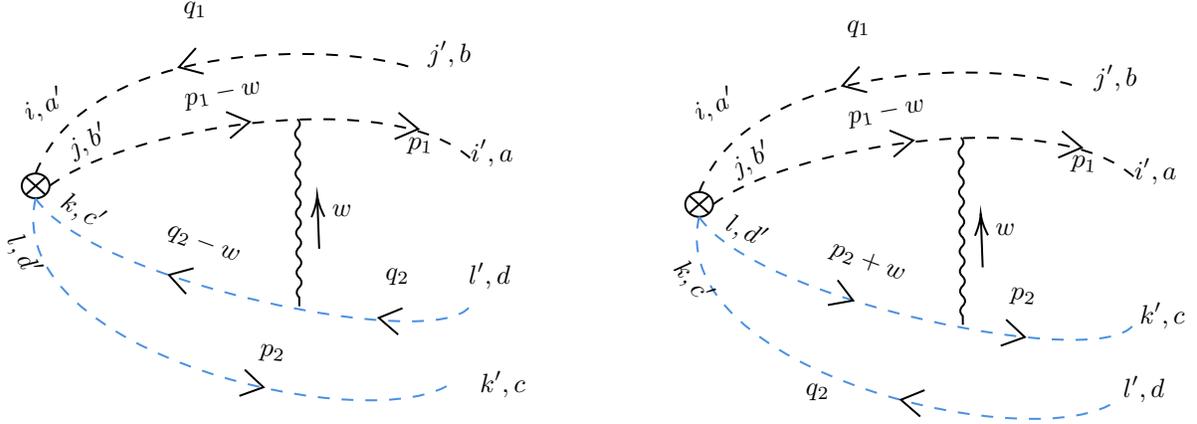
\begin{figure}
    \centering
    \begin{tikzpicture}[x=0.75pt,y=0.75pt,yscale=-1,xscale=1]
    
    \draw   (18,106.75) .. controls (18,103.3) and (21.13,100.5) .. (25,100.5) .. controls (28.87,100.5) and (32,103.3) .. (32,106.75) .. controls (32,110.2) and (28.87,113) .. (25,113) .. controls (21.13,113) and (18,110.2) .. (18,106.75) -- cycle ; \draw   (20.05,102.33) -- (29.95,111.17) ; \draw   (29.95,102.33) -- (20.05,111.17) ;
    \draw  [dash pattern={on 4.5pt off 4.5pt}]  (25,100.5) .. controls (52,31.5) and (181,34.5) .. (216,47.5) ;
    \draw [color={rgb, 255:red, 74; green, 144; blue, 226 }  ,draw opacity=1 ] [dash pattern={on 4.5pt off 4.5pt}]  (25,113) .. controls (7,191.5) and (196,235.5) .. (236,205.5) ;
    \draw  [dash pattern={on 4.5pt off 4.5pt}]  (32,106.75) .. controls (72,76.75) and (205,55.5) .. (245,93.5) ;
    \draw [color={rgb, 255:red, 74; green, 144; blue, 226 }  ,draw opacity=1 ] [dash pattern={on 4.5pt off 4.5pt}]  (25,113) .. controls (52,154.75) and (227,191.5) .. (244,167.5) ;
    \draw   (100.88,161.27) -- (92.22,151.7) -- (104.68,148.32) ;
    \draw   (121.03,69.42) -- (132.94,74.41) -- (123.1,82.76) ;
    \draw   (109.72,50.12) -- (97.23,46.84) -- (105.82,37.2) ;
    \draw   (130.48,199.52) -- (139.89,208.37) -- (127.75,212.74) ;
    \draw   (207.83,181.8) -- (198.07,173.36) -- (210.02,168.48) ;
    \draw   (208.02,69.78) -- (217.94,78.04) -- (206.09,83.14) ;
    \draw    (157,73) .. controls (158.68,74.65) and (158.7,76.32) .. (157.05,78) .. controls (155.4,79.69) and (155.42,81.35) .. (157.11,83) .. controls (158.79,84.65) and (158.81,86.32) .. (157.16,88) .. controls (155.51,89.68) and (155.53,91.35) .. (157.21,93) .. controls (158.89,94.65) and (158.91,96.32) .. (157.26,98) .. controls (155.61,99.69) and (155.63,101.35) .. (157.32,103) .. controls (159,104.65) and (159.02,106.32) .. (157.37,108) .. controls (155.72,109.68) and (155.74,111.35) .. (157.42,113) .. controls (159.11,114.65) and (159.13,116.31) .. (157.48,118) .. controls (155.83,119.68) and (155.85,121.35) .. (157.53,123) .. controls (159.21,124.65) and (159.23,126.32) .. (157.58,128) .. controls (155.93,129.68) and (155.95,131.35) .. (157.63,133) .. controls (159.32,134.65) and (159.34,136.31) .. (157.69,138) .. controls (156.04,139.68) and (156.06,141.35) .. (157.74,143) .. controls (159.42,144.65) and (159.44,146.32) .. (157.79,148) .. controls (156.14,149.69) and (156.16,151.35) .. (157.85,153) .. controls (159.53,154.65) and (159.55,156.32) .. (157.9,158) .. controls (156.25,159.68) and (156.27,161.34) .. (157.95,162.99) -- (158,167.5) -- (158,167.5) ;
    \draw    (168,138.5) -- (167.07,113.5) ;
    \draw [shift={(167,111.5)}, rotate = 87.88] [color={rgb, 255:red, 0; green, 0; blue, 0 }  ][line width=0.75]    (10.93,-3.29) .. controls (6.95,-1.4) and (3.31,-0.3) .. (0,0) .. controls (3.31,0.3) and (6.95,1.4) .. (10.93,3.29)   ;
    \draw   (352.64,116.11) .. controls (352.64,112.66) and (355.78,109.86) .. (359.64,109.86) .. controls (363.51,109.86) and (366.64,112.66) .. (366.64,116.11) .. controls (366.64,119.56) and (363.51,122.36) .. (359.64,122.36) .. controls (355.78,122.36) and (352.64,119.56) .. (352.64,116.11) -- cycle ; \draw   (354.69,111.69) -- (364.59,120.53) ; \draw   (364.59,111.69) -- (354.69,120.53) ;
    \draw  [dash pattern={on 4.5pt off 4.5pt}]  (359.64,109.86) .. controls (386.64,40.86) and (515.64,43.86) .. (550.64,56.86) ;
    \draw [color={rgb, 255:red, 74; green, 144; blue, 226 }  ,draw opacity=1 ] [dash pattern={on 4.5pt off 4.5pt}]  (359.64,122.36) .. controls (341.64,200.86) and (530.64,244.86) .. (570.64,214.86) ;
    \draw  [dash pattern={on 4.5pt off 4.5pt}]  (366.64,116.11) .. controls (406.64,86.11) and (539.64,64.86) .. (579.64,102.86) ;
    \draw [color={rgb, 255:red, 74; green, 144; blue, 226 }  ,draw opacity=1 ] [dash pattern={on 4.5pt off 4.5pt}]  (359.64,122.36) .. controls (386.64,164.11) and (561.64,200.86) .. (578.64,176.86) ;
    \draw   (429.48,154.32) -- (437.26,164.61) -- (424.56,166.89) ;
    \draw   (455.67,78.78) -- (467.58,83.77) -- (457.74,92.12) ;
    \draw   (444.36,59.48) -- (431.88,56.2) -- (440.46,46.56) ;
    \draw   (514.46,174.22) -- (523.86,183.06) -- (511.72,187.44) ;
    \draw   (471.14,223.16) -- (461.38,214.71) -- (473.33,209.84) ;
    \draw   (542.67,79.14) -- (552.58,87.4) -- (540.73,92.5) ;
    \draw    (491.64,82.36) .. controls (493.33,84.01) and (493.34,85.68) .. (491.69,87.36) .. controls (490.04,89.05) and (490.06,90.71) .. (491.75,92.36) .. controls (493.43,94.01) and (493.45,95.68) .. (491.8,97.36) .. controls (490.15,99.04) and (490.17,100.71) .. (491.85,102.36) .. controls (493.54,104.01) and (493.56,105.67) .. (491.91,107.36) .. controls (490.26,109.04) and (490.28,110.71) .. (491.96,112.36) .. controls (493.64,114.01) and (493.66,115.68) .. (492.01,117.36) .. controls (490.36,119.05) and (490.38,120.71) .. (492.07,122.36) .. controls (493.75,124.01) and (493.77,125.68) .. (492.12,127.36) .. controls (490.47,129.04) and (490.49,130.71) .. (492.17,132.36) .. controls (493.85,134.01) and (493.87,135.68) .. (492.22,137.36) .. controls (490.57,139.05) and (490.59,140.71) .. (492.28,142.35) .. controls (493.96,144) and (493.98,145.67) .. (492.33,147.35) .. controls (490.68,149.03) and (490.7,150.7) .. (492.38,152.35) .. controls (494.07,154) and (494.09,155.66) .. (492.44,157.35) .. controls (490.79,159.03) and (490.81,160.7) .. (492.49,162.35) .. controls (494.17,164) and (494.19,165.67) .. (492.54,167.35) .. controls (490.89,169.03) and (490.91,170.7) .. (492.59,172.35) -- (492.64,176.86) -- (492.64,176.86) ;
    \draw    (502.64,147.86) -- (501.72,122.86) ;
    \draw [shift={(501.64,120.86)}, rotate = 87.88] [color={rgb, 255:red, 0; green, 0; blue, 0 }  ][line width=0.75]    (10.93,-3.29) .. controls (6.95,-1.4) and (3.31,-0.3) .. (0,0) .. controls (3.31,0.3) and (6.95,1.4) .. (10.93,3.29)   ;
    
    \draw (15.06,63.48) node [anchor=north west][inner sep=0.75pt]  [rotate=-336.36] [align=left] {$\displaystyle i ,a'$};
    \draw (97.72,57.16) node [anchor=north west][inner sep=0.75pt]  [rotate=-350.14] [align=left] {$\displaystyle p_{1} -w$};
    \draw (243,82) node [anchor=north west][inner sep=0.75pt]   [align=left] {$\displaystyle i' ,a$};
    \draw (242,144) node [anchor=north west][inner sep=0.75pt]   [align=left] {$\displaystyle l' ,d$};
    \draw (200,147) node [anchor=north west][inner sep=0.75pt]   [align=left] {$\displaystyle q_{2}$};
    \draw (211.81,80.29) node [anchor=north west][inner sep=0.75pt]  [rotate=-4.73] [align=left] {$\displaystyle p_{1}$};
    \draw (20.54,127.26) node [anchor=north west][inner sep=0.75pt]  [rotate=-62.2] [align=left] {$\displaystyle l ,d'$};
    \draw (173,115) node [anchor=north west][inner sep=0.75pt]   [align=left] {$\displaystyle w$};
    \draw (92.7,122.49) node [anchor=north west][inner sep=0.75pt]  [rotate=-20.13] [align=left] {$\displaystyle q_{2} -w$};
    \draw (42.11,106.05) node [anchor=north west][inner sep=0.75pt]  [rotate=-31.63] [align=left] {$\displaystyle k ,c'$};
    \draw (37.05,82.23) node [anchor=north west][inner sep=0.75pt]  [rotate=-333.14] [align=left] {$\displaystyle j ,b'$};
    \draw (137.15,184.96) node [anchor=north west][inner sep=0.75pt]  [rotate=-4.73] [align=left] {$\displaystyle p_{2}$};
    \draw (98,12.33) node [anchor=north west][inner sep=0.75pt]   [align=left] {$\displaystyle q_{1}$};
    \draw (353.04,63.5) node [anchor=north west][inner sep=0.75pt]  [rotate=-336.36] [align=left] {$\displaystyle i ,a'$};
    \draw (432.36,66.52) node [anchor=north west][inner sep=0.75pt]  [rotate=-350.14] [align=left] {$\displaystyle p_{1} -w$};
    \draw (577.64,91.36) node [anchor=north west][inner sep=0.75pt]   [align=left] {$\displaystyle i' ,a$};
    \draw (571.98,201.36) node [anchor=north west][inner sep=0.75pt]   [align=left] {$\displaystyle l' ,d$};
    \draw (411.98,205.02) node [anchor=north west][inner sep=0.75pt]   [align=left] {$\displaystyle q_{2}$};
    \draw (546.46,89.65) node [anchor=north west][inner sep=0.75pt]  [rotate=-4.73] [align=left] {$\displaystyle p_{1}$};
    \draw (376.52,117.02) node [anchor=north west][inner sep=0.75pt]  [rotate=-26.82] [align=left] {$\displaystyle l ,d'$};
    \draw (507.64,124.36) node [anchor=north west][inner sep=0.75pt]   [align=left] {$\displaystyle w$};
    \draw (427.35,131.85) node [anchor=north west][inner sep=0.75pt]  [rotate=-20.13] [align=left] {$\displaystyle p_{2} +w$};
    \draw (354.87,139.77) node [anchor=north west][inner sep=0.75pt]  [rotate=-51.5] [align=left] {$\displaystyle k ,c'$};
    \draw (371.7,91.59) node [anchor=north west][inner sep=0.75pt]  [rotate=-333.14] [align=left] {$\displaystyle j ,b'$};
    \draw (515.79,156.32) node [anchor=north west][inner sep=0.75pt]  [rotate=-4.73] [align=left] {$\displaystyle p_{2}$};
    \draw (432.64,21.69) node [anchor=north west][inner sep=0.75pt]   [align=left] {$\displaystyle q_{1}$};
    \draw (247.67,198.67) node [anchor=north west][inner sep=0.75pt]   [align=left] {$\displaystyle k' ,c$};
    \draw (221.33,32) node [anchor=north west][inner sep=0.75pt]   [align=left] {$\displaystyle j' ,b$};
    \draw (580.33,164) node [anchor=north west][inner sep=0.75pt]   [align=left] {$\displaystyle k' ,c$};
    \draw (557.33,44) node [anchor=north west][inner sep=0.75pt]   [align=left] {$\displaystyle j' ,b$};

    \end{tikzpicture}
    
    \caption{\label{fig:scalar meson gluon exchange diagrams} Example diagrams for gluon exchange between scalar mesons.}
    \label{fig:my_label}
\end{figure}

The leading contribution involving gluons comes from the vertex illustrated in Figure \ref{fig:Feynman rule scalar QCD} along with its Feynman rule. 
The diagrams that contribute in the leading order can be arranged in pairs, such as the diagrams presented in Figure \ref{fig:scalar meson gluon exchange diagrams}. In each pair, in one diagram the exchange is between scalar lines with the same charge flow direction and in the other between lines with opposite directions.
In the left diagram in Figure \ref{fig:scalar meson gluon exchange diagrams}, the color factor is 

\begin{equation}
    \delta_{a' b} \delta_{c d'} (t^A)_{a b'} (t^A)_{c' d} \approx 
    \delta_{a' b} \delta_{c d'} \delta_{a d} \delta_{c' b'},
\end{equation}
which, after contracting the color indices within the mesons $\delta_{b' a'} \delta_{d' c'}$, gives $\delta_{c b} \delta_{a d}$.
In the right diagram the color factor is:
\begin{equation}
    \delta_{a' b} \delta_{c' d} (t^A)_{a b'} (t^A)_{c d'} = 
    \delta_{a' b} \delta_{c' d} \delta_{a d'} \delta_{c b'} \overset{\cdot \delta_{b' a'} \delta_{d' c'}}{\goto}
    \delta_{c b} \delta_{a d},
\end{equation}
so the two diagrams eventually have the same color factor. The flavor factors are trivially identical between the diagrams.
The remaining component is the momentum loop factor, in which only the ultraviolet-divergent term contributes to the anomalous dimension. This is the leading order term for large loop momentum $w$, and has opposite values between the two diagrams. This can be seen from the vertex factors, which are $(2p_1-w)_{\mu}(2q_2-w)^{\mu} \sim w^2$ and $(2p_1-w)_{\mu}(2p_2+w)^{\mu} \sim -w^2$, for the left and right diagrams, respectively.

In this way, we see the divergent terms of the two diagrams cancel out, as is the case for all other pairs of diagrams. Thus, contributions to the difference of anomalous dimensions from this type of diagrams is of the next order in the 't Hooft gauge coupling: $ O(g^4 N) =O\left(\frac{\lambda^2}{N} \right)$.

\begin{figure}

\begin{equation*} 
\mathord{
\begin{tikzpicture}[x=0.75pt,y=0.75pt,yscale=-.65,xscale=.65, baseline = -215pt]

\draw [color={rgb, 255:red, 0; green, 0; blue, 0 }  ,draw opacity=1 ][line width=1.5]  [dash pattern={on 5.63pt off 4.5pt}]  (137,361) -- (266.8,524.41) ;
\draw [color={rgb, 255:red, 0; green, 0; blue, 0 }  ,draw opacity=1 ][line width=1.5]  [dash pattern={on 5.63pt off 4.5pt}]  (286,363) -- (129.8,515.41) ;
\draw   (173.93,390.81) -- (178.05,411.86) -- (157.44,405.91) ;
\draw   (229.93,462.81) -- (234.05,483.86) -- (213.44,477.91) ;
\draw   (178.76,485.08) -- (157.35,486.33) -- (166.01,466.7) ;
\draw   (254.14,410.35) -- (233.04,414.21) -- (239.23,393.67) ;

\draw (184.73,374.27) node [anchor=north west][inner sep=0.75pt]  [rotate=-45.38] [align=left] {i,a};
\draw (247.73,446.27) node [anchor=north west][inner sep=0.75pt]  [rotate=-45.38] [align=left] {$\displaystyle \ell $,d};
\draw (215.13,385.57) node [anchor=north west][inner sep=0.75pt]  [rotate=-318.75] [align=left] {k,c};
\draw (140.17,462.28) node [anchor=north west][inner sep=0.75pt]  [rotate=-316.04] [align=left] {j,b};

\end{tikzpicture}
}
=     -i\Tilde{f}(\delta_{ij}\delta_{ab}\delta_{kl}\delta_{cd}+\delta_{il}\delta_{ad}\delta_{kj}\delta_{cb}) -i\Tilde{h}(\delta_{ij}\delta_{bc}\delta_{kl}\delta_{ad}+\delta_{il}\delta_{cd}\delta_{kj}\delta_{ab}) -2i(\Tilde{f}+\Tilde{h})\delta_{abcd}\delta_{ijkl}
\end{equation*}
\caption{\label{fig:Feynman Rules Scalar} Feynman rule for $\phi^4$ interaction.}
\end{figure}

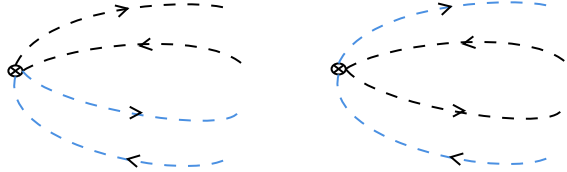
\begin{figure}[h]
\centering
\begin{tikzpicture}[x=0.75pt,y=0.75pt,yscale=-1,xscale=1]

\draw   (17,74.53) .. controls (17,73) and (18.6,71.76) .. (20.57,71.76) .. controls (22.54,71.76) and (24.14,73) .. (24.14,74.53) .. controls (24.14,76.06) and (22.54,77.3) .. (20.57,77.3) .. controls (18.6,77.3) and (17,76.06) .. (17,74.53) -- cycle ; \draw   (18.05,72.57) -- (23.09,76.49) ; \draw   (23.09,72.57) -- (18.05,76.49) ;
\draw  [dash pattern={on 4.5pt off 4.5pt}]  (20.57,71.76) .. controls (34.33,41.16) and (110.54,37.39) .. (128.38,43.16) ;
\draw [color={rgb, 255:red, 74; green, 144; blue, 226 }  ,draw opacity=1 ] [dash pattern={on 4.5pt off 4.5pt}]  (20.57,77.3) .. controls (11.39,112.11) and (107.74,131.62) .. (128.13,118.32) ;
\draw  [dash pattern={on 4.5pt off 4.5pt}]  (24.14,74.53) .. controls (44.53,61.23) and (114.11,53.8) .. (134.5,70.65) ;
\draw [color={rgb, 255:red, 74; green, 144; blue, 226 }  ,draw opacity=1 ] [dash pattern={on 4.5pt off 4.5pt}]  (24.14,74.53) .. controls (37.9,93.04) and (123.8,106.57) .. (132.46,95.93) ;
\draw   (78.21,92.61) -- (83.82,95.6) -- (78.21,98.6) ;
\draw   (89.66,63.98) -- (83.56,61.84) -- (88.52,58.08) ;
\draw   (70.72,41.49) -- (77.05,43.05) -- (72.58,47.25) ;
\draw   (81.92,123) -- (77.22,119) -- (83.46,117.16) ;
\draw   (180,73.53) .. controls (180,72) and (181.6,70.76) .. (183.57,70.76) .. controls (185.54,70.76) and (187.14,72) .. (187.14,73.53) .. controls (187.14,75.06) and (185.54,76.3) .. (183.57,76.3) .. controls (181.6,76.3) and (180,75.06) .. (180,73.53) -- cycle ; \draw   (181.05,71.57) -- (186.09,75.49) ; \draw   (186.09,71.57) -- (181.05,75.49) ;
\draw [color={rgb, 255:red, 74; green, 144; blue, 226 }  ,draw opacity=1 ] [dash pattern={on 4.5pt off 4.5pt}]  (183.57,70.76) .. controls (197.33,40.16) and (273.54,36.39) .. (291.38,42.16) ;
\draw [color={rgb, 255:red, 74; green, 144; blue, 226 }  ,draw opacity=1 ] [dash pattern={on 4.5pt off 4.5pt}]  (183.57,76.3) .. controls (174.39,111.11) and (270.74,130.62) .. (291.13,117.32) ;
\draw  [dash pattern={on 4.5pt off 4.5pt}]  (187.14,73.53) .. controls (207.53,60.23) and (277.11,52.8) .. (297.5,69.65) ;
\draw [color={rgb, 255:red, 0; green, 0; blue, 0 }  ,draw opacity=1 ] [dash pattern={on 4.5pt off 4.5pt}]  (187.14,73.53) .. controls (200.9,92.04) and (286.8,105.57) .. (295.46,94.93) ;
\draw   (241.21,91.61) -- (246.82,94.6) -- (241.21,97.6) ;
\draw   (252.66,62.98) -- (246.56,60.84) -- (251.52,57.08) ;
\draw   (233.72,40.49) -- (240.05,42.05) -- (235.58,46.25) ;
\draw   (244.92,122) -- (240.22,118) -- (246.46,116.16) ;

\end{tikzpicture}
\caption{\label{fig:scalar tree level} Tree-level diagrams for the scalar correlation function. Left diagram is unpermuted, right is permuted.}
\end{figure}

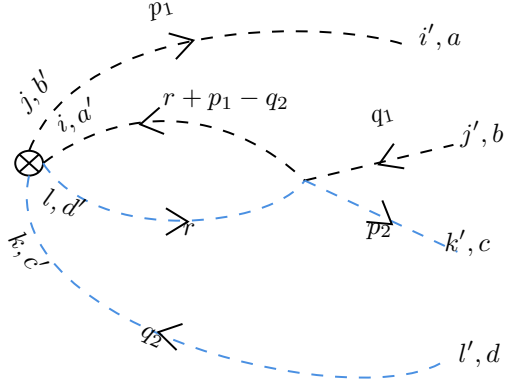
\begin{figure}
\centering
\begin{tikzpicture}[x=0.75pt,y=0.75pt,yscale=-1,xscale=1]

\draw   (66,110.75) .. controls (66,107.3) and (69.13,104.5) .. (73,104.5) .. controls (76.87,104.5) and (80,107.3) .. (80,110.75) .. controls (80,114.2) and (76.87,117) .. (73,117) .. controls (69.13,117) and (66,114.2) .. (66,110.75) -- cycle ; \draw   (68.05,106.33) -- (77.95,115.17) ; \draw   (77.95,106.33) -- (68.05,115.17) ;
\draw  [dash pattern={on 4.5pt off 4.5pt}]  (73,104.5) .. controls (100,35.5) and (229,38.5) .. (264,51.5) ;
\draw [color={rgb, 255:red, 74; green, 144; blue, 226 }  ,draw opacity=1 ] [dash pattern={on 4.5pt off 4.5pt}]  (73,117) .. controls (55,195.5) and (244,239.5) .. (284,209.5) ;
\draw  [dash pattern={on 4.5pt off 4.5pt}]  (80,110.75) .. controls (120,80.75) and (172,81.5) .. (212,119.5) ;
\draw [color={rgb, 255:red, 74; green, 144; blue, 226 }  ,draw opacity=1 ] [dash pattern={on 4.5pt off 4.5pt}]  (80,110.75) .. controls (107,152.5) and (195,143.5) .. (212,119.5) ;
\draw  [dash pattern={on 4.5pt off 4.5pt}]  (212,119.5) -- (287,101.5) ;
\draw [color={rgb, 255:red, 74; green, 144; blue, 226 }  ,draw opacity=1 ] [dash pattern={on 4.5pt off 4.5pt}]  (212,119.5) -- (289,155.5) ;
\draw   (142,133.5) -- (153,140.25) -- (142,147) ;
\draw   (141.05,95.99) -- (129.08,91.17) -- (138.8,82.68) ;
\draw   (143.38,45.24) -- (155.8,48.76) -- (147.03,58.24) ;
\draw   (147.36,205.06) -- (138.14,196.02) -- (150.37,191.89) ;
\draw   (251.13,129.55) -- (255.91,141.54) -- (243.06,140.37) ;
\draw   (262.18,113.35) -- (249.43,111.38) -- (256.96,100.9) ;

\draw (82.5,85.18) node [anchor=north west][inner sep=0.75pt]  [rotate=-334.55] [align=left] {$\displaystyle i ,a'$};
\draw (149,140) node [anchor=north west][inner sep=0.75pt]   [align=left] {$\displaystyle r$};
\draw (139,72) node [anchor=north west][inner sep=0.75pt]   [align=left] {$\displaystyle r+p_{1} -q_{2}$};
\draw (268,39) node [anchor=north west][inner sep=0.75pt]   [align=left] {$\displaystyle i' ,a$};
\draw (288,199) node [anchor=north west][inner sep=0.75pt]   [align=left] {$\displaystyle l' ,d$};
\draw (242,139) node [anchor=north west][inner sep=0.75pt]   [align=left] {$\displaystyle p_{2}$};
\draw (242.82,83.7) node [anchor=north west][inner sep=0.75pt]  [rotate=-344.74] [align=left] {$\displaystyle q_{1}$};
\draw (61.75,76.4) node [anchor=north west][inner sep=0.75pt]  [rotate=-317.06] [align=left] {$\displaystyle j ,b'$};
\draw (130.82,29.7) node [anchor=north west][inner sep=0.75pt]  [rotate=-344.74] [align=left] {$\displaystyle p_{1}$};
\draw (289,88) node [anchor=north west][inner sep=0.75pt]   [align=left] {$\displaystyle j' ,b$};
\draw (281,143) node [anchor=north west][inner sep=0.75pt]   [align=left] {$\displaystyle k' ,c$};
\draw (130.62,190.36) node [anchor=north west][inner sep=0.75pt]  [rotate=-25.65] [align=left] {$\displaystyle q_{2}$};
\draw (82.45,119.68) node [anchor=north west][inner sep=0.75pt]  [rotate=-21.64] [align=left] {$\displaystyle l ,d'$};
\draw (71.2,140.78) node [anchor=north west][inner sep=0.75pt]  [rotate=-55.09] [align=left] {$\displaystyle k ,c'$};

\end{tikzpicture}
\caption{\label{fig:diagram3} Representative diagram for the scalar meson correction from $\phi^4$ interaction.}
\end{figure}

\paragraph{$\phi^4$ interaction diagrams.}

The $\phi^4$ interactions can be described using a single comprehensive vertex Feynman rule, illustrated in Figure \ref{fig:Feynman Rules Scalar}.
In Figure \ref{fig:Feynman Rules Scalar} we introduced the many-index $\delta$ symbol, which is 1 if all indices are equal and 0 otherwise.
Without loss of generality (in the large N limit), we can assume all external legs have different flavor indices, e.g. $(i',j',k',l') = (1,2,3,4)$ (this numbering is for convenience, although $i',k'$ are components of a different representation than $j',l'$).
Then the flavor indices will have to match them -- either in the unpermuted order: $(j,i,l,k) = (1,2,3,4)$, or in the permuted order: $(j,i,l,k) = (1,4,3,2)$ (permuting the other pair as well returns to a configuration equivalent to the original).
Thus, we can omit the flavor $\delta$ symbols and keep the relevant information by referring to the diagrams as either unpermuted or permuted.
In particular, in these cases the last term of Figure \ref{fig:Feynman Rules Scalar} always vanishes.

In what follows, we need the contribution of the tree-level diagrams, depicted in Figure \ref{fig:scalar tree level}.
The momentum factor in both of them is $\frac{i}{p_1^2} \frac{i}{q_1^2} \frac{i}{p_2^2} \frac{i}{q_2^2} = \frac{1}{p_1^2 p_2^2 q_1^2 q_2^2}$.
In addition there are the color factors -- for the unpermuted diagram this is $\delta_{a b'} \delta_{b a'} \delta_{c d'} \delta_{d c'} \delta_{a' b'} \delta_{c' d'} = \delta_{a b} \delta_{c d}$, and for the permuted diagram we get $\delta_{a b'} \delta_{b c'} \delta_{c b'} \delta_{d a'} \delta_{a' b'} \delta_{c d} = \delta_{a d} \delta_{c b}$.

Next, we turn to evaluate the 1-loop correction. A representative (unpermuted) diagram is depicted in Figure \ref{fig:diagram3}. We denote the vertex factor as $-i\kappa \equiv -i \left( 
\tilde{f} \delta_{b a'} \delta_{c d'} + \tilde{h} \delta_{c b} \delta_{a' d'} \right)$.
The diagram then gives:
\begin{equation}
    -\frac{\kappa M^{-2\epsilon}}{(4\pi)^2 p_1^2 p_2^2 q_1^2 q_2^2 \epsilon}.
\end{equation}
We can see that the dependence on the momenta is the same as the tree-level diagrams to leading order in $\epsilon$. Then the difference between different one-loop diagrams will manifest in the color factors, contained inside $\kappa$.
After inclusion of the color factors from the free legs $\delta_{a b'} \delta_{d c'}$ and contraction with $\delta_{a' b'} \delta_{c' d'}$, $\kappa$ becomes $\Tilde{f}\delta_{a b} \delta_{c d} + \Tilde{h} \delta_{a d} \delta_{c b}$.
A similar contraction for the other diagrams shows that the $\Tilde{f}$ terms recover the color factor of the tree level diagram, and the $\Tilde{h}$ terms switch between the unpermuted and permuted diagrams. Since the bi-meson operator is symmetrized, they get equal coefficients, and in the end we get a positive multiple of $(\Tilde{f}+\Tilde{h})(\delta_{a b} \delta_{c d} +\delta_{a d} \delta_{c b})$ instead of $\kappa$.
There is also a symmetry factor in some diagrams. There are 4 diagrams, 2 of which have the same charge flow direction in the two loop propagators -- giving a symmetry factor of $2$. Then the sum over diagrams gives a factor of 3. 
The overall correction to the renormalization function is thus
\begin{equation}
    \delta \left( \frac{Z_{(\phi^* \phi)^2}}{Z^2_{\phi^* \phi}} \right) = -\frac{3(\Tilde{f}+\Tilde{h}) M^{-2\epsilon}}{(4\pi)^2 \epsilon} <0.
\end{equation}
From here we find that \cite{paper7}
\begin{equation}
    \gamma_{(\phi^* \phi)^2} - 2\gamma_{\phi^* \phi} = \frac{\partial}{\partial \log M} \log \left( \frac{Z_{(\phi^* \phi)^2}}{Z^2_{\phi^* \phi}} \right) = -2\epsilon \log \left( \frac{Z_{(\phi^* \phi)^2}}{Z^2_{\phi^* \phi}} \right) >0,
\end{equation}
supporting the CCC.

\subsection{Fermionic double mesons}\label{section 4.2}

To compute the anomalous dimensions of these operators, we compute their correlation function with 2 fermion operators and 2 antifermion operators:
\begin{equation}
    \left<\left( \bar{\psi}_{\dot{\alpha} b'j'} \psi_{\alpha a'i'} \bar{\psi}_{\dot{\beta} d'l'} \psi_{\beta c'k'} \right)(p) \bar{\psi}_{\dot{\gamma} ai}(-p_1) \psi_{\gamma bj}(p_2) \bar{\psi}_{\dot{\delta} ck}(-p_3) \psi_{\delta dl}(p_4) \right> \delta_{a'b'}\delta_{c'd'},
\end{equation}
where $a,b,c,d,a',b',c',d'$ are color indices, $i,j,k,l,i',j',k',l'$ are flavor indices, and $\alpha, \dot{\alpha}, \beta, \dot{\beta}, \gamma, \dot{\gamma}, \delta, \dot{\delta}$ are spinor indices.
The spinor indices are abbreviated for most of this subsection as $A=\alpha\dot{\gamma},B=\gamma\dot{\alpha},C=\beta\dot{\delta},D=\delta\dot{\beta},A'=\beta\dot{\gamma},C'=\alpha\dot{\delta}$.
As in the scalar case, we can assume the flavor indices of the
4 legs of the bi-meson are $1,2,3,4$ respectively (even for $\bar{\psi}$
operators and odd for $\psi$). Since we are interested in a symmetrized
representation, this includes the meson with the possible permutations
$1\swap3,2\swap4$ (permuting the particles completely and not just
the flavor index). Also as in the scalar case, it's enough to consider
the permutation $2\swap4$ (with an exception we discuss later).

\begin{figure}
\centering
\begin{tikzpicture}[x=0.75pt,y=0.75pt,yscale=-1,xscale=1]

\draw   (17,74.53) .. controls (17,73) and (18.6,71.76) .. (20.57,71.76) .. controls (22.54,71.76) and (24.14,73) .. (24.14,74.53) .. controls (24.14,76.06) and (22.54,77.3) .. (20.57,77.3) .. controls (18.6,77.3) and (17,76.06) .. (17,74.53) -- cycle ; \draw   (18.05,72.57) -- (23.09,76.49) ; \draw   (23.09,72.57) -- (18.05,76.49) ;
\draw    (20.57,71.76) .. controls (34.33,41.16) and (110.54,37.39) .. (128.38,43.16) ;
\draw [color={rgb, 255:red, 74; green, 144; blue, 226 }  ,draw opacity=1 ]   (20.57,77.3) .. controls (11.39,112.11) and (107.74,131.62) .. (128.13,118.32) ;
\draw    (24.14,74.53) .. controls (44.53,61.23) and (114.11,53.8) .. (134.5,70.65) ;
\draw [color={rgb, 255:red, 74; green, 144; blue, 226 }  ,draw opacity=1 ]   (24.14,74.53) .. controls (37.9,93.04) and (123.8,106.57) .. (132.46,95.93) ;
\draw   (78.21,92.61) -- (83.82,95.6) -- (78.21,98.6) ;
\draw   (89.66,63.98) -- (83.56,61.84) -- (88.52,58.08) ;
\draw   (70.72,41.49) -- (77.05,43.05) -- (72.58,47.25) ;
\draw   (81.92,123) -- (77.22,119) -- (83.46,117.16) ;
\draw   (204.25,27.31) -- (225.5,27.31) -- (225.5,48.5) ;
\draw   (164.01,48.5) -- (164.53,27.3) -- (204.77,27.32) ;

\draw   (204.25,36.16) -- (215.5,36.16) -- (215.5,48.5) ;
\draw   (174.01,48.5) -- (174.35,36.16) -- (204.59,36.17) ;

\draw   (234.75,15.43) -- (244.5,15.43) -- (244.5,48.5) ;
\draw   (193.01,48.57) -- (193.45,15.43) -- (235.19,15.43) ;

\draw   (233.25,22.35) -- (234.5,22.35) -- (234.5,48.5) ;
\draw   (202.01,48.5) -- (202.28,22.35) -- (233.53,22.36) ;

\draw   (315,72.53) .. controls (315,71) and (316.6,69.76) .. (318.57,69.76) .. controls (320.54,69.76) and (322.14,71) .. (322.14,72.53) .. controls (322.14,74.06) and (320.54,75.3) .. (318.57,75.3) .. controls (316.6,75.3) and (315,74.06) .. (315,72.53) -- cycle ; \draw   (316.05,70.57) -- (321.09,74.49) ; \draw   (321.09,70.57) -- (316.05,74.49) ;
\draw [color={rgb, 255:red, 74; green, 144; blue, 226 }  ,draw opacity=1 ]   (318.57,69.76) .. controls (332.33,39.16) and (408.54,35.39) .. (426.38,41.16) ;
\draw [color={rgb, 255:red, 74; green, 144; blue, 226 }  ,draw opacity=1 ]   (318.57,75.3) .. controls (309.39,110.11) and (405.74,129.62) .. (426.13,116.32) ;
\draw    (322.14,72.53) .. controls (342.53,59.23) and (412.11,51.8) .. (432.5,68.65) ;
\draw [color={rgb, 255:red, 0; green, 0; blue, 0 }  ,draw opacity=1 ]   (322.14,72.53) .. controls (335.9,91.04) and (421.8,104.57) .. (430.46,93.93) ;
\draw   (376.21,90.61) -- (381.82,93.6) -- (376.21,96.6) ;
\draw   (387.66,61.98) -- (381.56,59.84) -- (386.52,56.08) ;
\draw   (368.72,39.49) -- (375.05,41.05) -- (370.58,45.25) ;
\draw   (379.92,121) -- (375.22,117) -- (381.46,115.16) ;
\draw   (506.25,28.31) -- (528.5,28.31) -- (528.5,49.5) ;
\draw   (466.01,49.5) -- (466.53,28.3) -- (506.77,28.32) ;

\draw   (520.67,37.16) -- (536.33,37.16) -- (536.33,49.5) ;
\draw   (475.51,49.5) -- (475.53,37.16) -- (521.19,37.17) ;

\draw   (536.75,16.43) -- (545,16.43) -- (545,49.5);
\draw   (494.01,49.57) -- (494.05,16.43) -- (537.19,16.43) ;

\draw   (513.67,43.76) -- (518.33,43.76) -- (518.33,49.5) ;
\draw   (503.01,49.5) -- (503.13,43.76) -- (513.79,43.76) ;

\draw (156,89) node [anchor=north west][inner sep=0.75pt]   [align=left] {$\displaystyle =( -)^{2+4} =( +)$};
\draw (152,48) node [anchor=north west][inner sep=0.75pt]   [align=left] {($\displaystyle \overline{\psi } \psi $)($\displaystyle \overline{\psi } \psi $)$\displaystyle \overline{\psi } \psi $$\displaystyle \overline{\psi } \psi $};
\draw (454,49) node [anchor=north west][inner sep=0.75pt]   [align=left] {($\displaystyle \overline{\psi } \psi $)($\displaystyle \overline{\psi } \psi $)$\displaystyle \overline{\psi } \psi $$\displaystyle \overline{\psi } \psi $};
\draw (459.33,89.67) node [anchor=north west][inner sep=0.75pt]   [align=left] {$\displaystyle =( -)^{2+3} =( -)$};

\end{tikzpicture}

\caption{\label{fig:fermion_contractions2} Fermion operator contractions - tree level.}
\end{figure}
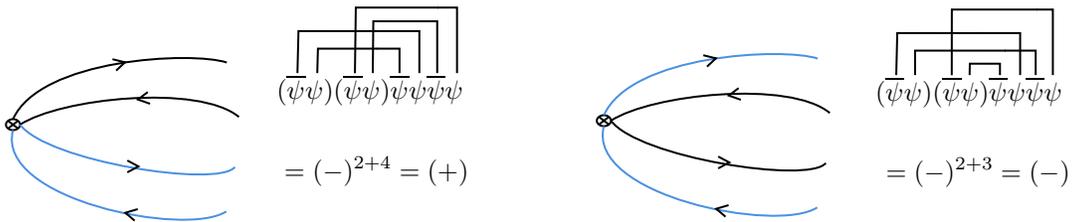

In a similar manner to the computation of the single fermionic meson operators, and as shown explicitly for the tree-level diagrams in Figure \ref{fig:fermion_contractions2}, we find that the fermionic operator contractions give in this case a (+) sign in the unpermuted (or doubly permuted) branch and a (--) sign in the singly permuted branches.
This means the tree level amplitude equals (after contraction of the color indices of the bi-meson operator, but before considering contractions of spinor structures):
\begin{equation}
\frac{1}{p_{1}^{2}p_{2}^{2}p_{3}^{2}p_{4}^{2}}\left(\cancel{p_{1}}_{\alpha\dot{\gamma}}\cancel{p_{2}}_{\gamma\dot{\alpha}}\cancel{p_{3}}_{\beta\dot{\delta}}\cancel{p_{4}}_{\delta\dot{\beta}}\delta_{ab}\delta_{cd} - \cancel{p_{1}}_{\beta\dot{\gamma}}\cancel{p_{2}}_{\gamma\dot{\alpha}}\cancel{p_{3}}_{\alpha\dot{\delta}}\cancel{p_{4}}_{\delta\dot{\beta}}\delta_{ad}\delta_{bc}\right) .
\end{equation}

Any quantum correction that keeps the flavors matched will keep the
fermionic lines the same (and might add some new ones detached from
the originals). Then going forward, we can calculate only the first
``branch'', involving $\delta_{ab}\delta_{cd}$, and obtain the
other by the permutation
\begin{equation}
\cancel{p_{1}}_{A}\cancel{p_{3}}_{C}\swap\cancel{p_{3}}_{C'}\cancel{p_{1}}_{A'}\text{ (spinor indices\&roles)},a\swap c\text{ (color indices)}.
\end{equation}

\subsubsection{1-loop level diagrams} 

\begin{figure}
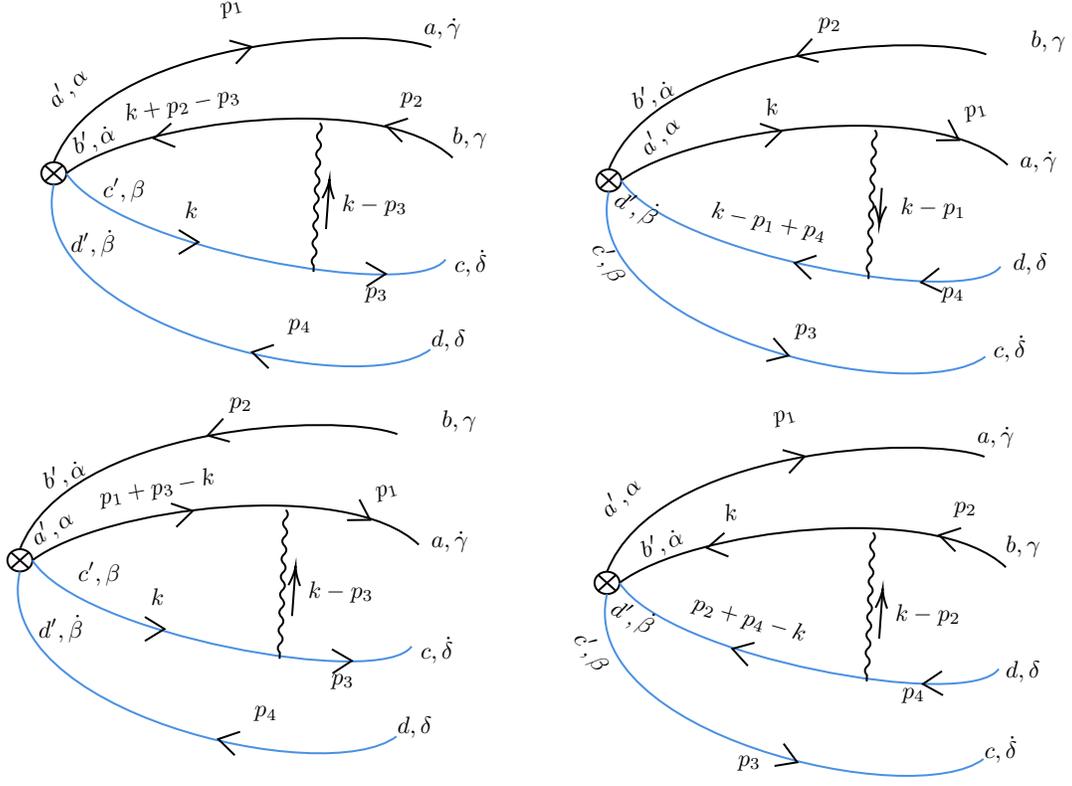

\centering

\caption{\label{fig:unpermuted diagrams} Unpermuted diagrams, in order I-IV.} 
\end{figure}

Here we only consider inter-meson diagrams. The intra-meson diagrams are derived from the ones computed for the single meson case; their leading contribution will be that computed for one meson, multiplied by the tree level contribution for the other meson.
There are 8 inter-meson diagrams in this order (including permutation), and they are presented in Figs. \ref{fig:unpermuted diagrams}, \ref{fig:permuted diagrams}.
Black (blue) fermion lines indicate the fields whose color indices are to be contracted together.

\begin{figure}
\centering
%
\caption{\label{fig:permuted diagrams}Permuted diagrams, in order I'-IV'.}
\end{figure}

The results for diagrams I-IV, respectively, are (up to $O(\epsilon^0)$ corrections):
\begin{align}
I=\frac{g^{2}}{4\left(4\pi\right)^{2}\epsilon}\left(p_{E}^{2}\right)^{-\epsilon}\frac{\cancel{p_{1}}_{A}\cancel{p_{4}}_{D}}{p_{1}^{2}p_{2}^{2}p_{3}^{2}p_{4}^{2}}\left(t^{I}\right)_{cc'}\left(t^{I}\right)_{b'b}\delta_{aa'}\delta_{dd'}\left[\cancel{p_{2}}\gamma^{\mu}\gamma^{\nu}\right]_{B}\left[\gamma_{\nu}\gamma_{\mu}\cancel{p_{3}}\right]_{C} \\
II = \frac{g^{2}}{4\left(4\pi\right)^{2}\epsilon}\left(p_{E}^{2}\right)^{-\epsilon}\frac{\cancel{p_{2}}_{B}\cancel{p_{3}}_{C}}{p_{1}^{2}p_{2}^{2}p_{3}^{2}p_{4}^{2}}\left(t^{I}\right)_{aa'}\left(t^{I}\right)_{d'd}\delta_{bb'}\delta_{cc'}\left[\cancel{p_{4}}\gamma^{\mu}\gamma^{\nu}\right]_{D}\left[\gamma_{\nu}\gamma_{\mu}\cancel{p_{1}}\right]_{A} \\
III = -\frac{g^{2}}{4\left(4\pi\right)^{2}\epsilon}\left(p_{E}^{2}\right)^{-\epsilon}\frac{\cancel{p_{2}}_{B}\cancel{p_{4}}_{D}}{p_{1}^{2}p_{2}^{2}p_{3}^{2}p_{4}^{2}}\left(t^{I}\right)_{aa'}\left(t^{I}\right)_{cc'}\delta_{bb'}\delta_{dd'}\left[\gamma^{\nu}\gamma^{\mu}\cancel{p_{3}}\right]_{C}\left[\gamma_{\nu}\gamma_{\mu}\cancel{p_{1}}\right]_{A} \\
IV = -\frac{g^{2}}{4\left(4\pi\right)^{2}\epsilon}\left(p_{E}^{2}\right)^{-\epsilon}\frac{\cancel{p_{1}}_{A}\cancel{p_{3}}_{C}}{p_{1}^{2}p_{2}^{2}p_{3}^{2}p_{4}^{2}}\left(t^{I}\right)_{b'b}\left(t^{I}\right)_{d'd}\delta_{aa'}\delta_{cc'}\left[\cancel{p_{4}}\gamma^{\mu}\gamma^{\nu}\right]_{D}\left[\cancel{p_{2}}\gamma_{\mu}\gamma_{\nu}\right]_{B}.
\end{align}

We compute explicitly the color index contractions for unpermuted diagram I:
\begin{equation}
\left(t^{I}\right)_{cc'}\left(t^{I}\right)_{b'b}\delta_{aa'}\delta_{dd'}\delta_{a'b'}\delta_{c'd'}=\left(t^{I}\right)_{b'b}\left(t^{I}\right)_{cc'}\delta_{c'd}\delta_{b'a}=
\left(t^{I}\right)_{ab}\left(t^{I}\right)_{cd} \approx \frac{1}{2} \delta_{ad}\delta_{bc}.
\end{equation}

A similar calculation yields the same result for diagrams II-IV.

\subsubsection{Fermion index contraction}

The contraction is done with the chiral spinor structures: $\left\{ 1\pm \gamma^5, (1\pm \gamma^5) \gamma^{\mu}, (1\pm \gamma^5) \gamma^{\mu \nu} \right\}$, since they also span the degrees of freedom quadratic in the fermions, and have simpler representations than their counterparts without chiral projections.
The change from section \ref{single fermionic mesons} is the chiral projection of tensors. It creates some redundancy, as the tensor structure $\gamma^5 \gamma^{\mu \nu}$ can be expressed using the one without $\gamma^5$, but this does not affect our results. 
In this part of the results we use the following identities for the $\gamma$ matrices (to leading order in $\epsilon$): 

\begin{align}\label{eq:7}
    & \gamma^{\mu} \gamma^{\nu} = \eta^{\mu \nu} + \gamma^{\mu \nu}; \indent
    \gamma^5 \gamma^{\mu}\gamma^{\nu} = \eta^{\mu \nu} \gamma^5 +\frac{i}{2} \epsilon^{\mu \nu \rho \sigma} \gamma_{\rho \sigma}; \indent
    \gamma^{\mu} \gamma^{\nu} \gamma^{\rho} = \eta^{\mu \nu} \gamma^{\rho} -\eta^{\mu \rho} \gamma^{\nu} +\eta^{\nu \rho} \gamma^{\mu} +i\epsilon^{\mu \nu \rho \sigma} \gamma^5 \gamma_{\sigma}; \nonumber\\
    & \gamma^{\mu} \gamma^{\nu} \gamma^{\rho} \gamma^{\sigma} = \eta^{\mu \nu}\eta^{\rho \sigma} +\eta^{\nu\rho}\eta^{\mu\sigma} -\eta^{\mu\rho}\eta^{\nu\sigma} +\eta^{\mu\nu}\gamma^{\rho\sigma} +\eta^{\mu\rho}\gamma^{\sigma\nu} +\eta^{\nu\rho}\gamma^{\mu\sigma} +\eta^{\mu\sigma}\gamma^{\nu\rho} +\eta^{\nu\sigma}\gamma^{\rho\mu} +\eta^{\rho\sigma}\gamma^{\mu\nu} +i\epsilon^{\mu\nu\rho\sigma}\gamma^{5}.
\end{align}

The identities (\ref{eq:7}) use the fact that in 4 dimensions, the algebra of $\gamma$ matrices (indeed, all $4\times 4$ matrices in the spinor indices) is spanned by
$\left\{ 1,\gamma^5, \gamma^{\mu}, \gamma^5 \gamma^{\mu}, \gamma^{\mu\nu} \right\}$.


\paragraph{Chiral scalars.}

We contract the spinor matrices in each diagram with (non-normalized)
chiral projection operators $\left(1\pm\gamma^{5}\right)_{\alpha\dot{\alpha}}\left(1\pm\gamma^{5}\right)_{\beta\dot{\beta}}$,
corresponding to the symmetric square of each chiral single scalar meson. 
This gives:

Tree level:
\begin{equation}
\cancel{p_{1}}_{\alpha\dot{\gamma}}\cancel{p_{2}}_{\gamma\dot{\alpha}}\cancel{p_{3}}_{\beta\dot{\delta}}\cancel{p_{4}}_{\delta\dot{\beta}}\goto\left[\cancel{p}_{2}\left(1\pm\gamma^{5}\right)\cancel{p}_{1}\right]_{\gamma\dot{\gamma}}\left[\cancel{p}_{4}\left(1\pm\gamma^{5}\right)\cancel{p}_{3}\right]_{\delta\dot{\delta}}.
\end{equation}

Diagram I:
\begin{equation}
\cancel{p_{1}}_{A}\cancel{p_{4}}_{D}\left[\cancel{p_{2}}\gamma^{\mu}\gamma^{\nu}\right]_{B}\left[\gamma_{\nu}\gamma_{\mu}\cancel{p_{3}}\right]_{C}\goto\left[\cancel{p_{2}}\gamma^{\mu}\gamma^{\nu}\left(1\pm\gamma^{5}\right)\cancel{p_{1}}\right]_{\gamma\dot{\gamma}}\left[\cancel{p_{4}}\left(1\pm\gamma^{5}\right)\gamma_{\nu}\gamma_{\mu}\cancel{p_{3}}\right]_{\delta\dot{\delta}}= \nonumber
\end{equation}
\begin{equation}
=\left[\cancel{p_{2}}\left(\eta^{\mu\nu}+\gamma^{\mu\nu}\pm\eta^{\mu\nu}\gamma^{5}\pm\frac{i}{2}\epsilon^{\mu\nu\rho\sigma}\gamma_{\rho\sigma}\right)\cancel{p_{1}}\right]_{\gamma\dot{\gamma}}\left[\cancel{p_{4}}\left(\eta_{\mu\nu}-\gamma_{\mu\nu}\pm\eta_{\mu\nu}\gamma^{5}\mp\frac{i}{2}\epsilon_{\mu\nu\rho'\sigma'}\gamma^{\rho'\sigma'}\right)\cancel{p_{3}}\right]_{\delta\dot{\delta}}= \nonumber
\end{equation}
\begin{align}
 & =d\left[\cancel{p_{2}}\left(1\pm\gamma^{5}\right)\cancel{p_{1}}\right]_{\gamma\dot{\gamma}}\left[\cancel{p_{4}}\left(1\pm\gamma^{5}\right)\cancel{p_{3}}\right]_{\delta\dot{\delta}}
 -\left[\cancel{p_{2}}\gamma^{\mu\nu}\cancel{p_{1}}\right]_{\gamma\dot{\gamma}}\left[\cancel{p_{4}}\gamma_{\mu\nu}\cancel{p_{3}}\right]_{\delta\dot{\delta}} \nonumber\\
 & +\frac{1}{4}\underbrace{\epsilon^{\mu\nu\rho\sigma}\epsilon_{\mu\nu\rho'\sigma'}}_{\approx-2\left(\delta_{\rho'}^{\rho}\delta_{\sigma'}^{\sigma}-\delta_{\rho'}^{\sigma}\delta_{\sigma'}^{\rho}\right)}\left[\cancel{p_{2}}\gamma_{\rho\sigma}\cancel{p_{1}}\right]_{\gamma\dot{\gamma}}\left[\cancel{p_{4}}\gamma^{\rho'\sigma'}\cancel{p_{3}}\right]_{\delta\dot{\delta}}
 \mp i\epsilon^{\mu\nu\rho\sigma}\left[\cancel{p_{2}}\gamma_{\rho\sigma}\cancel{p_{1}}\right]_{\gamma\dot{\gamma}}\left[\cancel{p_{4}}\gamma_{\mu\nu}\cancel{p_{3}}\right]_{\delta\dot{\delta}}= \nonumber
\end{align}
\begin{equation}
=d\left[\cancel{p_{2}}\left(1\pm\gamma^{5}\right)\cancel{p_{1}}\right]_{\gamma\dot{\gamma}}\left[\cancel{p_{4}}\left(1\pm\gamma^{5}\right)\cancel{p_{3}}\right]_{\delta\dot{\delta}}-2\left[\cancel{p_{2}}\gamma^{\mu\nu}\cancel{p_{1}}\right]_{\gamma\dot{\gamma}}\left[\cancel{p_{4}}\gamma_{\mu\nu}\cancel{p_{3}}\right]_{\delta\dot{\delta}}\mp i\epsilon_{\mu\nu\rho\sigma}\left[\cancel{p_{2}}\gamma^{\mu\nu}\cancel{p_{1}}\right]_{\gamma\dot{\gamma}}\left[\cancel{p_{4}}\gamma^{\rho\sigma}\cancel{p_{3}}\right]_{\delta\dot{\delta}}.
\end{equation}

Diagram II:
\begin{equation}
\cancel{p_{2}}_{B}\cancel{p_{3}}_{C}\left[\cancel{p_{4}}\gamma^{\mu}\gamma^{\nu}\right]_{D}\left[\gamma_{\nu}\gamma_{\mu}\cancel{p_{1}}\right]_{A}\goto\left[\cancel{p_{4}}\gamma^{\mu}\gamma^{\nu}\left(1\pm\gamma^{5}\right)\cancel{p_{3}}\right]_{\delta\dot{\delta}}\left[\cancel{p_{2}}\left(1\pm\gamma^{5}\right)\gamma_{\nu}\gamma_{\mu}\cancel{p_{1}}\right]_{\gamma\dot{\gamma}}= \nonumber
\end{equation}
\begin{equation}
=\cdots=d\left[\cancel{p_{2}}\left(1\pm\gamma^{5}\right)\cancel{p_{1}}\right]_{\gamma\dot{\gamma}}\left[\cancel{p_{4}}\left(1\pm\gamma^{5}\right)\cancel{p_{3}}\right]_{\delta\dot{\delta}}-2\left[\cancel{p_{2}}\gamma^{\mu\nu}\cancel{p_{1}}\right]_{\gamma\dot{\gamma}}\left[\cancel{p_{4}}\gamma_{\mu\nu}\cancel{p_{3}}\right]_{\delta\dot{\delta}}\mp i\epsilon_{\mu\nu\rho\sigma}\left[\cancel{p_{2}}\gamma^{\mu\nu}\cancel{p_{1}}\right]_{\gamma\dot{\gamma}}\left[\cancel{p_{4}}\gamma^{\rho\sigma}\cancel{p_{3}}\right]_{\delta\dot{\delta}}.
\end{equation}

Diagram III:
\begin{equation}
\cancel{p_{2}}_{B}\cancel{p_{4}}_{D}\left[\gamma^{\nu}\gamma^{\mu}\cancel{p_{3}}\right]_{C}\left[\gamma_{\nu}\gamma_{\mu}\cancel{p_{1}}\right]_{A}\goto\left[\cancel{p_{4}}\left(1\pm\gamma^{5}\right)\gamma^{\nu}\gamma^{\mu}\cancel{p_{3}}\right]_{\delta\dot{\delta}}\left[\cancel{p_{2}}\left(1\pm\gamma^{5}\right)\gamma_{\nu}\gamma_{\mu}\cancel{p_{1}}\right]_{\gamma\dot{\gamma}}= \nonumber
\end{equation}
\begin{equation}
=\left[\cancel{p_{2}}\left(\eta^{\mu\nu}-\gamma^{\mu\nu}\pm\eta^{\mu\nu}\gamma^{5}\mp\frac{i}{2}\epsilon^{\mu\nu\rho\sigma}\gamma_{\rho\sigma}\right)\cancel{p_{1}}\right]_{\gamma\dot{\gamma}}\left[\cancel{p_{4}}\left(\eta_{\mu\nu}-\gamma_{\mu\nu}\pm\eta_{\mu\nu}\gamma^{5}\mp\frac{i}{2}\epsilon_{\mu\nu\rho'\sigma'}\gamma^{\rho'\sigma'}\right)\cancel{p_{3}}\right]_{\delta\dot{\delta}} =\cdots = \nonumber
\end{equation}
\begin{equation}
=d\left[\cancel{p_{2}}\left(1\pm\gamma^{5}\right)\cancel{p_{1}}\right]_{\gamma\dot{\gamma}}\left[\cancel{p_{4}}\left(1\pm\gamma^{5}\right)\cancel{p_{3}}\right]_{\delta\dot{\delta}}+2\left[\cancel{p_{2}}\gamma^{\mu\nu}\cancel{p_{1}}\right]_{\gamma\dot{\gamma}}\left[\cancel{p_{4}}\gamma_{\mu\nu}\cancel{p_{3}}\right]_{\delta\dot{\delta}}\pm i\epsilon_{\mu\nu\rho\sigma}\left[\cancel{p_{2}}\gamma^{\mu\nu}\cancel{p_{1}}\right]_{\gamma\dot{\gamma}}\left[\cancel{p_{4}}\gamma^{\rho\sigma}\cancel{p_{3}}\right]_{\delta\dot{\delta}}.
\end{equation}

Diagram IV:
\begin{equation}
\cancel{p_{1}}_{A}\cancel{p_{3}}_{C}\left[\cancel{p_{4}}\gamma^{\mu}\gamma^{\nu}\right]_{D}\left[\cancel{p_{2}}\gamma_{\mu}\gamma_{\nu}\right]_{B}\goto\left[\cancel{p_{4}}\gamma^{\mu}\gamma^{\nu}\left(1\pm\gamma^{5}\right)\cancel{p_{3}}\right]_{\delta\dot{\delta}}\left[\cancel{p_{2}}\gamma_{\mu}\gamma_{\nu}\left(1\pm\gamma^{5}\right)\cancel{p_{1}}\right]_{\gamma\dot{\gamma}}= \nonumber
\end{equation}
\begin{equation}
=\cdots=d\left[\cancel{p_{2}}\left(1\pm\gamma^{5}\right)\cancel{p_{1}}\right]_{\gamma\dot{\gamma}}\left[\cancel{p_{4}}\left(1\pm\gamma^{5}\right)\cancel{p_{3}}\right]_{\delta\dot{\delta}}+2\left[\cancel{p_{2}}\gamma^{\mu\nu}\cancel{p_{1}}\right]_{\gamma\dot{\gamma}}\left[\cancel{p_{4}}\gamma_{\mu\nu}\cancel{p_{3}}\right]_{\delta\dot{\delta}}\pm i\epsilon_{\mu\nu\rho\sigma}\left[\cancel{p_{2}}\gamma^{\mu\nu}\cancel{p_{1}}\right]_{\gamma\dot{\gamma}}\left[\cancel{p_{4}}\gamma^{\rho\sigma}\cancel{p_{3}}\right]_{\delta\dot{\delta}}.
\end{equation}

\paragraph{Chiral tensors.}

The chiral tensors transform in the same representations of the global symmetry as the chiral scalars. As shown in section \ref{section 4.1}, they have a higher anomalous dimension than the scalars, and so are not important on their own to determining the minimal anomalous dimension of the representations. However, there can be mixing between their bi-mesons and the scalar bi-mesons, and so they must be considered in this context. The relevant Lorentz representation is then the one that can mix with the scalars - i.e. the scalar representation.
So, we contract the amplitudes with the chiral tensor fermion structure, with contracted Lorentz indices:
$\left[\left(1\pm\gamma^{5}\right)\gamma^{\rho\sigma}\right]_{\alpha\dot{\alpha}}\left[\left(1\pm\gamma^{5}\right)\gamma_{\rho\sigma}\right]_{\beta\dot{\beta}}$. This gives:

Tree level:

\begin{equation}
\cancel{p_{1}}_{\alpha\dot{\gamma}}\cancel{p_{2}}_{\gamma\dot{\alpha}}\cancel{p_{3}}_{\beta\dot{\delta}}\cancel{p_{4}}_{\delta\dot{\beta}}\goto\left[\cancel{p}_{2}\left(1\pm\gamma^{5}\right)\gamma^{\rho\sigma}\cancel{p}_{1}\right]_{\gamma\dot{\gamma}}\left[\cancel{p}_{4}\left(1\pm\gamma^{5}\right)\gamma_{\rho\sigma}\cancel{p}_{3}\right]_{\delta\dot{\delta}}.
\end{equation}
Diagram I:
\begin{align}
\cancel{p_{1}}_{A}\cancel{p_{4}}_{D}\left[\cancel{p_{2}}\gamma^{\mu}\gamma^{\nu}\right]_{B}\left[\gamma_{\nu}\gamma_{\mu}\cancel{p_{3}}\right]_{C}
\goto\left[\cancel{p}_{2}\gamma^{\mu}\gamma^{\nu}\left(1\pm\gamma^{5}\right)\gamma^{\rho\sigma}\cancel{p}_{1}\right]_{\gamma\dot{\gamma}}\left[\cancel{p}_{4}\left(1\pm\gamma^{5}\right)\gamma_{\rho\sigma}\gamma_{\nu}\gamma_{\mu}\cancel{p}_{3}\right]_{\delta\dot{\delta}}=\nonumber \\
=\left[\cancel{p}_{2}\left(\eta^{\mu\nu}+\gamma^{\mu\nu}\right)\left(1\pm\gamma^{5}\right)\gamma^{\rho\sigma}\cancel{p}_{1}\right]_{\gamma\dot{\gamma}}\left[\cancel{p}_{4}\left(1\pm\gamma^{5}\right)\gamma_{\rho\sigma}\left(\eta_{\mu\nu}-\gamma_{\mu\nu}\right)\cancel{p}_{3}\right]_{\delta\dot{\delta}}=\nonumber \\
=d\left[\cancel{p}_{2}\left(1\pm\gamma^{5}\right)\gamma^{\rho\sigma}\cancel{p}_{1}\right]_{\gamma\dot{\gamma}}\left[\cancel{p}_{4}\left(1\pm\gamma^{5}\right)\gamma_{\rho\sigma}\cancel{p}_{3}\right]_{\delta\dot{\delta}}
-\left[\cancel{p}_{2}\gamma^{\mu\nu}\left(1\pm\gamma^{5}\right)\gamma^{\rho\sigma}\cancel{p}_{1}\right]_{\gamma\dot{\gamma}}\left[\cancel{p}_{4}\left(1\pm\gamma^{5}\right)\gamma_{\rho\sigma}\gamma_{\mu\nu}\cancel{p}_{3}\right]_{\delta\dot{\delta}},
\end{align}
and the second term here needs simplification. First we note that
$\gamma^{\mu\nu}\gamma^{\rho\sigma}$ is the part of $\gamma^{\mu}\gamma^{\nu}\gamma^{\rho}\gamma^{\sigma}$
which is antisymmetric in each of the swaps $\mu\swap\nu,\rho\swap\sigma$,
and so we have:
\begin{align}
 & \gamma^{\mu\nu}\left(1\pm\gamma^{5}\right)\gamma^{\rho\sigma}=\left(1\pm\gamma^{5}\right)\gamma^{\mu\nu}\gamma^{\rho\sigma}= \nonumber \\
 & =\left(1\pm\gamma^{5}\right)\left(\eta^{\nu\rho}\eta^{\mu\sigma}-\eta^{\mu\rho}\eta^{\nu\sigma}+\eta^{\mu\rho}\gamma^{\sigma\nu}+\eta^{\nu\rho}\gamma^{\mu\sigma}+\eta^{\mu\sigma}\gamma^{\nu\rho}+\eta^{\nu\sigma}\gamma^{\rho\mu}+i\epsilon^{\mu\nu\rho\sigma}\gamma^{5}\right),
\end{align}
and similarly:
\begin{align}
& \left(1\pm\gamma^{5}\right)\gamma_{\rho\sigma}\gamma_{\mu\nu} =\left(1\pm\gamma^{5}\right)\left(\eta_{\sigma\mu}\eta_{\rho\nu}-\eta_{\rho\mu}\eta_{\sigma\nu}+\eta_{\rho\mu}\gamma_{\nu\sigma}+\eta_{\sigma\mu}\gamma_{\rho\nu}+\eta_{\rho\nu}\gamma_{\sigma\mu}+\eta_{\sigma\nu}\gamma_{\mu\rho}+i\epsilon_{\rho\sigma\mu\nu}\gamma^{5}\right).
\end{align}
Putting it together, this second term is (for $d=4$):
\begin{align}
& \left[\cancel{p}_{2}\gamma^{\mu\nu}\left(1\pm\gamma^{5}\right)\gamma^{\rho\sigma}\cancel{p}_{1}\right]_{\gamma\dot{\gamma}}\left[\cancel{p}_{4}\left(1\pm\gamma^{5}\right)\gamma_{\rho\sigma}\gamma_{\mu\nu}\cancel{p}_{3}\right]_{\delta\dot{\delta}} =\nonumber \\
& =\left[\cancel{p}_{2}\left(1\pm\gamma^{5}\right)\left(\eta^{\nu\rho}\eta^{\mu\sigma}-\eta^{\mu\rho}\eta^{\nu\sigma}+\eta^{\mu\rho}\gamma^{\sigma\nu}+\eta^{\nu\rho}\gamma^{\mu\sigma}+\eta^{\mu\sigma}\gamma^{\nu\rho}+\eta^{\nu\sigma}\gamma^{\rho\mu}+i\epsilon^{\mu\nu\rho\sigma}\gamma^{5}\right)\cancel{p}_{1}\right]_{\gamma\dot{\gamma}} \nonumber \\
& \cdot\left[\cancel{p}_{4}\left(1\pm\gamma^{5}\right)\left(\eta_{\sigma\mu}\eta_{\rho\nu}-\eta_{\rho\mu}\eta_{\sigma\nu}+\eta_{\rho\mu}\gamma_{\nu\sigma}+\eta_{\sigma\mu}\gamma_{\rho\nu}+\eta_{\rho\nu}\gamma_{\sigma\mu}+\eta_{\sigma\nu}\gamma_{\mu\rho}+i\epsilon_{\rho\sigma\mu\nu}\gamma^{5}\right)\cancel{p}_{3}\right]_{\delta\dot{\delta}}= \nonumber
\end{align}
\begin{align}
& =\left(\eta^{\nu\rho}\eta^{\mu\sigma}-\eta^{\mu\rho}\eta^{\nu\sigma}\right)\left(\eta_{\sigma\mu}\eta_{\rho\nu}-\eta_{\rho\mu}\eta_{\sigma\nu}\right)\left[\cancel{p}_{2}\left(1\pm\gamma^{5}\right)\cancel{p}_{1}\right]_{\gamma\dot{\gamma}}\left[\cancel{p}_{4}\left(1\pm\gamma^{5}\right)\cancel{p}_{3}\right]_{\delta\dot{\delta}} \nonumber \\
& +\left[\cancel{p}_{2}\left(1\pm\gamma^{5}\right)\left(\eta^{\mu\rho}\gamma^{\sigma\nu}+\eta^{\nu\rho}\gamma^{\mu\sigma}+\eta^{\mu\sigma}\gamma^{\nu\rho}+\eta^{\nu\sigma}\gamma^{\rho\mu}\right)\cancel{p}_{1}\right]_{\gamma\dot{\gamma}}\cdot\nonumber \\
& \cdot\left[\cancel{p}_{4}\left(1\pm\gamma^{5}\right)\left(\eta_{\rho\mu}\gamma_{\nu\sigma}+\eta_{\sigma\mu}\gamma_{\rho\nu}+\eta_{\rho\nu}\gamma_{\sigma\mu}+\eta_{\sigma\nu}\gamma_{\mu\rho}\right)\cancel{p}_{3}\right]_{\delta\dot{\delta}} \nonumber \\
& -\underbrace{\epsilon^{\mu\nu\rho\sigma}\epsilon_{\rho\sigma\mu\nu}}_{=\epsilon^{\mu\nu\rho\sigma}\epsilon_{\mu\nu\rho\sigma}=-24}\left[\cancel{p}_{2}\left(1\pm\gamma^{5}\right)\gamma^{5}\cancel{p}_{1}\right]_{\gamma\dot{\gamma}}\left[\cancel{p}_{4}\left(1\pm\gamma^{5}\right)\gamma^{5}\cancel{p}_{3}\right]_{\delta\dot{\delta}}= \nonumber
\end{align}
\begin{align}
 & =24\left[\cancel{p}_{2}\left(1\pm\gamma^{5}\right)\cancel{p}_{1}\right]_{\gamma\dot{\gamma}}\left[\cancel{p}_{4}\left(1\pm\gamma^{5}\right)\cancel{p}_{3}\right]_{\delta\dot{\delta}}
 -8\left[\cancel{p}_{2}\left(1\pm\gamma^{5}\right)\gamma^{\sigma\nu}\cancel{p}_{1}\right]_{\gamma\dot{\gamma}}\left[\cancel{p}_{4}\left(1\pm\gamma^{5}\right)\gamma_{\sigma\nu}\cancel{p}_{3}\right]_{\delta\dot{\delta}} \nonumber \\
 & +24\left[\cancel{p}_{2}\left(1\pm\gamma^{5}\right)\gamma^{5}\cancel{p}_{1}\right]_{\gamma\dot{\gamma}}\left[\cancel{p}_{4}\left(1\pm\gamma^{5}\right)\gamma^{5}\cancel{p}_{3}\right]_{\delta\dot{\delta}}=\cdots
\end{align}
Since $(1\pm \gamma^5)\gamma^5 = \pm (1\pm\gamma^5)$,  the last term here is 
\begin{equation}
24\left[\cancel{p}_{2}\left(1\pm\gamma^{5}\right)\gamma^{5}\cancel{p}_{1}\right]_{\gamma\dot{\gamma}}\left[\cancel{p}_{4}\left(1\pm\gamma^{5}\right)\gamma^{5}\cancel{p}_{3}\right]_{\delta\dot{\delta}} 
=24\left[\cancel{p}_{2}\left(1\pm\gamma^{5}\right)\cancel{p}_{1}\right]_{\gamma\dot{\gamma}}\left[\cancel{p}_{4}\left(1\pm\gamma^{5}\right)\cancel{p}_{3}\right]_{\delta\dot{\delta}},
\end{equation}
so the total contribution of the diagram is:
\begin{equation}
=12\left[\cancel{p}_{2}\left(1\pm\gamma^{5}\right)\gamma^{\sigma\nu}\cancel{p}_{1}\right]_{\gamma\dot{\gamma}}\left[\cancel{p}_{4}\left(1\pm\gamma^{5}\right)\gamma_{\sigma\nu}\cancel{p}_{3}\right]_{\delta\dot{\delta}}
-48\left[\cancel{p}_{2}\left(1\pm\gamma^{5}\right)\cancel{p}_{1}\right]_{\gamma\dot{\gamma}}\left[\cancel{p}_{4}\left(1\pm\gamma^{5}\right)\cancel{p}_{3}\right]_{\delta\dot{\delta}}.
\end{equation}
Diagram II:
\begin{align}
\cancel{p_{2}}_{B}\cancel{p_{3}}_{C}\left[\cancel{p_{4}}\gamma^{\mu}\gamma^{\nu}\right]_{D}\left[\gamma_{\nu}\gamma_{\mu}\cancel{p_{1}}\right]_{A}
\goto\left[\cancel{p}_{2}\left(1\pm\gamma^{5}\right)\gamma^{\rho\sigma}\gamma_{\nu}\gamma_{\mu}\cancel{p}_{1}\right]_{\gamma\dot{\gamma}}\left[\cancel{p}_{4}\gamma^{\mu}\gamma^{\nu}\left(1\pm\gamma^{5}\right)\gamma_{\rho\sigma}\cancel{p}_{3}\right]_{\delta\dot{\delta}}= \nonumber\\
=\left[\cancel{p}_{2}\left(1\pm\gamma^{5}\right)\gamma^{\rho\sigma}\left(\eta^{\mu\nu}-\gamma^{\mu\nu}\right)\cancel{p}_{1}\right]_{\gamma\dot{\gamma}}\left[\cancel{p}_{4}\left(\eta_{\mu\nu}+\gamma_{\mu\nu}\right)\left(1\pm\gamma^{5}\right)\gamma_{\rho\sigma}\cancel{p}_{3}\right]_{\delta\dot{\delta}}= \nonumber\\
=\cdots\approx12\left[\cancel{p}_{2}\left(1\pm\gamma^{5}\right)\gamma^{\sigma\nu}\cancel{p}_{1}\right]_{\gamma\dot{\gamma}}\left[\cancel{p}_{4}\left(1\pm\gamma^{5}\right)\gamma_{\sigma\nu}\cancel{p}_{3}\right]_{\delta\dot{\delta}}
-48\left[\cancel{p}_{2}\left(1\pm\gamma^{5}\right)\cancel{p}_{1}\right]_{\gamma\dot{\gamma}}\left[\cancel{p}_{4}\left(1\pm\gamma^{5}\right)\cancel{p}_{3}\right]_{\delta\dot{\delta}}.
\end{align}
Diagram III:
\begin{align}
\cancel{p_{2}}_{B}\cancel{p_{4}}_{D}\left[\gamma^{\nu}\gamma^{\mu}\cancel{p_{3}}\right]_{C}\left[\gamma_{\nu}\gamma_{\mu}\cancel{p_{1}}\right]_{A}
\goto\left[\cancel{p}_{2}\left(1\pm\gamma^{5}\right)\gamma^{\rho\sigma}\gamma_{\nu}\gamma_{\mu}\cancel{p}_{1}\right]_{\gamma\dot{\gamma}}\left[\cancel{p}_{4}\left(1\pm\gamma^{5}\right)\gamma_{\rho\sigma}\gamma^{\nu}\gamma^{\mu}\cancel{p}_{3}\right]_{\delta\dot{\delta}}= \nonumber\\
=\left[\cancel{p}_{2}\left(1\pm\gamma^{5}\right)\gamma^{\rho\sigma}\left(\eta^{\mu\nu}-\gamma^{\mu\nu}\right)\cancel{p}_{1}\right]_{\gamma\dot{\gamma}}\left[\cancel{p}_{4}\left(1\pm\gamma^{5}\right)\gamma_{\rho\sigma}\left(\eta_{\mu\nu}-\gamma_{\mu\nu}\right)\cancel{p}_{3}\right]_{\delta\dot{\delta}}=\nonumber\\
=4\left[\cancel{p}_{2}\left(1\pm\gamma^{5}\right)\gamma^{\rho\sigma}\cancel{p}_{1}\right]_{\gamma\dot{\gamma}}\left[\cancel{p}_{4}\left(1\pm\gamma^{5}\right)\gamma_{\rho\sigma}\cancel{p}_{3}\right]_{\delta\dot{\delta}}
+\left[\cancel{p}_{2}\left(1\pm\gamma^{5}\right)\gamma^{\rho\sigma}\gamma^{\mu\nu}\cancel{p}_{1}\right]_{\gamma\dot{\gamma}}\left[\cancel{p}_{4}\left(1\pm\gamma^{5}\right)\gamma_{\rho\sigma}\gamma_{\mu\nu}\cancel{p}_{3}\right]_{\delta\dot{\delta}} & =\cdots
\end{align}
This time the two tensors $\gamma^{\mu\nu},\gamma^{\rho\sigma}$ appear
in the same order in both matix elements. This is almost identical
to the previous result, and only changes the sign of the terms of
the form $\eta^{\mu\rho}\gamma^{\nu\sigma}$ etc., in one of the factors.
Then the second term gives:
\begin{align}
24\left[\cancel{p}_{2}\left(1\pm\gamma^{5}\right)\cancel{p}_{1}\right]_{\gamma\dot{\gamma}}\left[\cancel{p}_{4}\left(1\pm\gamma^{5}\right)\cancel{p}_{3}\right]_{\delta\dot{\delta}}
+8\left[\cancel{p}_{2}\left(1\pm\gamma^{5}\right)\gamma^{\sigma\nu}\cancel{p}_{1}\right]_{\gamma\dot{\gamma}}\left[\cancel{p}_{4}\left(1\pm\gamma^{5}\right)\gamma_{\sigma\nu}\cancel{p}_{3}\right]_{\delta\dot{\delta}}+\nonumber \\
+24\left[\cancel{p}_{2}\left(1\pm\gamma^{5}\right)\cancel{p}_{1}\right]_{\gamma\dot{\gamma}}\left[\cancel{p}_{4}\left(1\pm\gamma^{5}\right)\cancel{p}_{3}\right]_{\delta\dot{\delta}},
\end{align}
 and the contribution of the diagram is 
\begin{equation}
12\left[\cancel{p}_{2}\left(1\pm\gamma^{5}\right)\gamma^{\sigma\nu}\cancel{p}_{1}\right]_{\gamma\dot{\gamma}}\left[\cancel{p}_{4}\left(1\pm\gamma^{5}\right)\gamma_{\sigma\nu}\cancel{p}_{3}\right]_{\delta\dot{\delta}}+48\left[\cancel{p}_{2}\left(1\pm\gamma^{5}\right)\cancel{p}_{1}\right]_{\gamma\dot{\gamma}}\left[\cancel{p}_{4}\left(1\pm\gamma^{5}\right)\cancel{p}_{3}\right]_{\delta\dot{\delta}}.
\end{equation}
Diagram IV similarly gives:
\begin{align}
\cancel{p_{1}}_{A}\cancel{p_{3}}_{C}\left[\cancel{p_{4}}\gamma^{\mu}\gamma^{\nu}\right]_{D}\left[\cancel{p_{2}}\gamma_{\mu}\gamma_{\nu}\right]_{B}
\goto\left[\cancel{p}_{2}\gamma_{\mu}\gamma_{\nu}\left(1\pm\gamma^{5}\right)\gamma^{\rho\sigma}\cancel{p}_{1}\right]_{\gamma\dot{\gamma}}\left[\cancel{p}_{4}\gamma^{\mu}\gamma^{\nu}\left(1\pm\gamma^{5}\right)\gamma_{\rho\sigma}\cancel{p}_{3}\right]_{\delta\dot{\delta}}=\nonumber \\
=\left[\cancel{p}_{2}\left(\eta^{\mu\nu}+\gamma^{\mu\nu}\right)\left(1\pm\gamma^{5}\right)\gamma^{\rho\sigma}\cancel{p}_{1}\right]_{\gamma\dot{\gamma}}\left[\cancel{p}_{4}\left(\eta_{\mu\nu}+\gamma_{\mu\nu}\right)\left(1\pm\gamma^{5}\right)\gamma_{\rho\sigma}\cancel{p}_{3}\right]_{\delta\dot{\delta}}=\nonumber \\
=\cdots=12\left[\cancel{p}_{2}\left(1\pm\gamma^{5}\right)\gamma^{\sigma\nu}\cancel{p}_{1}\right]_{\gamma\dot{\gamma}}\left[\cancel{p}_{4}\left(1\pm\gamma^{5}\right)\gamma_{\sigma\nu}\cancel{p}_{3}\right]_{\delta\dot{\delta}}+48\left[\cancel{p}_{2}\left(1\pm\gamma^{5}\right)\cancel{p}_{1}\right]_{\gamma\dot{\gamma}}\left[\cancel{p}_{4}\left(1\pm\gamma^{5}\right)\cancel{p}_{3}\right]_{\delta\dot{\delta}}.
\end{align}

\paragraph{Chiral vectors.}

Now we contract the amplitudes with the chiral vector spinor structure:
$\left[\left(1\pm\gamma^{5}\right)\gamma^{\rho}\right]_{\alpha\dot{\alpha}}\left[\left(1\pm\gamma^{5}\right)\gamma^{\sigma}\right]_{\beta\dot{\beta}}$,
where $\rho,\sigma$ are for now independent. Note that in this case,
the combination of the two permutations generally does not yield the
original configuration, because of the contraction with possibly
different $\gamma$ matrices. Thus, it is necessary to explicitly
symmetrize by both permutations $\psi_{1} \swap\psi_{3}$
and $\overline{\psi}_{2} \swap\overline{\psi}_{4}$,
which is equivalent to symmetrizing by the permutations $\psi_{1} \swap\psi_{3}$ and $\gamma^{\rho}\swap\gamma^{\sigma}$.
It follows that we can keep treating only the two branches we discussed before, but impose that the indices will eventually need to be symmetric.

Tree level:
\begin{equation}
\cancel{p_{1}}_{\alpha\dot{\gamma}}\cancel{p_{2}}_{\gamma\dot{\alpha}}\cancel{p_{3}}_{\beta\dot{\delta}}\cancel{p_{4}}_{\delta\dot{\beta}}\goto\left[\cancel{p}_{2}\left(1\pm\gamma^{5}\right)\gamma^{\rho}\cancel{p}_{1}\right]_{\gamma\dot{\gamma}}\left[\cancel{p}_{4}\left(1\pm\gamma^{5}\right)\gamma^{\sigma}\cancel{p}_{3}\right]_{\delta\dot{\delta}}.
\end{equation}
Diagram I:
\begin{align}
\cancel{p_{1}}_{A}\cancel{p_{4}}_{D}\left[\cancel{p_{2}}\gamma^{\mu}\gamma^{\nu}\right]_{B}\left[\gamma_{\nu}\gamma_{\mu}\cancel{p_{3}}\right]_{C} & \goto\left[\cancel{p_{2}}\gamma^{\mu}\gamma^{\nu}\left(1\pm\gamma^{5}\right)\gamma^{\rho}\cancel{p_{1}}\right]_{\gamma\dot{\gamma}}\left[\cancel{p_{4}}\left(1\pm\gamma^{5}\right)\gamma^{\sigma}\gamma_{\nu}\gamma_{\mu}\cancel{p_{3}}\right]_{\delta\dot{\delta}}= \nonumber\\
 & =\left[\cancel{p_{2}}\left(1\pm\gamma^{5}\right)\gamma^{\mu}\gamma^{\nu}\gamma^{\rho}\cancel{p_{1}}\right]_{\gamma\dot{\gamma}}\left[\cancel{p_{4}}\left(1\pm\gamma^{5}\right)\gamma^{\sigma}\gamma_{\nu}\gamma_{\mu}\cancel{p_{3}}\right]_{\delta\dot{\delta}}=\cdots
\end{align}
Now we simplify using the formula (\ref{eq:7}) for the product of 3 $\gamma$ matrics to get:
\begin{align}
\cdots & =\left[\cancel{p_{2}}\left(1\pm\gamma^{5}\right)\left(\eta^{\mu\nu}\gamma^{\rho}-\eta^{\mu\rho}\gamma^{\nu}+\eta^{\nu\rho}\gamma^{\mu}+i\gamma^{5}\gamma_{\tau}\epsilon^{\mu\nu\rho\tau}\right)\cancel{p_{1}}\right]_{\gamma\dot{\gamma}} \nonumber \\
 & \cdot\left[\cancel{p_{4}}\left(1\pm\gamma^{5}\right)\left(\delta_{\nu}^{\sigma}\gamma_{\mu}-\delta_{\mu}^{\sigma}\gamma_{\nu}+\eta_{\nu\mu}\gamma^{\sigma}+i\gamma^{5}\gamma_{\omega}\epsilon_{\ \nu\mu}^{\sigma\ \ \omega}\right)\cancel{p_{3}}\right]_{\delta\dot{\delta}}= \nonumber\\
 & =\eta^{\mu\sigma}\left[\cancel{p_{2}}\left(1\pm\gamma^{5}\right)\gamma^{\rho}\cancel{p_{1}}\right]_{\gamma\dot{\gamma}}\left[\cancel{p_{4}}\left(1\pm\gamma^{5}\right)\gamma_{\mu}\cancel{p_{3}}\right]_{\delta\dot{\delta}}-\eta^{\sigma\nu}\left[\cancel{p_{2}}\left(1\pm\gamma^{5}\right)\gamma^{\rho}\cancel{p_{1}}\right]_{\gamma\dot{\gamma}}\left[\cancel{p_{4}}\left(1\pm\gamma^{5}\right)\gamma_{\nu}\cancel{p_{3}}\right]_{\delta\dot{\delta}} \nonumber \\
 & +d\left[\cancel{p_{2}}\left(1\pm\gamma^{5}\right)\gamma^{\rho}\cancel{p_{1}}\right]_{\gamma\dot{\gamma}}\left[\cancel{p_{4}}\left(1\pm\gamma^{5}\right)\gamma^{\sigma}\cancel{p_{3}}\right]_{\delta\dot{\delta}}-\left[\cancel{p_{2}}\left(1\pm\gamma^{5}\right)\gamma^{\sigma}\cancel{p_{1}}\right]_{\gamma\dot{\gamma}}\left[\cancel{p_{4}}\left(1\pm\gamma^{5}\right)\gamma^{\rho}\cancel{p_{3}}\right]_{\delta\dot{\delta}} \nonumber \\
 & +\eta^{\sigma\rho}\left[\cancel{p_{2}}\left(1\pm\gamma^{5}\right)\gamma^{\nu}\cancel{p_{1}}\right]_{\gamma\dot{\gamma}}\left[\cancel{p_{4}}\left(1\pm\gamma^{5}\right)\gamma_{\nu}\cancel{p_{3}}\right]_{\delta\dot{\delta}}-\left[\cancel{p_{2}}\left(1\pm\gamma^{5}\right)\gamma^{\rho}\cancel{p_{1}}\right]_{\gamma\dot{\gamma}}\left[\cancel{p_{4}}\left(1\pm\gamma^{5}\right)\gamma^{\sigma}\cancel{p_{3}}\right]_{\delta\dot{\delta}} \nonumber \\
 & +\eta^{\sigma\rho}\left[\cancel{p_{2}}\left(1\pm\gamma^{5}\right)\gamma^{\mu}\cancel{p_{1}}\right]_{\gamma\dot{\gamma}}\left[\cancel{p_{4}}\left(1\pm\gamma^{5}\right)\gamma_{\mu}\cancel{p_{3}}\right]_{\delta\dot{\delta}}-\left[\cancel{p_{2}}\left(1\pm\gamma^{5}\right)\gamma^{\sigma}\cancel{p_{1}}\right]_{\gamma\dot{\gamma}}\left[\cancel{p_{4}}\left(1\pm\gamma^{5}\right)\gamma^{\rho}\cancel{p_{3}}\right]_{\delta\dot{\delta}} \nonumber \\
 & +\delta_{\mu}^{\rho}\left[\cancel{p_{2}}\left(1\pm\gamma^{5}\right)\gamma^{\mu}\cancel{p_{1}}\right]_{\gamma\dot{\gamma}}\left[\cancel{p_{4}}\left(1\pm\gamma^{5}\right)\gamma^{\sigma}\cancel{p_{3}}\right]_{\delta\dot{\delta}}+\cancel{2i\epsilon^{\mu\sigma\rho\tau}\left[\cancel{p_{2}}\left(1\pm\gamma^{5}\right)\gamma^{5}\gamma_{\tau}\cancel{p_{1}}\right]_{\gamma\dot{\gamma}}\left[\cancel{p_{4}}\left(1\pm\gamma^{5}\right)\gamma_{\mu}\cancel{p_{3}}\right]_{\delta\dot{\delta}}} \nonumber \\
 & +\cancel{2i\epsilon_{\ \ \mu}^{\sigma\rho\ \omega}\left[\cancel{p_{2}}\left(1\pm\gamma^{5}\right)\gamma^{\mu}\cancel{p_{1}}\right]_{\gamma\dot{\gamma}}\left[\cancel{p_{4}}\left(1\pm\gamma^{5}\right)\gamma^{5}\gamma_{\omega}\cancel{p_{3}}\right]_{\delta\dot{\delta}}}+ \epsilon^{\mu\nu\rho\tau}\epsilon_{\mu\nu}^{\ \ \sigma\omega}\left[\cancel{p_{2}}\left(1\pm\gamma^{5}\right)\gamma^{5}\gamma_{\tau}\cancel{p_{1}}\right]_{\gamma\dot{\gamma}}\left[\cancel{p_{4}}\left(1\pm\gamma^{5}\right)\gamma^{5}\gamma_{\omega}\cancel{p_{3}}\right]_{\delta\dot{\delta}}\overset{d\approx4}{\approx} \nonumber \\
 & \nonumber \\
& \approx 4\left[\cancel{p_{2}}\left(1\pm\gamma^{5}\right)\gamma^{\rho}\cancel{p_{1}}\right]_{\gamma\dot{\gamma}}\left[\cancel{p_{4}}\left(1\pm\gamma^{5}\right)\gamma^{\sigma}\cancel{p_{3}}\right]_{\delta\dot{\delta}}.
\end{align}
Diagram II:
\begin{align}
\cancel{p_{2}}_{B}\cancel{p_{3}}_{C}\left[\cancel{p_{4}}\gamma^{\mu}\gamma^{\nu}\right]_{D}\left[\gamma_{\nu}\gamma_{\mu}\cancel{p_{1}}\right]_{A} & \goto\left[\cancel{p}_{2}\left(1\pm\gamma^{5}\right)\gamma^{\rho}\gamma_{\nu}\gamma_{\mu}\cancel{p}_{1}\right]_{\gamma\dot{\gamma}}\left[\cancel{p}_{4}\gamma^{\mu}\gamma^{\nu}\left(1\pm\gamma^{5}\right)\gamma^{\sigma}\cancel{p}_{3}\right]_{\delta\dot{\delta}}= \nonumber \\
 & =\left[\cancel{p}_{2}\left(1\pm\gamma^{5}\right)\gamma^{\rho}\gamma^{\nu}\gamma^{\mu}\cancel{p}_{1}\right]_{\gamma\dot{\gamma}}\left[\cancel{p}_{4}\left(1\pm\gamma^{5}\right)\gamma_{\mu}\gamma_{\nu}\gamma^{\sigma}\cancel{p}_{3}\right]_{\delta\dot{\delta}}=\cdots \nonumber
 \end{align}
 \begin{align}
 & =\left[\cancel{p}_{2}\left(1\pm\gamma^{5}\right)\left(\eta^{\rho\nu}\gamma^{\mu}-\eta^{\rho\mu}\gamma^{\nu}+\eta^{\nu\mu}\gamma^{\rho}+i\gamma^{5}\gamma_{\tau}\epsilon^{\rho\nu\mu\tau}\right)\cancel{p}_{1}\right]_{\gamma\dot{\gamma}}
 \left[\cancel{p}_{4}\left(1\pm\gamma^{5}\right)\left(\eta_{\mu\nu}\gamma^{\sigma}-\delta_{\mu}^{\sigma}\gamma_{\nu}+\delta_{\nu}^{\sigma}\gamma_{\mu}+i\gamma^{5}\gamma^{\omega}\epsilon_{\mu\nu\ \omega}^{\ \ \sigma}\right)\cancel{p}_{3}\right]_{\delta\dot{\delta}}= \nonumber\\
 & =\left[\cancel{p_{2}}\left(1\pm\gamma^{5}\right)\left(\eta^{\mu\nu}\gamma^{\rho}-\eta^{\mu\rho}\gamma^{\nu}+\eta^{\nu\rho}\gamma^{\mu}-i\gamma^{5}\gamma_{\tau}\epsilon^{\mu\nu\rho\tau}\right)\cancel{p_{1}}\right]_{\gamma\dot{\gamma}}
 \left[\cancel{p_{4}}\left(1\pm\gamma^{5}\right)\left(\delta_{\nu}^{\sigma}\gamma_{\mu}-\delta_{\mu}^{\sigma}\gamma_{\nu}+\eta_{\nu\mu}\gamma^{\sigma}-i\gamma^{5}\gamma_{\omega}\epsilon_{\ \nu\mu}^{\sigma\ \ \omega}\right)\cancel{p_{3}}\right]_{\delta\dot{\delta}},
\end{align}
meaning we get a relative $\left(-\right)$ sign for every term in
the diagram I expression that includes an additional $\gamma^{5}$.
In the same presentation order as that calculation, the last term
gets $\left(-\right)^{2}=\left(+\right),$ and the two before it get
a $\left(-\right)$ sign. Finally we have:
\begin{equation}
\cdots=4\left[\cancel{p_{2}}\left(1\pm\gamma^{5}\right)\gamma^{\rho}\cancel{p_{1}}\right]_{\gamma\dot{\gamma}}\left[\cancel{p_{4}}\left(1\pm\gamma^{5}\right)\gamma^{\sigma}\cancel{p_{3}}\right]_{\delta\dot{\delta}}.
\end{equation}
Diagram III:
\begin{equation}
    \cancel{p_{2}}_{B}\cancel{p_{4}}_{D}\left[\gamma^{\nu}\gamma^{\mu}\cancel{p_{3}}\right]_{C}\left[\gamma_{\nu}\gamma_{\mu}\cancel{p_{1}}\right]_{A} \goto\left[\cancel{p}_{2}\left(1\pm\gamma^{5}\right)\gamma^{\rho}\gamma^{\nu}\gamma^{\mu}\cancel{p}_{1}\right]_{\gamma\dot{\gamma}}\left[\cancel{p}_{4}\left(1\pm\gamma^{5}\right)\gamma^{\sigma}\gamma_{\nu}\gamma_{\mu}\cancel{p}_{3}\right]_{\delta\dot{\delta}}= \nonumber
\end{equation}
\begin{align}
 & =\left[\cancel{p}_{2}\left(1\pm\gamma^{5}\right)\left(\eta^{\rho\nu}\gamma^{\mu}-\eta^{\rho\mu}\gamma^{\nu}+\eta^{\nu\mu}\gamma^{\rho}+i\gamma^{5}\gamma_{\tau}\epsilon^{\rho\nu\mu\tau}\right)\cancel{p}_{1}\right]_{\gamma\dot{\gamma}}
 \left[\cancel{p}_{4}\left(1\pm\gamma^{5}\right)\left(\delta_{\nu}^{\sigma}\gamma_{\mu}-\delta_{\mu}^{\sigma}\gamma_{\nu}+\eta_{\mu\nu}\gamma^{\sigma}+i\gamma^{5}\gamma_{\omega}\epsilon_{\ \nu\mu}^{\sigma\ \ \omega}\right)\cancel{p}_{3}\right]_{\delta\dot{\delta}} = \nonumber\\
 & =\left[\cancel{p_{2}}\left(1\pm\gamma^{5}\right)\left(\eta^{\mu\nu}\gamma^{\rho}-\eta^{\mu\rho}\gamma^{\nu}+\eta^{\nu\rho}\gamma^{\mu}-i\gamma^{5}\gamma_{\tau}\epsilon^{\mu\nu\rho\tau}\right)\cancel{p_{1}}\right]_{\gamma\dot{\gamma}}
 \left[\cancel{p_{4}}\left(1\pm\gamma^{5}\right)\left(\delta_{\nu}^{\sigma}\gamma_{\mu}-\delta_{\mu}^{\sigma}\gamma_{\nu}+\eta_{\nu\mu}\gamma^{\sigma}+i\gamma^{5}\gamma_{\omega}\epsilon_{\ \nu\mu}^{\sigma\ \ \omega}\right)\cancel{p_{3}}\right]_{\delta\dot{\delta}}=\cdots
\end{align}
This time only the first $\gamma^{5}\gamma_{\tau}\epsilon^{\mu\nu\rho\tau}$
gets a $\left(-\right)$ sign, so we get:
\begin{align}
 & \cdots\approx4\left[\cancel{p_{2}}\left(1\pm\gamma^{5}\right)\gamma^{\rho}\cancel{p_{1}}\right]_{\gamma\dot{\gamma}}\left[\cancel{p_{4}}\left(1\pm\gamma^{5}\right)\gamma^{\sigma}\cancel{p_{3}}\right]_{\delta\dot{\delta}}-2\left[\cancel{p_{2}}\left(1\pm\gamma^{5}\right)\gamma^{\sigma}\cancel{p_{1}}\right]_{\gamma\dot{\gamma}}\left[\cancel{p_{4}}\left(1\pm\gamma^{5}\right)\gamma^{\rho}\cancel{p_{3}}\right]_{\delta\dot{\delta}} \nonumber \\
 & +2\eta^{\sigma\rho}\left[\cancel{p_{2}}\left(1\pm\gamma^{5}\right)\gamma^{\mu}\cancel{p_{1}}\right]_{\gamma\dot{\gamma}}\left[\cancel{p_{4}}\left(1\pm\gamma^{5}\right)\gamma_{\mu}\cancel{p_{3}}\right]_{\delta\dot{\delta}}\mp 2i\epsilon^{\mu\sigma\rho\tau}\left[\cancel{p_{2}}\left(1\pm\gamma^{5}\right)\gamma_{\tau}\cancel{p_{1}}\right]_{\gamma\dot{\gamma}}\left[\cancel{p_{4}}\left(1\pm\gamma^{5}\right)\gamma_{\mu}\cancel{p_{3}}\right]_{\delta\dot{\delta}} \nonumber \\
 & \pm2i\epsilon^{\sigma\rho\mu\tau}\left[\cancel{p_{2}}\left(1\pm\gamma^{5}\right)\gamma_{\mu}\cancel{p_{1}}\right]_{\gamma\dot{\gamma}}\left[\cancel{p_{4}}\left(1\pm\gamma^{5}\right)\gamma_{\tau}\cancel{p_{3}}\right]_{\delta\dot{\delta}}+2\eta^{\rho\sigma}\left[\cancel{p_{2}}\left(1\pm\gamma^{5}\right)\gamma_{\tau}\cancel{p_{1}}\right]_{\gamma\dot{\gamma}}\left[\cancel{p_{4}}\left(1\pm\gamma^{5}\right)\gamma^{\tau}\cancel{p_{3}}\right]_{\delta\dot{\delta}} \nonumber \\
 & -2\left[\cancel{p_{2}}\left(1\pm\gamma^{5}\right)\gamma^{\sigma}\cancel{p_{1}}\right]_{\gamma\dot{\gamma}}\left[\cancel{p_{4}}\left(1\pm\gamma^{5}\right)\gamma^{\rho}\cancel{p_{3}}\right]_{\delta\dot{\delta}}= \nonumber\\
 & =4\left[\cancel{p_{2}}\left(1\pm\gamma^{5}\right)\gamma^{\rho}\cancel{p_{1}}\right]_{\gamma\dot{\gamma}}\left[\cancel{p_{4}}\left(1\pm\gamma^{5}\right)\gamma^{\sigma}\cancel{p_{3}}\right]_{\delta\dot{\delta}}-4\left[\cancel{p_{2}}\left(1\pm\gamma^{5}\right)\gamma^{\sigma}\cancel{p_{1}}\right]_{\gamma\dot{\gamma}}\left[\cancel{p_{4}}\left(1\pm\gamma^{5}\right)\gamma^{\rho}\cancel{p_{3}}\right]_{\delta\dot{\delta}}\nonumber \\
 & +4\eta^{\sigma\rho}\left[\cancel{p_{2}}\left(1\pm\gamma^{5}\right)\gamma^{\mu}\cancel{p_{1}}\right]_{\gamma\dot{\gamma}}\left[\cancel{p_{4}}\left(1\pm\gamma^{5}\right)\gamma_{\mu}\cancel{p_{3}}\right]_{\delta\dot{\delta}} \mp 4i\epsilon^{\mu\sigma\rho\tau}\left[\cancel{p_{2}}\left(1\pm\gamma^{5}\right)\gamma_{\tau}\cancel{p_{1}}\right]_{\gamma\dot{\gamma}}\left[\cancel{p_{4}}\left(1\pm\gamma^{5}\right)\gamma_{\mu}\cancel{p_{3}}\right]_{\delta\dot{\delta}}.
\end{align}
Diagram IV:
\begin{align}
 & \cancel{p_{1}}_{A}\cancel{p_{3}}_{C}\left[\cancel{p_{4}}\gamma^{\mu}\gamma^{\nu}\right]_{D}\left[\cancel{p_{2}}\gamma_{\mu}\gamma_{\nu}\right]_{B}
 \goto\left[\cancel{p}_{2}\gamma_{\mu}\gamma_{\nu}\left(1\pm\gamma^{5}\right)\gamma^{\rho}\cancel{p}_{1}\right]_{\gamma\dot{\gamma}}\left[\cancel{p}_{4}\gamma^{\mu}\gamma^{\nu}\left(1\pm\gamma^{5}\right)\gamma^{\sigma}\cancel{p}_{3}\right]_{\delta\dot{\delta}}= \nonumber\\
 & =\left[\cancel{p}_{2}\left(1\pm\gamma^{5}\right)\gamma_{\mu}\gamma_{\nu}\gamma^{\rho}\cancel{p}_{1}\right]_{\gamma\dot{\gamma}}\left[\cancel{p}_{4}\left(1\pm\gamma^{5}\right)\gamma^{\mu}\gamma^{\nu}\gamma^{\sigma}\cancel{p}_{3}\right]_{\delta\dot{\delta}}= \nonumber\\
 & =\left[\cancel{p_{2}}\left(1\pm\gamma^{5}\right)\left(\eta^{\mu\nu}\gamma^{\rho}-\eta^{\mu\rho}\gamma^{\nu}+\eta^{\nu\rho}\gamma^{\mu}+i\gamma^{5}\gamma_{\tau}\epsilon^{\mu\nu\rho\tau}\right)\cancel{p_{1}}\right]_{\gamma\dot{\gamma}}
 \left[\cancel{p_{4}}\left(1\pm\gamma^{5}\right)\left(\delta_{\nu}^{\sigma}\gamma_{\mu}-\delta_{\mu}^{\sigma}\gamma_{\nu}+\eta_{\nu\mu}\gamma^{\sigma}-i\gamma^{5}\gamma_{\omega}\epsilon_{\ \nu\mu}^{\sigma\ \ \omega}\right)\cancel{p_{3}}\right]_{\delta\dot{\delta}}=\cdots= \nonumber\\
 & =4\left[\cancel{p_{2}}\left(1\pm\gamma^{5}\right)\gamma^{\rho}\cancel{p_{1}}\right]_{\gamma\dot{\gamma}}\left[\cancel{p_{4}}\left(1\pm\gamma^{5}\right)\gamma^{\sigma}\cancel{p_{3}}\right]_{\delta\dot{\delta}}-4\left[\cancel{p_{2}}\left(1\pm\gamma^{5}\right)\gamma^{\sigma}\cancel{p_{1}}\right]_{\gamma\dot{\gamma}}\left[\cancel{p_{4}}\left(1\pm\gamma^{5}\right)\gamma^{\rho}\cancel{p_{3}}\right]_{\delta\dot{\delta}}\nonumber \\
 & +4\eta^{\sigma\rho}\left[\cancel{p_{2}}\left(1\pm\gamma^{5}\right)\gamma^{\mu}\cancel{p_{1}}\right]_{\gamma\dot{\gamma}}\left[\cancel{p_{4}}\left(1\pm\gamma^{5}\right)\gamma_{\mu}\cancel{p_{3}}\right]_{\delta\dot{\delta}} \pm 4i\epsilon^{\mu\sigma\rho\tau}\left[\cancel{p_{2}}\left(1\pm\gamma^{5}\right)\gamma_{\tau}\cancel{p_{1}}\right]_{\gamma\dot{\gamma}}\left[\cancel{p_{4}}\left(1\pm\gamma^{5}\right)\gamma_{\mu}\cancel{p_{3}}\right]_{\delta\dot{\delta}}.
\end{align}

\subsubsection{Resummation}

\paragraph{Chiral scalar bi-meson}
\subparagraph{Intra-meson contributions}

The correlation function for each branch (permuted or unpermuted)
consists of 
\begin{equation}
\text{Tree level}+4\cdot\text{prop. corr.}+2\cdot\text{gluon exchange,}
\end{equation}
which becomes 
\begin{equation}
\frac{1}{p_{1}^{2}p_{2}^{2}p_{3}^{2}p_{4}^{2}}\left[ 1+\frac{4g^{2}C_{F}}{(4\pi)^{2}\epsilon}\left(p_{E}^{2}\right)^{-\epsilon}\right]
\left\{ \delta_{ab}\delta_{cd}\left[\cancel{p_{2}}\left(1\pm\gamma^{5}\right)\cancel{p_{1}}\right]_{\gamma\dot{\gamma}}\left[\cancel{p_{4}}\left(1\pm\gamma^{5}\right)\cancel{p_{3}}\right]_{\delta\dot{\delta}} -\delta_{ad}\delta_{bc}\left[\cancel{p_{2}}\left(1\pm\gamma^{5}\right)\cancel{p_{3}}\right]_{\gamma\dot{\delta}}\left[\cancel{p_{4}}\left(1\pm\gamma^{5}\right)\cancel{p_{1}}\right]_{\delta\dot{\gamma}}\right\}. 
\end{equation}

\subparagraph{Inter-meson contributions}

First, we look at the one-loop unpermuted contribution, and divide it into convenient building blocks:
\begin{itemize}
\item Spinor structure: we have from diagrams I,II
\begin{equation}
4\left[\cancel{p_{2}}\left(1\pm\gamma^{5}\right)\cancel{p_{1}}\right]_{\gamma\dot{\gamma}}\left[\cancel{p_{4}}\left(1\pm\gamma^{5}\right)\cancel{p_{3}}\right]_{\delta\dot{\delta}}-2\left[\cancel{p_{2}}\gamma^{\mu\nu}\cancel{p_{1}}\right]_{\gamma\dot{\gamma}}\left[\cancel{p_{4}}\gamma_{\mu\nu}\cancel{p_{3}}\right]_{\delta\dot{\delta}}\mp i\epsilon_{\mu\nu\rho\sigma}\left[\cancel{p_{2}}\gamma^{\mu\nu}\cancel{p_{1}}\right]_{\gamma\dot{\gamma}}\left[\cancel{p_{4}}\gamma^{\rho\sigma}\cancel{p_{3}}\right]_{\delta\dot{\delta}},
\end{equation}
 and from diagrams III,IV
\begin{equation}
4\left[\cancel{p_{2}}\left(1\pm\gamma^{5}\right)\cancel{p_{1}}\right]_{\gamma\dot{\gamma}}\left[\cancel{p_{4}}\left(1\pm\gamma^{5}\right)\cancel{p_{3}}\right]_{\delta\dot{\delta}}+2\left[\cancel{p_{2}}\gamma^{\mu\nu}\cancel{p_{1}}\right]_{\gamma\dot{\gamma}}\left[\cancel{p_{4}}\gamma_{\mu\nu}\cancel{p_{3}}\right]_{\delta\dot{\delta}}\pm i\epsilon_{\mu\nu\rho\sigma}\left[\cancel{p_{2}}\gamma^{\mu\nu}\cancel{p_{1}}\right]_{\gamma\dot{\gamma}}\left[\cancel{p_{4}}\gamma^{\rho\sigma}\cancel{p_{3}}\right]_{\delta\dot{\delta}}.
\end{equation}
\item All 4 diagrams have the same color structure $\frac{1}{2} \delta_{ad}\delta_{bc}$. The remaining coefficient is
\begin{equation}
\pm\frac{g^{2}}{4\left(4\pi\right)^{2}\epsilon}\frac{\left(p_{E}^{2}\right)^{-\epsilon}}{p_{1}^{2}p_{2}^{2}p_{3}^{2}p_{4}^{2}},
\end{equation}
with a $(+)$ sign for diagrams I,II and a $(-)$ sign for diagrams
III,IV.
\end{itemize}
Summing these 4 diagrams gives:
\begin{equation}
-\delta_{ad}\delta_{bc}\frac{g^{2}}{\left(4\pi\right)^{2}\epsilon}\frac{\left(p_{E}^{2}\right)^{-\epsilon}}{p_{1}^{2}p_{2}^{2}p_{3}^{2}p_{4}^{2}}
\left(\left[\cancel{p_{2}}\gamma^{\mu\nu}\cancel{p_{1}}\right]_{\gamma\dot{\gamma}}\left[\cancel{p_{4}}\gamma_{\mu\nu}\cancel{p_{3}}\right]_{\delta\dot{\delta}}\pm\frac{i}{2}\epsilon_{\mu\nu\rho\sigma}\left[\cancel{p_{2}}\gamma^{\mu\nu}\cancel{p_{1}}\right]_{\gamma\dot{\gamma}}\left[\cancel{p_{4}}\gamma^{\rho\sigma}\cancel{p_{3}}\right]_{\delta\dot{\delta}}\right)=\cdots 
\end{equation}
We see that this is a linear combination of the structures we had in 2-tensor bi-meson operators.
We will show that it is in fact proportional to the structure appearing in the respective chiral
2-tensor operators, and find the proportionality constant. First we note
that:
\begin{equation}
\gamma^{5}\gamma^{\mu\nu}=\frac{i}{2}\epsilon^{\mu\nu\tau\omega}\gamma_{\tau\omega}
\Rightarrow\left(1\pm\gamma^{5}\right)\gamma^{\mu\nu}=\gamma^{\mu\nu}\pm\frac{i}{2}\epsilon^{\mu\nu\tau\omega}\gamma_{\tau\omega} \nonumber
\end{equation}
\begin{equation}
\Rightarrow\left[\left(1\pm\gamma^{5}\right)\gamma^{\mu\nu}\right]_{\alpha\dot{\alpha}}\left[\left(1\pm\gamma^{5}\right)\gamma^{\rho\sigma}\right]_{\beta\dot{\beta}}=\left[\gamma^{\mu\nu}\pm\frac{i}{2}\epsilon^{\mu\nu\tau\omega}\gamma_{\tau\omega}\right]_{\alpha\dot{\alpha}}\left[\gamma^{\rho\sigma}\pm\frac{i}{2}\epsilon^{\rho\sigma\kappa\lambda}\gamma_{\kappa\lambda}\right]_{\beta\dot{\beta}}.
\end{equation}
Contracting this with $\eta_{\mu\rho}\eta_{\nu\sigma}$ then gives
\begin{equation}
\left[\gamma^{\mu\nu}\pm\frac{i}{2}\epsilon^{\mu\nu\tau\omega}\gamma_{\tau\omega}\right]_{\alpha\dot{\alpha}}\left[\gamma_{\mu\nu}\pm\frac{i}{2}\epsilon_{\mu\nu}^{\ \ \kappa\lambda}\gamma_{\kappa\lambda}\right]_{\beta\dot{\beta}}= \nonumber
\end{equation}
\begin{equation}
=\left[\gamma^{\mu\nu}\right]_{\alpha\dot{\alpha}}\left[\gamma_{\mu\nu}\right]_{\beta\dot{\beta}}\pm\frac{i}{2}\epsilon^{\mu\nu\tau\omega}\left[\gamma_{\tau\omega}\right]_{\alpha\dot{\alpha}}\left[\gamma_{\mu\nu}\right]_{\beta\dot{\beta}}\pm\frac{i}{2}\epsilon_{\mu\nu}^{\ \ \kappa\lambda}\left[\gamma^{\mu\nu}\right]_{\alpha\dot{\alpha}}\left[\gamma_{\kappa\lambda}\right]_{\beta\dot{\beta}}-\frac{1}{4}\underbrace{\epsilon^{\mu\nu\tau\omega}\epsilon_{\mu\nu}^{\ \ \kappa\lambda}}_{\approx-2\left(\eta^{\tau\kappa}\eta^{\omega\lambda}-\eta^{\tau\lambda}\eta^{\omega\kappa}\right)}\left[\gamma_{\tau\omega}\right]_{\alpha\dot{\alpha}}\left[\gamma_{\kappa\lambda}\right]_{\beta\dot{\beta}}= \nonumber
\end{equation}
\begin{equation}
=2\left[\gamma^{\mu\nu}\right]_{\alpha\dot{\alpha}}\left[\gamma_{\mu\nu}\right]_{\beta\dot{\beta}}\pm i\epsilon^{\mu\nu\tau\omega}\left[\gamma_{\tau\omega}\right]_{\alpha\dot{\alpha}}\left[\gamma_{\mu\nu}\right]_{\beta\dot{\beta}},
\end{equation}
and so the proportionality constant is $\frac{1}{2}$. Thus: 
\begin{equation}
\cdots= - \delta_{ad}\delta_{bc} \frac{g^{2}}{2\left(4\pi\right)^{2}\epsilon}\frac{\left(p_{E}^{2}\right)^{-\epsilon}}{p_{1}^{2}p_{2}^{2}p_{3}^{2}p_{4}^{2}}\left[\cancel{p}_{2}\left(1\pm\gamma^{5}\right)\gamma^{\mu\nu}\cancel{p}_{1}\right]_{\gamma\dot{\gamma}}\left[\cancel{p}_{4}\left(1\pm\gamma^{5}\right)\gamma_{\mu\nu}\cancel{p}_{3}\right]_{\delta\dot{\delta}}.
\end{equation}

\subparagraph{Final amplitude}

In total, the correlation function is:
\begin{align}
& \left\langle M_{ss}^{\pm}(p)\bar{\psi}(-p_{1})\psi(p_{2}) \bar{\psi}(-p_{3})\psi(p_{4})\right\rangle
=\frac{1}{p_{1}^{2}p_{2}^{2}p_{3}^{2}p_{4}^{2}}\left\{ 1+\frac{4g^{2}C_{F}}{(4\pi)^{2}\epsilon}\left(p_{E}^{2}\right)^{-\epsilon}\right\} \nonumber \\
& \cdot\left\{ \delta_{ab}\delta_{cd}\left[\cancel{p_{2}}\left(1\pm\gamma^{5}\right)\cancel{p_{1}}\right]_{\gamma\dot{\gamma}}\left[\cancel{p_{4}}\left(1\pm\gamma^{5}\right)\cancel{p_{3}}\right]_{\delta\dot{\delta}} -\delta_{ad}\delta_{bc}\left[\cancel{p_{2}}\left(1\pm\gamma^{5}\right)\cancel{p_{3}}\right]_{\gamma\dot{\delta}}\left[\cancel{p_{4}}\left(1\pm\gamma^{5}\right)\cancel{p_{1}}\right]_{\delta\dot{\gamma}}\right\} \nonumber \\
& -\frac{g^{2}}{2\left(4\pi\right)^{2}\epsilon}\frac{\left(p_{E}^{2}\right)^{-\epsilon}}{p_{1}^{2}p_{2}^{2}p_{3}^{2}p_{4}^{2}}  \delta_{ad}\delta_{bc}
\left[\cancel{p}_{2}\left(1\pm\gamma^{5}\right)\gamma^{\mu\nu}\cancel{p}_{1}\right]_{\gamma\dot{\gamma}}\left[\cancel{p}_{4}\left(1\pm\gamma^{5}\right)\gamma_{\mu\nu}\cancel{p}_{3}\right]_{\delta\dot{\delta}} \nonumber \\
& +\frac{g^{2}}{2\left(4\pi\right)^{2}\epsilon}\frac{\left(p_{E}^{2}\right)^{-\epsilon}}{p_{1}^{2}p_{2}^{2}p_{3}^{2}p_{4}^{2}} \delta_{ab}\delta_{cd}
\left[\cancel{p_{2}}\left(1\pm\gamma^{5}\right)\gamma^{\mu\nu}\cancel{p_{3}}\right]_{\gamma\dot{\delta}}\left[\cancel{p_{4}}\left(1\pm\gamma^{5}\right)\gamma_{\mu\nu}\cancel{p_{1}}\right]_{\delta\dot{\gamma}}.
\end{align}
Deducting the fermion renormalization function correction gives:
\begin{align}
& \left\langle M_{ss}^{\pm}(p)\left[\bar{\psi}\right](-p_{1})\left[\psi\right](p_{2})\left[\bar{\psi}\right](-p_{3})\left[\psi\right](p_{4})\right\rangle
=\frac{1}{p_{1}^{2}p_{2}^{2}p_{3}^{2}p_{4}^{2}}\left\{ 1+\frac{6g^{2}C_{F}}{(4\pi)^{2}\epsilon}\left(p_{E}^{2}\right)^{-\epsilon}\right\} \nonumber \\
& \cdot\left\{ \delta_{ab}\delta_{cd}\left[\cancel{p_{2}}\left(1\pm\gamma^{5}\right)\cancel{p_{1}}\right]_{\gamma\dot{\gamma}}\left[\cancel{p_{4}}\left(1\pm\gamma^{5}\right)\cancel{p_{3}}\right]_{\delta\dot{\delta}} -\delta_{ad}\delta_{bc}\left[\cancel{p_{2}}\left(1\pm\gamma^{5}\right)\cancel{p_{3}}\right]_{\gamma\dot{\delta}}\left[\cancel{p_{4}}\left(1\pm\gamma^{5}\right)\cancel{p_{1}}\right]_{\delta\dot{\gamma}}\right\} \nonumber \\
& -\frac{g^{2}}{2\left(4\pi\right)^{2}\epsilon}\frac{\left(p_{E}^{2}\right)^{-\epsilon}}{p_{1}^{2}p_{2}^{2}p_{3}^{2}p_{4}^{2}} \delta_{ad}\delta_{bc} 
\left[\cancel{p}_{2}\left(1\pm\gamma^{5}\right)\gamma^{\mu\nu}\cancel{p}_{1}\right]_{\gamma\dot{\gamma}}\left[\cancel{p}_{4}\left(1\pm\gamma^{5}\right)\gamma_{\mu\nu}\cancel{p}_{3}\right]_{\delta\dot{\delta}} \nonumber \\
& +\frac{g^{2}}{2\left(4\pi\right)^{2}\epsilon}\frac{\left(p_{E}^{2}\right)^{-\epsilon}}{p_{1}^{2}p_{2}^{2}p_{3}^{2}p_{4}^{2}} \delta_{ab}\delta_{cd} 
\left[\cancel{p_{2}}\left(1\pm\gamma^{5}\right)\gamma^{\mu\nu}\cancel{p_{3}}\right]_{\gamma\dot{\delta}}\left[\cancel{p_{4}}\left(1\pm\gamma^{5}\right)\gamma_{\mu\nu}\cancel{p_{1}}\right]_{\delta\dot{\gamma}}.
\end{align}

\paragraph{Chiral tensor bi-meson}

The intra-meson contributions are:
\begin{align}
& \frac{1}{p_{1}^{2}p_{2}^{2}p_{3}^{2}p_{4}^{2}}\left\{ 1-\frac{4g^{2}C_{F}}{(4\pi)^{2}\epsilon}\left(p_{E}^{2}\right)^{-\epsilon}\right\} \nonumber \\
& \cdot\left\{ \delta_{ab}\delta_{cd}\left[\cancel{p}_{2}\left(1\pm\gamma^{5}\right)\gamma^{\mu\nu}\cancel{p}_{1}\right]_{\gamma\dot{\gamma}}\left[\cancel{p}_{4}\left(1\pm\gamma^{5}\right)\gamma_{\mu\nu}\cancel{p}_{3}\right]_{\delta\dot{\delta}} -\delta_{ad}\delta_{bc}\left[\cancel{p_{2}}\left(1\pm\gamma^{5}\right)\gamma^{\mu\nu}\cancel{p_{3}}\right]_{\gamma\dot{\delta}}\left[\cancel{p_{4}}\left(1\pm\gamma^{5}\right)\gamma_{\mu\nu}\cancel{p_{1}}\right]_{\delta\dot{\gamma}}\right\}.
\end{align}

\subparagraph{2-meson contributions}

As in the scalar case, we consider first the unpermuted diagrams.
Everything is as in that case, except for the spinor structure, which
is 
\begin{equation}
12\left[\cancel{p}_{2}\left(1\pm\gamma^{5}\right)\gamma^{\mu\nu}\cancel{p}_{1}\right]_{\gamma\dot{\gamma}}\left[\cancel{p}_{4}\left(1\pm\gamma^{5}\right)\gamma_{\mu\nu}\cancel{p}_{3}\right]_{\delta\dot{\delta}}-48\left[\cancel{p}_{2}\left(1\pm\gamma^{5}\right)\cancel{p}_{1}\right]_{\gamma\dot{\gamma}}\left[\cancel{p}_{4}\left(1\pm\gamma^{5}\right)\cancel{p}_{3}\right]_{\delta\dot{\delta}}
\end{equation}
for diagrams I,II and 
\begin{equation}
12\left[\cancel{p}_{2}\left(1\pm\gamma^{5}\right)\gamma^{\mu\nu}\cancel{p}_{1}\right]_{\gamma\dot{\gamma}}\left[\cancel{p}_{4}\left(1\pm\gamma^{5}\right)\gamma_{\mu\nu}\cancel{p}_{3}\right]_{\delta\dot{\delta}}+48\left[\cancel{p}_{2}\left(1\pm\gamma^{5}\right)\cancel{p}_{1}\right]_{\gamma\dot{\gamma}}\left[\cancel{p}_{4}\left(1\pm\gamma^{5}\right)\cancel{p}_{3}\right]_{\delta\dot{\delta}}
\end{equation}
for diagrams III,IV. Summing them gives:
\begin{equation}
=-\frac{24g^{2}}{\left(4\pi\right)^{2}\epsilon}\frac{\left(p_{E}^{2}\right)^{-\epsilon}}{p_{1}^{2}p_{2}^{2}p_{3}^{2}p_{4}^{2}} \delta_{ad}\delta_{bc}  \left[\cancel{p}_{2}\left(1\pm\gamma^{5}\right)\cancel{p}_{1}\right]_{\gamma\dot{\gamma}}\left[\cancel{p}_{4}\left(1\pm\gamma^{5}\right)\cancel{p}_{3}\right]_{\delta\dot{\delta}},
\end{equation}

In total, the correlation function is:
\begin{align}
& \left\langle M_{tt}^{\pm}(p)\bar{\psi}(-p_{1})\psi(p_{2})\bar{\psi}(-p_{3})\psi(p_{4})\right\rangle = \frac{1}{p_{1}^{2}p_{2}^{2}p_{3}^{2}p_{4}^{2}}\left\{ 1-\frac{4g^{2}C_{F}}{(4\pi)^{2}\epsilon}\left(p_{E}^{2}\right)^{-\epsilon}\right\} \nonumber \\
& \cdot\left\{ \delta_{ab}\delta_{cd}\left[\cancel{p}_{2}\left(1\pm\gamma^{5}\right)\gamma^{\mu\nu}\cancel{p}_{1}\right]_{\gamma\dot{\gamma}}\left[\cancel{p}_{4}\left(1\pm\gamma^{5}\right)\gamma_{\mu\nu}\cancel{p}_{3}\right]_{\delta\dot{\delta}} -\delta_{ad}\delta_{bc}\left[\cancel{p_{2}}\left(1\pm\gamma^{5}\right)\gamma^{\mu\nu}\cancel{p_{3}}\right]_{\gamma\dot{\delta}}\left[\cancel{p_{4}}\left(1\pm\gamma^{5}\right)\gamma_{\mu\nu}\cancel{p_{1}}\right]_{\delta\dot{\gamma}}\right\} \nonumber \\
& -\frac{24g^{2}}{\left(4\pi\right)^{2}\epsilon}\frac{\left(p_{E}^{2}\right)^{-\epsilon}}{p_{1}^{2}p_{2}^{2}p_{3}^{2}p_{4}^{2}} \delta_{ad}\delta_{bc}  \left[\cancel{p}_{2}\left(1\pm\gamma^{5}\right)\cancel{p}_{1}\right]_{\gamma\dot{\gamma}}\left[\cancel{p}_{4}\left(1\pm\gamma^{5}\right)\cancel{p}_{3}\right]_{\delta\dot{\delta}}
+\frac{24g^{2}}{\left(4\pi\right)^{2}\epsilon}\frac{\left(p_{E}^{2}\right)^{-\epsilon}}{p_{1}^{2}p_{2}^{2}p_{3}^{2}p_{4}^{2}}  \delta_{ab}\delta_{cd}  \left[\cancel{p}_{2}\left(1\pm\gamma^{5}\right)\cancel{p}_{3}\right]_{\gamma\dot{\delta}}\left[\cancel{p}_{4}\left(1\pm\gamma^{5}\right)\cancel{p}_{1}\right]_{\delta\dot{\gamma}}.
\end{align}
After deduction of the fermion renormalization correction this
becomes:
\begin{align}
& \left\langle M_{tt}^{\pm}(p)\left[\bar{\psi}\right](-p_{1})\left[\psi\right](p_{2})\left[\bar{\psi}\right](-p_{3})\left[\psi\right](p_{4})\right\rangle =
\frac{1}{p_{1}^{2}p_{2}^{2}p_{3}^{2}p_{4}^{2}}\left\{ 1-\frac{2g^{2}C_{F}}{(4\pi)^{2}\epsilon}\left(p_{E}^{2}\right)^{-\epsilon}\right\} \nonumber \\
& \cdot \left\{ \delta_{ab}\delta_{cd}\left[\cancel{p}_{2}\left(1\pm\gamma^{5}\right)\gamma^{\mu\nu}\cancel{p}_{1}\right]_{\gamma\dot{\gamma}}\left[\cancel{p}_{4}\left(1\pm\gamma^{5}\right)\gamma_{\mu\nu}\cancel{p}_{3}\right]_{\delta\dot{\delta}} -\delta_{ad}\delta_{bc}\left[\cancel{p_{2}}\left(1\pm\gamma^{5}\right)\gamma^{\mu\nu}\cancel{p_{3}}\right]_{\gamma\dot{\delta}}\left[\cancel{p_{4}}\left(1\pm\gamma^{5}\right)\gamma_{\mu\nu}\cancel{p_{1}}\right]_{\delta\dot{\gamma}}\right\} \nonumber \\
& -\frac{24g^{2}}{\left(4\pi\right)^{2}\epsilon}\frac{\left(p_{E}^{2}\right)^{-\epsilon}}{p_{1}^{2}p_{2}^{2}p_{3}^{2}p_{4}^{2}} \delta_{ad}\delta_{bc} \left[\cancel{p}_{2}\left(1\pm\gamma^{5}\right)\cancel{p}_{1}\right]_{\gamma\dot{\gamma}}\left[\cancel{p}_{4}\left(1\pm\gamma^{5}\right)\cancel{p}_{3}\right]_{\delta\dot{\delta}}
+\frac{24g^{2}}{\left(4\pi\right)^{2}\epsilon}\frac{\left(p_{E}^{2}\right)^{-\epsilon}}{p_{1}^{2}p_{2}^{2}p_{3}^{2}p_{4}^{2}}  \delta_{ab}\delta_{cd}  \left[\cancel{p}_{2}\left(1\pm\gamma^{5}\right)\cancel{p}_{3}\right]_{\gamma\dot{\delta}}\left[\cancel{p}_{4}\left(1\pm\gamma^{5}\right)\cancel{p}_{1}\right]_{\delta\dot{\gamma}}.
\end{align}

\paragraph{Chiral vector bi-meson}
\subparagraph{Intra-meson contributions}

In this case, there is no single meson contribution to the renormalization
function. The amplitude (already corrected for the external fermion
renormalization) is just the tree level:
\begin{align}
 & \frac{1}{p_{1}^{2}p_{2}^{2}p_{3}^{2}p_{4}^{2}}\left\{ \delta_{ab}\delta_{cd}\left[\cancel{p}_{2}\left(1\pm\gamma^{5}\right)\gamma^{\rho}\cancel{p}_{1}\right]_{\gamma\dot{\gamma}}\left[\cancel{p}_{4}\left(1\pm\gamma^{5}\right)\gamma^{\sigma}\cancel{p}_{3}\right]_{\delta\dot{\delta}}-\delta_{ad}\delta_{bc}\left[\cancel{p_{2}}\left(1\pm\gamma^{5}\right)\gamma^{\rho}\cancel{p_{3}}\right]_{\gamma\dot{\delta}}\left[\cancel{p_{4}}\left(1\pm\gamma^{5}\right)\gamma^{\sigma}\cancel{p_{1}}\right]_{\delta\dot{\gamma}}\right\}.
\end{align}

\subparagraph{Inter-meson contributions}

Again, everything is the same as in the scalar case, except for the
spinor structure. This structure is:
\begin{equation}
4\left[\cancel{p_{2}}\left(1\pm\gamma^{5}\right)\gamma^{\rho}\cancel{p_{1}}\right]_{\gamma\dot{\gamma}}\left[\cancel{p_{4}}\left(1\pm\gamma^{5}\right)\gamma^{\sigma}\cancel{p_{3}}\right]_{\delta\dot{\delta}}
\end{equation}
for diagrams I,II, 
\begin{align}
 & 4\left[\cancel{p_{2}}\left(1\pm\gamma^{5}\right)\gamma^{\rho}\cancel{p_{1}}\right]_{\gamma\dot{\gamma}}\left[\cancel{p_{4}}\left(1\pm\gamma^{5}\right)\gamma^{\sigma}\cancel{p_{3}}\right]_{\delta\dot{\delta}}-4\left[\cancel{p_{2}}\left(1\pm\gamma^{5}\right)\gamma^{\sigma}\cancel{p_{1}}\right]_{\gamma\dot{\gamma}}\left[\cancel{p_{4}}\left(1\pm\gamma^{5}\right)\gamma^{\rho}\cancel{p_{3}}\right]_{\delta\dot{\delta}}\nonumber \\
 & +4\eta^{\sigma\rho}\left[\cancel{p_{2}}\left(1\pm\gamma^{5}\right)\gamma^{\mu}\cancel{p_{1}}\right]_{\gamma\dot{\gamma}}\left[\cancel{p_{4}}\left(1\pm\gamma^{5}\right)\gamma_{\mu}\cancel{p_{3}}\right]_{\delta\dot{\delta}} \mp 4i\epsilon^{\mu\sigma\rho\tau}\left[\cancel{p_{2}}\left(1\pm\gamma^{5}\right)\gamma_{\tau}\cancel{p_{1}}\right]_{\gamma\dot{\gamma}}\left[\cancel{p_{4}}\left(1\pm\gamma^{5}\right)\gamma_{\mu}\cancel{p_{3}}\right]_{\delta\dot{\delta}}
\end{align}
for diagram III and 
\begin{align}
 & 4\left[\cancel{p_{2}}\left(1\pm\gamma^{5}\right)\gamma^{\rho}\cancel{p_{1}}\right]_{\gamma\dot{\gamma}}\left[\cancel{p_{4}}\left(1\pm\gamma^{5}\right)\gamma^{\sigma}\cancel{p_{3}}\right]_{\delta\dot{\delta}}-4\left[\cancel{p_{2}}\left(1\pm\gamma^{5}\right)\gamma^{\sigma}\cancel{p_{1}}\right]_{\gamma\dot{\gamma}}\left[\cancel{p_{4}}\left(1\pm\gamma^{5}\right)\gamma^{\rho}\cancel{p_{3}}\right]_{\delta\dot{\delta}}\nonumber \\
 & +4\eta^{\sigma\rho}\left[\cancel{p_{2}}\left(1\pm\gamma^{5}\right)\gamma^{\mu}\cancel{p_{1}}\right]_{\gamma\dot{\gamma}}\left[\cancel{p_{4}}\left(1\pm\gamma^{5}\right)\gamma_{\mu}\cancel{p_{3}}\right]_{\delta\dot{\delta}} \pm 4i\epsilon^{\mu\sigma\rho\tau}\left[\cancel{p_{2}}\left(1\pm\gamma^{5}\right)\gamma_{\tau}\cancel{p_{1}}\right]_{\gamma\dot{\gamma}}\left[\cancel{p_{4}}\left(1\pm\gamma^{5}\right)\gamma_{\mu}\cancel{p_{3}}\right]_{\delta\dot{\delta}}
\end{align}
for diagram IV. Summing them gives:
\begin{align}
 & \frac{g^{2}}{\left(4\pi\right)^{2}\epsilon} \delta_{ad}\delta_{bc} \frac{\left(p_{E}^{2}\right)^{-\epsilon}}{p_{1}^{2}p_{2}^{2}p_{3}^{2}p_{4}^{2}}
 \left\{ \left[\cancel{p_{2}}\left(1\pm\gamma^{5}\right)\gamma^{\sigma}\cancel{p_{1}}\right]_{\gamma\dot{\gamma}}\left[\cancel{p_{4}}\left(1\pm\gamma^{5}\right)\gamma^{\rho}\cancel{p_{3}}\right]_{\delta\dot{\delta}}-\eta^{\sigma\rho}\left[\cancel{p_{2}}\left(1\pm\gamma^{5}\right)\gamma^{\mu}\cancel{p_{1}}\right]_{\gamma\dot{\gamma}}\left[\cancel{p_{4}}\left(1\pm\gamma^{5}\right)\gamma_{\mu}\cancel{p_{3}}\right]_{\delta\dot{\delta}}\right\} .
\end{align}

In total, the (partially renormalized) correlation function is:
\begin{align}
& \left\langle M_{vv}^{\pm\mu\nu}(p)\left[\bar{\psi}\right](-p_{1})\left[\psi\right](p_{2})\left[\bar{\psi}\right](-p_{3})\left[\psi\right](p_{4})\right\rangle = \nonumber \\
& = \frac{1}{p_{1}^{2}p_{2}^{2}p_{3}^{2}p_{4}^{2}} \left\{ \delta_{ab}\delta_{cd}\left[\cancel{p}_{2}\left(1\pm\gamma^{5}\right)\gamma^{\mu}\cancel{p}_{1}\right]_{\gamma\dot{\gamma}}\left[\cancel{p}_{4}\left(1\pm\gamma^{5}\right)\gamma^{\nu}\cancel{p}_{3}\right]_{\delta\dot{\delta}} -\delta_{ad}\delta_{bc}\left[\cancel{p_{2}}\left(1\pm\gamma^{5}\right)\gamma^{\mu}\cancel{p_{3}}\right]_{\gamma\dot{\delta}}\left[\cancel{p_{4}}\left(1\pm\gamma^{5}\right)\gamma^{\nu}\cancel{p_{1}}\right]_{\delta\dot{\gamma}}\right\} \nonumber \\
& +\frac{g^{2}}{\left(4\pi\right)^{2}\epsilon}\frac{\left(p_{E}^{2}\right)^{-\epsilon}}{p_{1}^{2}p_{2}^{2}p_{3}^{2}p_{4}^{2}} \delta_{ad}\delta_{bc}
\left\{ \left[\cancel{p_{2}}\left(1\pm\gamma^{5}\right)\gamma^{\nu}\cancel{p_{1}}\right]_{\gamma\dot{\gamma}}\left[\cancel{p_{4}}\left(1\pm\gamma^{5}\right)\gamma^{\mu}\cancel{p_{3}}\right]_{\delta\dot{\delta}}-\eta^{\mu\nu}\left[\cancel{p_{2}}\left(1\pm\gamma^{5}\right)\gamma^{\rho}\cancel{p_{1}}\right]_{\gamma\dot{\gamma}}\left[\cancel{p_{4}}\left(1\pm\gamma^{5}\right)\gamma_{\rho}\cancel{p_{3}}\right]_{\delta\dot{\delta}}\right\} -\nonumber \\
& -\frac{g^{2}}{\left(4\pi\right)^{2}\epsilon}\frac{\left(p_{E}^{2}\right)^{-\epsilon}}{p_{1}^{2}p_{2}^{2}p_{3}^{2}p_{4}^{2}} \delta_{ab}\delta_{cd} 
\left\{ \left[\cancel{p_{2}}\left(1\pm\gamma^{5}\right)\gamma^{\nu}\cancel{p_{3}}\right]_{\gamma\dot{\delta}}\left[\cancel{p_{4}}\left(1\pm\gamma^{5}\right)\gamma^{\mu}\cancel{p_{1}}\right]_{\delta\dot{\gamma}}-\eta^{\mu\nu}\left[\cancel{p_{2}}\left(1\pm\gamma^{5}\right)\gamma^{\rho}\cancel{p_{3}}\right]_{\gamma\dot{\delta}}\left[\cancel{p_{4}}\left(1\pm\gamma^{5}\right)\gamma_{\rho}\cancel{p_{1}}\right]_{\delta\dot{\gamma}}\right\}.
\end{align}

\subparagraph{Separation into irreps.}

The vectors bi-meson with free indices is a Lorentz 2-tensor, and as
such can be decomposed into 3 irreps.:

\textit{Trace (scalar).}
For this we contract the indices above with $\eta_{\mu\nu}$, giving:
\begin{equation*}
\left\langle M_{vv}^{\pm,tr}(p)\left[\bar{\psi}\right](-p_{1})\left[\psi\right](p_{2})\left[\bar{\psi}\right](-p_{3})\left[\psi\right](p_{4})\right\rangle =
\end{equation*}
\begin{align}
=\frac{1}{p_{1}^{2}p_{2}^{2}p_{3}^{2}p_{4}^{2}}\left[\cancel{p}_{2}\left(1\pm\gamma^{5}\right)\gamma^{\mu}\cancel{p}_{1}\right]_{\gamma\dot{\gamma}}\left[\cancel{p}_{4}\left(1\pm\gamma^{5}\right)\gamma_{\mu}\cancel{p}_{3}\right]_{\delta\dot{\delta}} \left\{ \delta_{ab}\delta_{cd} -3\delta_{ad}\delta_{bc}\frac{g^{2}\left(p_{E}^{2}\right)^{-\epsilon}}{\left(4\pi\right)^{2}\epsilon}\right\} \nonumber\\
-\frac{1}{p_{1}^{2}p_{2}^{2}p_{3}^{2}p_{4}^{2}}\left[\cancel{p_{2}}\left(1\pm\gamma^{5}\right)\gamma^{\mu}\cancel{p_{3}}\right]_{\gamma\dot{\delta}}\left[\cancel{p_{4}}\left(1\pm\gamma^{5}\right)\gamma_{\mu}\cancel{p_{1}}\right]_{\delta\dot{\gamma}} \left\{ \delta_{ad}\delta_{bc} -3\delta_{ab}\delta_{cd}\frac{g^{2}\left(p_{E}^{2}\right)^{-\epsilon}}{\left(4\pi\right)^{2}\epsilon}\right\}.
\end{align}

\textit{Traceless symmetric.}
We denote this irrep. by $(\mu\nu)$, and project on it by taking 
\begin{equation}
\left(\cdots\right)_{(\mu\nu)}=\frac{1}{2}\left[\left(\cdots\right)_{\mu\nu}+\left(\cdots\right)_{\nu\mu}\right]-\frac{1}{d}\eta_{\mu\nu}\left(\cdots\right)_{\rho}^{\rho}.
\end{equation}
Under this operation, the last term in each correction (the one
$\propto\eta^{\mu\nu}$) vanishes, and the expressions with 
\[
\left[\cancel{p}_{2}\left(1\pm\gamma^{5}\right)\gamma^{\mu}\cancel{p}_{1}\right]_{\gamma\dot{\gamma}}\left[\cancel{p}_{4}\left(1\pm\gamma^{5}\right)\gamma^{\nu}\cancel{p}_{3}\right]_{\delta\dot{\delta}},\left[\cancel{p}_{2}\left(1\pm\gamma^{5}\right)\gamma^{\nu}\cancel{p}_{1}\right]_{\gamma\dot{\gamma}}\left[\cancel{p}_{4}\left(1\pm\gamma^{5}\right)\gamma^{\mu}\cancel{p}_{3}\right]_{\delta\dot{\delta}}
\]
are identified (and similarly with the permuted terms). Thus, we get:
\begin{align}
& \left\langle M_{vv}^{\pm(\mu\nu)}(p)\left[\bar{\psi}\right](-p_{1})\left[\psi\right](p_{2})\left[\bar{\psi}\right](-p_{3})\left[\psi\right](p_{4})\right\rangle =\nonumber \\
& =\frac{1}{p_{1}^{2}p_{2}^{2}p_{3}^{2}p_{4}^{2}}\left[\cancel{p}_{2}\left(1\pm\gamma^{5}\right)\gamma^{(\mu}\cancel{p}_{1}\right]_{\gamma\dot{\gamma}}\left[\cancel{p}_{4}\left(1\pm\gamma^{5}\right)\gamma^{\nu)}\cancel{p}_{3}\right]_{\delta\dot{\delta}} \left\{ \delta_{ab}\delta_{cd} +\delta_{ad}\delta_{bc}\frac{g^{2}\left(p_{E}^{2}\right)^{-\epsilon}}{\left(4\pi\right)^{2}\epsilon}\right\} \nonumber \\
& -\frac{1}{p_{1}^{2}p_{2}^{2}p_{3}^{2}p_{4}^{2}}\left[\cancel{p_{2}}\left(1\pm\gamma^{5}\right)\gamma^{(\mu}\cancel{p_{3}}\right]_{\gamma\dot{\delta}}\left[\cancel{p_{4}}\left(1\pm\gamma^{5}\right)\gamma^{\nu)}\cancel{p_{1}}\right]_{\delta\dot{\gamma}} \left\{ \delta_{ad}\delta_{bc} +\delta_{ab}\delta_{cd}\frac{g^{2}\left(p_{E}^{2}\right)^{-\epsilon}}{\left(4\pi\right)^{2}\epsilon}\right\}.
\end{align}

The third representation is the antisymmetric one, but as mentioned above, the permutation of both the fermions and the antifermions is equivalent to that of the Lorentz indices. Therefore, the antisymmetric representation is orthogonal to the symmetrized bi-meson operator and need not be considered here.

\subsubsection{Fierz identities}

Some of the quantities involve a combination of color and 
spinor structures that does not enable directly comparing them for extraction of the renormalization function. 
We can remedy this using the Fierz identities.
We use the identities provided in \cite{paper15}. We need to account
for the difference in normalization used here, relative to \cite{paper15}:
\begin{equation}
\sigma^{\mu\nu} =i\gamma^{\mu\nu}; \indent
P_{R,L} =\frac{1}{2}\left(1\pm\gamma^{5}\right).
\end{equation}
Note that the chiral projection operator appears in all terms equally
and so its relative normalization always cancels out.

\paragraph{Scalars and tensors }

We get for the scalars:
\begin{align}
& \delta_{ab}\delta_{cd}\left[\cancel{p_{2}}\left(1\pm\gamma^{5}\right)\cancel{p_{3}}\right]_{\gamma\dot{\delta}}\left[\cancel{p_{4}}\left(1\pm\gamma^{5}\right)\cancel{p_{1}}\right]_{\delta\dot{\gamma}} = \nonumber\\
& =\delta_{ab}\delta_{cd}\left(\frac{1}{2}\left[\cancel{p_{2}}\left(1\pm\gamma^{5}\right)\cancel{p_{1}}\right]_{\gamma\dot{\gamma}}\left[\cancel{p_{4}}\left(1\pm\gamma^{5}\right)\cancel{p_{3}}\right]_{\delta\dot{\delta}}-\frac{1}{8}\left[\cancel{p_{2}}\left(1\pm\gamma^{5}\right)\gamma^{\mu\nu}\cancel{p_{1}}\right]_{\gamma\dot{\gamma}}\left[\cancel{p_{4}}\left(1\pm\gamma^{5}\right)\gamma_{\mu\nu}\cancel{p_{3}}\right]_{\delta\dot{\delta}}\right),
\end{align}
 and similarly
\begin{align}
& \delta_{ad}\delta_{bc}\left[\cancel{p_{2}}\left(1\pm\gamma^{5}\right)\cancel{p_{1}}\right]_{\gamma\dot{\gamma}}\left[\cancel{p_{4}}\left(1\pm\gamma^{5}\right)\cancel{p_{3}}\right]_{\delta\dot{\delta}} =\nonumber \\
& =\delta_{ad}\delta_{bc}\left(\frac{1}{2}\left[\cancel{p_{2}}\left(1\pm\gamma^{5}\right)\cancel{p_{3}}\right]_{\gamma\dot{\delta}}\left[\cancel{p_{4}}\left(1\pm\gamma^{5}\right)\cancel{p_{1}}\right]_{\delta\dot{\gamma}}-\frac{1}{8}\left[\cancel{p_{2}}\left(1\pm\gamma^{5}\right)\gamma^{\mu\nu}\cancel{p_{3}}\right]_{\gamma\dot{\delta}}\left[\cancel{p_{4}}\left(1\pm\gamma^{5}\right)\gamma_{\mu\nu}\cancel{p_{1}}\right]_{\delta\dot{\gamma}}\right).
\end{align}
For the tensors:
\begin{align}
& \delta_{ab}\delta_{cd}\left[\cancel{p}_{2}\left(1\pm\gamma^{5}\right)\gamma^{\mu\nu}\cancel{p}_{3}\right]_{\gamma\dot{\delta}}\left[\cancel{p}_{4}\left(1\pm\gamma^{5}\right)\gamma_{\mu\nu}\cancel{p}_{1}\right]_{\delta\dot{\gamma}}
= \nonumber\\
& =-\delta_{ab}\delta_{cd}\left(6\left[\cancel{p}_{2}\left(1\pm\gamma^{5}\right)\cancel{p}_{1}\right]_{\gamma\dot{\gamma}}\left[\cancel{p}_{4}\left(1\pm\gamma^{5}\right)\cancel{p}_{3}\right]_{\delta\dot{\delta}}+\frac{1}{2}\left[\cancel{p}_{2}\left(1\pm\gamma^{5}\right)\gamma^{\mu\nu}\cancel{p}_{1}\right]_{\gamma\dot{\gamma}}\left[\cancel{p}_{4}\left(1\pm\gamma^{5}\right)\gamma_{\mu\nu}\cancel{p}_{3}\right]_{\delta\dot{\delta}}\right),\nonumber \\
& \delta_{ad}\delta_{bc}\left[\cancel{p}_{2}\left(1\pm\gamma^{5}\right)\gamma^{\mu\nu}\cancel{p}_{1}\right]_{\gamma\dot{\gamma}}\left[\cancel{p}_{4}\left(1\pm\gamma^{5}\right)\gamma_{\mu\nu}\cancel{p}_{3}\right]_{\delta\dot{\delta}} = \nonumber\\
& =-\delta_{ad}\delta_{bc}\left(6\left[\cancel{p}_{2}\left(1\pm\gamma^{5}\right)\cancel{p}_{3}\right]_{\gamma\dot{\delta}}\left[\cancel{p}_{4}\left(1\pm\gamma^{5}\right)\cancel{p}_{1}\right]_{\delta\dot{\gamma}}+\frac{1}{2}\left[\cancel{p}_{2}\left(1\pm\gamma^{5}\right)\gamma^{\mu\nu}\cancel{p}_{3}\right]_{\gamma\dot{\delta}}\left[\cancel{p}_{4}\left(1\pm\gamma^{5}\right)\gamma_{\mu\nu}\cancel{p}_{1}\right]_{\delta\dot{\gamma}}\right).
\end{align}
We can insert these results back into the partially-renormalized correlation
functions and get:
\begin{align}
& \left\langle M_{ss}^{\pm}(p)\left[\bar{\psi}\right](-p_{1})\left[\psi\right](p_{2})\left[\bar{\psi}\right](-p_{3})\left[\psi\right](p_{4})\right\rangle
=\frac{1}{p_{1}^{2}p_{2}^{2}p_{3}^{2}p_{4}^{2}}\left[1+\frac{3g^{2}}{(4\pi)^{2}\epsilon}\left(p_{E}^{2}\right)^{-\epsilon}\left(2C_{F}-1\right)\right] \nonumber \\
& \cdot\left\{ \delta_{ab}\delta_{cd}\left[\cancel{p_{2}}\left(1\pm\gamma^{5}\right)\cancel{p_{1}}\right]_{\gamma\dot{\gamma}}\left[\cancel{p_{4}}\left(1\pm\gamma^{5}\right)\cancel{p_{3}}\right]_{\delta\dot{\delta}}-\delta_{ad}\delta_{bc}\left[\cancel{p_{2}}\left(1\pm\gamma^{5}\right)\cancel{p_{3}}\right]_{\gamma\dot{\delta}}\left[\cancel{p_{4}}\left(1\pm\gamma^{5}\right)\cancel{p_{1}}\right]_{\delta\dot{\gamma}}\right\} -
\frac{g^{2}}{4\left(4\pi\right)^{2}\epsilon}\frac{\left(p_{E}^{2}\right)^{-\epsilon}}{p_{1}^{2}p_{2}^{2}p_{3}^{2}p_{4}^{2}}\nonumber \\
& \cdot \left\{ \delta_{ab}\delta_{cd}\left[\cancel{p_{2}}\left(1\pm\gamma^{5}\right)\gamma^{\mu\nu}\cancel{p_{1}}\right]_{\gamma\dot{\gamma}}\left[\cancel{p_{4}}\left(1\pm\gamma^{5}\right)\gamma_{\mu\nu}\cancel{p_{3}}\right]_{\delta\dot{\delta}} -\delta_{ad}\delta_{bc}\left[\cancel{p_{2}}\left(1\pm\gamma^{5}\right)\gamma^{\mu\nu}\cancel{p_{3}}\right]_{\gamma\dot{\delta}}\left[\cancel{p_{4}}\left(1\pm\gamma^{5}\right)\gamma_{\mu\nu}\cancel{p_{1}}\right]_{\delta\dot{\gamma}}\right\},
\end{align}
and
\begin{align}
& \left\langle M_{tt}^{\pm}(p)\left[\bar{\psi}\right](-p_{1})\left[\psi\right](p_{2})\left[\bar{\psi}\right](-p_{3})\left[\psi\right](p_{4})\right\rangle
=\frac{1}{p_{1}^{2}p_{2}^{2}p_{3}^{2}p_{4}^{2}}\left[1-\frac{g^{2}}{(4\pi)^{2}\epsilon}\left(p_{E}^{2}\right)^{-\epsilon}\left(2C_{F}+3\right)\right] \nonumber \\
& \cdot\left\{ \delta_{ab}\delta_{cd}\left[\cancel{p}_{2}\left(1\pm\gamma^{5}\right)\gamma^{\mu\nu}\cancel{p}_{1}\right]_{\gamma\dot{\gamma}}\left[\cancel{p}_{4}\left(1\pm\gamma^{5}\right)\gamma_{\mu\nu}\cancel{p}_{3}\right]_{\delta\dot{\delta}} -\delta_{ad}\delta_{bc}\left[\cancel{p_{2}}\left(1\pm\gamma^{5}\right)\gamma^{\mu\nu}\cancel{p_{3}}\right]_{\gamma\dot{\delta}}\left[\cancel{p_{4}}\left(1\pm\gamma^{5}\right)\gamma_{\mu\nu}\cancel{p_{1}}\right]_{\delta\dot{\gamma}}\right\} \nonumber \\
& +\frac{\left(p_{E}^{2}\right)^{-\epsilon}}{p_{1}^{2}p_{2}^{2}p_{3}^{2}p_{4}^{2}}\frac{12g^{2}}{\left(4\pi\right)^{2}\epsilon}
\left\{ \delta_{ab}\delta_{cd}\left[\cancel{p_{2}}\left(1\pm\gamma^{5}\right)\cancel{p_{1}}\right]_{\gamma\dot{\gamma}}\left[\cancel{p_{4}}\left(1\pm\gamma^{5}\right)\cancel{p_{3}}\right]_{\delta\dot{\delta}}-\delta_{ad}\delta_{bc}\left[\cancel{p}_{2}\left(1\pm\gamma^{5}\right)\cancel{p}_{3}\right]_{\gamma\dot{\delta}}\left[\cancel{p}_{4}\left(1\pm\gamma^{5}\right)\cancel{p}_{1}\right]_{\delta\dot{\gamma}}\right\}.
\end{align}
The one-loop structures are now proportional to tree-level structures as expected, giving a consistency check on our computation.

\paragraph{Vectors}
\subparagraph{General vector identity}

\cite{paper15} gives a formula for the scalar (trace) irrep. of the chiral 2-vector meson,
but not for the traceless symmetric one -- we'll need to construct it. First, we
recall that the vectors and pseudovectors project each pair of fermions
on the representation $\left(N^{2}-1,1\right)+\left(1,N^{2}-1\right)+2\cdot\left(1,1\right)$.
We focus on the former two irreps., which
the chiral projection then reduces to either one. In a single meson,
the only elements that give these irreps. are the vectors and pseudovectors.
Since the bi-meson operators are 2 copies of the same projection, this
remains true in the 2-meson case, and these are the elements that
we need to compute. This is done most easily with the basis elements
$\Gamma^{\mu\pm}=\frac{1}{\sqrt{2}}\left(1\pm\gamma^{5}\right)\gamma^{\mu}$,
whose dual basis elements are $\Gamma_{\mu\pm}=\frac{1}{\sqrt{2}}\left(1\mp\gamma^{5}\right)\gamma_{\mu}$.
This leaves us with the identity:
\begin{equation}\label{eq:Fierz}
\frac{1}{2}\left[\cancel{p_{2}}\left(1\pm\gamma^{5}\right)\gamma^{\mu}\cancel{p_{3}}\right]_{\gamma\dot{\delta}}\left[\cancel{p_{4}}\left(1\pm\gamma^{5}\right)\gamma^{\nu}\cancel{p_{1}}\right]_{\delta\dot{\gamma}}=\frac{1}{2}C_{\pm\ \rho\sigma}^{\mu\nu}\left[\cancel{p}_{2}\left(1\pm\gamma^{5}\right)\gamma^{\rho}\cancel{p}_{1}\right]_{\gamma\dot{\gamma}}\left[\cancel{p}_{4}\left(1\pm\gamma^{5}\right)\gamma^{\sigma}\cancel{p}_{3}\right]_{\delta\dot{\delta}},
\end{equation}
and we can find the coefficients using equation (33) from \cite{paper15}:
\begin{align}
 & C_{\pm\ \rho\sigma}^{\mu\nu} =\frac{1}{16}tr\left[\Gamma^{\mu\pm}\Gamma_{\sigma\pm}\Gamma^{\nu\pm}\Gamma_{\rho\pm}\right] =\frac{1}{16}tr\left[\frac{1}{\sqrt{2}}\left(1\pm\gamma^{5}\right)\gamma^{\mu}\frac{1}{\sqrt{2}}\left(1\mp\gamma^{5}\right)\gamma_{\sigma}\frac{1}{\sqrt{2}}\left(1\pm\gamma^{5}\right)\gamma^{\nu}\frac{1}{\sqrt{2}}\left(1\mp\gamma^{5}\right)\gamma_{\rho}\right]= \nonumber\\
 & =\frac{1}{16}tr\left[\left(1\pm\gamma^{5}\right)\gamma^{\mu}\gamma_{\sigma}\left(1\pm\gamma^{5}\right)\gamma^{\nu}\gamma_{\rho}\right]
 =\frac{1}{8}tr\left[\left(1\pm\gamma^{5}\right)\gamma^{\mu}\gamma_{\sigma}\gamma^{\nu}\gamma_{\rho}\right] =\frac{1}{2}\left(\delta_{\sigma}^{\mu}\delta_{\rho}^{\nu}-\eta^{\mu\nu}\eta_{\rho\sigma}+\delta_{\rho}^{\mu}\delta_{\sigma}^{\nu}\mp i\epsilon_{\ \sigma\ \rho}^{\mu\ \nu}\right).
\end{align}
Inserting this back into (\ref{eq:Fierz}) gives:
\begin{align} \label{eq:-2}
 & \left[\cancel{p_{2}}\left(1\pm\gamma^{5}\right)\gamma^{\mu}\cancel{p_{3}}\right]_{\gamma\dot{\delta}}\left[\cancel{p_{4}}\left(1\pm\gamma^{5}\right)\gamma^{\nu}\cancel{p_{1}}\right]_{\delta\dot{\gamma}}=\nonumber \\
 & =\frac{1}{2}\left(\delta_{\sigma}^{\mu}\delta_{\rho}^{\nu}-\eta^{\mu\nu}\eta_{\rho\sigma}+\delta_{\rho}^{\mu}\delta_{\sigma}^{\nu}\mp i\epsilon_{\ \sigma\ \rho}^{\mu\ \nu}\right)\left[\cancel{p}_{2}\left(1\pm\gamma^{5}\right)\gamma^{\rho}\cancel{p}_{1}\right]_{\gamma\dot{\gamma}}\left[\cancel{p}_{4}\left(1\pm\gamma^{5}\right)\gamma^{\sigma}\cancel{p}_{3}\right]_{\delta\dot{\delta}}= \nonumber\\
 & =\frac{1}{2}\left[\cancel{p}_{2}\left(1\pm\gamma^{5}\right)\gamma^{\nu}\cancel{p}_{1}\right]_{\gamma\dot{\gamma}}\left[\cancel{p}_{4}\left(1\pm\gamma^{5}\right)\gamma^{\mu}\cancel{p}_{3}\right]_{\delta\dot{\delta}}+\frac{1}{2}\left[\cancel{p}_{2}\left(1\pm\gamma^{5}\right)\gamma^{\mu}\cancel{p}_{1}\right]_{\gamma\dot{\gamma}}\left[\cancel{p}_{4}\left(1\pm\gamma^{5}\right)\gamma^{\nu}\cancel{p}_{3}\right]_{\delta\dot{\delta}} \nonumber \\
 & -\frac{1}{2}\eta^{\mu\nu}\left[\cancel{p}_{2}\left(1\pm\gamma^{5}\right)\gamma^{\rho}\cancel{p}_{1}\right]_{\gamma\dot{\gamma}}\left[\cancel{p}_{4}\left(1\pm\gamma^{5}\right)\gamma_{\rho}\cancel{p}_{3}\right]_{\delta\dot{\delta}}\mp\frac{i}{2}\epsilon_{\ \sigma\ \rho}^{\mu\ \nu}\left[\cancel{p}_{2}\left(1\pm\gamma^{5}\right)\gamma^{\rho}\cancel{p}_{1}\right]_{\gamma\dot{\gamma}}\left[\cancel{p}_{4}\left(1\pm\gamma^{5}\right)\gamma^{\sigma}\cancel{p}_{3}\right]_{\delta\dot{\delta}}.
\end{align}
As a sanity check, we verify we can recover the contracted identity
by contracing both sides with $\eta_{\mu\nu}$:
\begin{align}
 & \left[\cancel{p_{2}}\left(1\pm\gamma^{5}\right)\gamma_{\mu}\cancel{p_{3}}\right]_{\gamma\dot{\delta}}\left[\cancel{p_{4}}\left(1\pm\gamma^{5}\right)\gamma^{\mu}\cancel{p_{1}}\right]_{\delta\dot{\gamma}}=\nonumber \\
 & =\frac{1}{2}\left[\cancel{p}_{2}\left(1\pm\gamma^{5}\right)\gamma^{\mu}\cancel{p}_{1}\right]_{\gamma\dot{\gamma}}\left[\cancel{p}_{4}\left(1\pm\gamma^{5}\right)\gamma_{\mu}\cancel{p}_{3}\right]_{\delta\dot{\delta}}\cdot2-\frac{d}{2}\left[\cancel{p}_{2}\left(1\pm\gamma^{5}\right)\gamma^{\rho}\cancel{p}_{1}\right]_{\gamma\dot{\gamma}}\left[\cancel{p}_{4}\left(1\pm\gamma^{5}\right)\gamma_{\rho}\cancel{p}_{3}\right]_{\delta\dot{\delta}}\overset{d\goto4}{\approx} \nonumber\\
 & \approx-\left[\cancel{p}_{2}\left(1\pm\gamma^{5}\right)\gamma^{\mu}\cancel{p}_{1}\right]_{\gamma\dot{\gamma}}\left[\cancel{p}_{4}\left(1\pm\gamma^{5}\right)\gamma_{\mu}\cancel{p}_{3}\right]_{\delta\dot{\delta}},
\end{align}
which agrees with the literature.

\subparagraph{Trace irrep.}

The amplitude becomes
\begin{equation}
 \left\langle M_{vv}^{\pm,tr}(p)\left[\bar{\psi}\right](-p_{1})\left[\psi\right](p_{2})\left[\bar{\psi}\right](-p_{3})\left[\psi\right](p_{4})\right\rangle =\nonumber
 \end{equation}
 \begin{align}
 & =\frac{1}{p_{1}^{2}p_{2}^{2}p_{3}^{2}p_{4}^{2}}\delta_{ab}\delta_{cd}\left[1-\frac{3g^{2}\left(p_{E}^{2}\right)^{-\epsilon}}{\left(4\pi\right)^{2}\epsilon} \right] \left[\cancel{p}_{2}\left(1\pm\gamma^{5}\right)\gamma^{\mu}\cancel{p}_{1}\right]_{\gamma\dot{\gamma}}\left[\cancel{p}_{4}\left(1\pm\gamma^{5}\right)\gamma_{\mu}\cancel{p}_{3}\right]_{\delta\dot{\delta}} \nonumber \\
 & -\frac{1}{p_{1}^{2}p_{2}^{2}p_{3}^{2}p_{4}^{2}}\delta_{ad}\delta_{bc}\left[1-\frac{3g^{2}\left(p_{E}^{2}\right)^{-\epsilon}}{\left(4\pi\right)^{2}\epsilon} \right]\left[\cancel{p_{2}}\left(1\pm\gamma^{5}\right)\gamma^{\mu}\cancel{p_{3}}\right]_{\gamma\dot{\delta}}\left[\cancel{p_{4}}\left(1\pm\gamma^{5}\right)\gamma_{\mu}\cancel{p_{1}}\right]_{\delta\dot{\gamma}}.
\end{align}
We can immediately see that the anomalous dimension of this bi-meson operator is positive, since as we saw for the scalar case, it has the opposite
sign to the correction of the renormalization function in our regularization scheme. 
This supports the conjecture, as the single-meson anomalous dimension vanishes.

\subparagraph{Traceless symmetric irrep.}

First we need to adjust the identity (\ref{eq:-2}): by substituting
$\mu\nu\goto(\mu\nu)$ we get simply:
\begin{align}
 & \Rightarrow\left[\cancel{p_{2}}\left(1\pm\gamma^{5}\right)\gamma^{(\mu}\cancel{p_{3}}\right]_{\gamma\dot{\delta}}\left[\cancel{p_{4}}\left(1\pm\gamma^{5}\right)\gamma^{\nu)}\cancel{p_{1}}\right]_{\delta\dot{\gamma}}=\left[\cancel{p}_{2}\left(1\pm\gamma^{5}\right)\gamma^{(\mu}\cancel{p}_{1}\right]_{\gamma\dot{\gamma}}\left[\cancel{p}_{4}\left(1\pm\gamma^{5}\right)\gamma^{\nu)}\cancel{p}_{3}\right]_{\delta\dot{\delta}},
\end{align}
and so the amplitude becomes
\begin{align}
 & \left\langle M_{vv}^{\pm(\mu\nu)}(p)\left[\bar{\psi}\right](-p_{1})\left[\psi\right](p_{2})\left[\bar{\psi}\right](-p_{3})\left[\psi\right](p_{4})\right\rangle =\nonumber \\
 & =\frac{1}{p_{1}^{2}p_{2}^{2}p_{3}^{2}p_{4}^{2}}\left[\cancel{p}_{2}\left(1\pm\gamma^{5}\right)\gamma^{(\mu}\cancel{p}_{1}\right]_{\gamma\dot{\gamma}}\left[\cancel{p}_{4}\left(1\pm\gamma^{5}\right)\gamma^{\nu)}\cancel{p}_{3}\right]_{\delta\dot{\delta}}\delta_{ab}\delta_{cd}\left[1-\frac{g^{2}\left(p_{E}^{2}\right)^{-\epsilon}}{\left(4\pi\right)^{2}\epsilon}\right] \nonumber \\
 & -\frac{1}{p_{1}^{2}p_{2}^{2}p_{3}^{2}p_{4}^{2}}\left[\cancel{p_{2}}\left(1\pm\gamma^{5}\right)\gamma^{(\mu}\cancel{p_{3}}\right]_{\gamma\dot{\delta}}\left[\cancel{p_{4}}\left(1\pm\gamma^{5}\right)\gamma^{\nu)}\cancel{p_{1}}\right]_{\delta\dot{\gamma}}\delta_{ad}\delta_{bc}\left[1-\frac{g^{2}\left(p_{E}^{2}\right)^{-\epsilon}}{\left(4\pi\right)^{2}\epsilon}\right].
\end{align}
Once again we can immediately see that the anomalous dimension is
positive, supporting the conjecture.

\subsubsection{Renormalization function and anomalous dimension calculation}

We see that for the irreps. of scalars and tensors, there is a mixing of operators at the 2-meson level, and the renormalization functions need to be diagonalized. We set a renormalization scale $M$ and note that renormalization will replace each factor of $\left(p_{E}^{2}\right)^{-\epsilon}$ by $M^{-2\epsilon}\left(1+O(\epsilon)\right)$.
Next we denote $\frac{g^{2}}{(4\pi)^{2}\epsilon}M^{-2\epsilon}\equiv\alpha$ for brevity.
The renormalization matrix is (recalling that $C_{F}=\frac{N^{2}-1}{2N}\approx \frac{N}{2}$):
\begin{equation}
\begin{pmatrix}1+3\left(2C_{F}-1\right)\alpha & 12\alpha\\
-\frac{\alpha}{4} & 1-\left(2C_{F}+3\right)\alpha
\end{pmatrix} 
=\begin{pmatrix}1+3(N-1)\alpha & 12\alpha\\
-\frac{\alpha}{4} & 1-(N+3)\alpha
\end{pmatrix} .
\end{equation}
Its eigenvalues are computed to be:
\begin{equation}
\lambda_{1,2} =1+(N-3)\alpha \pm 2N\alpha +O\left(\frac{\alpha}{N} \right).
\end{equation}
On the other hand, the squared renormalization matrix for the 1-meson
operators is simply:
\begin{equation}
\begin{pmatrix}1+6C_{F}\alpha\\
 & 1-2C_{F}\alpha
\end{pmatrix}
=\begin{pmatrix}1+3\alpha N \\
 & 1-\alpha N
\end{pmatrix},
\end{equation}
and its eigenvalues are the diagonal entries. The differentiation
by $\log (M)$ flips the sign of the correction, so the smallest anomalous
dimension will relate to the largest eigenvalue of the renormalization
matrix. Therefore, to test the conjecture, one needs to compare the
larger eigenvalues between the two matrices, i.e. compare 
\begin{equation}
\lambda_{1}=1+3(N-1)\alpha,
\indent \lambda_{1}^{(1)}=1+3\alpha N.
\end{equation}
Subtracting them, we get:
\begin{equation}
\lambda_{1}-\lambda_{1}^{(1)} = -3\alpha+O\left(\frac{\alpha}{N}\right)\overset{\alpha>0}{<}0,
\end{equation}
supporting the conjecture. Note that in all cases we see that the leading 2-meson correction to the anomalous dimension is $O\left(\alpha\right)$, in contrast to the overall leading correction to the renormalization function which is $O\left(\alpha N\right)$, as expected from the large $N$ limit.

\subsection{Mixed double mesons}

\begin{figure}
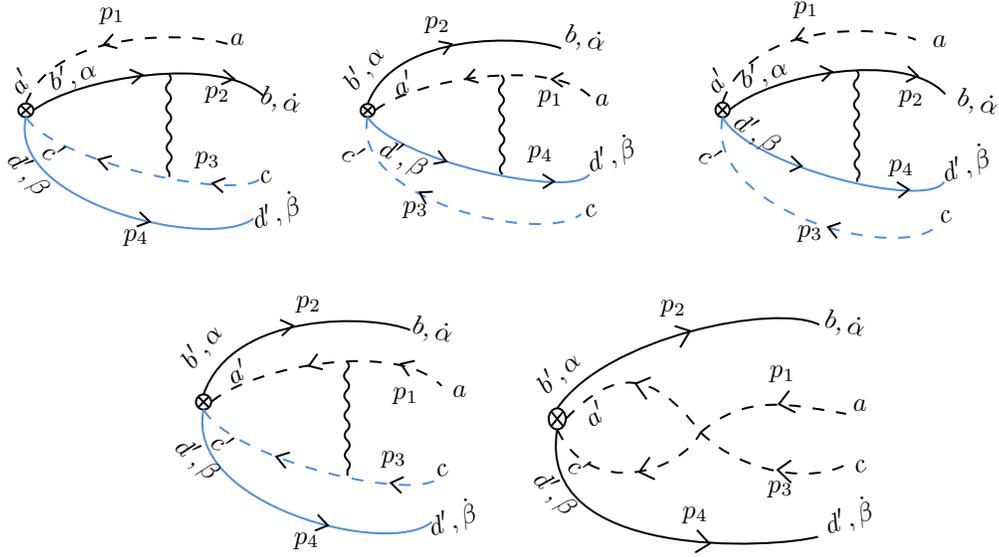

\centering


\caption{\label{fig:double mixed meson diagrams}Double mixed meson diagrams I-V.}
\end{figure}

Here our basic operator is of the form $\phi^* \psi$, and we consider the correlation function:

\begin{equation}
\left\langle \left(\phi_{ia'}^{*}\psi_{jb'\alpha}\phi_{kc'}^{*}\psi_{ld'\beta}\right)(p)\phi_{i'a}(p_{1})\overline{\psi}_{j'b\dot{\alpha}}(-p_{2})\phi_{k'c}(p_{3})\overline{\psi}_{l'd\dot{\beta}}(-p_{4})\right\rangle \delta_{a'b'}\delta_{c'd'} .
\end{equation}
We assume for simplicity $i\neq k,j\neq l$ (with a symmetrization of both pairs of indices), and that the other flavor indices configure
in a way that gives a non-vanishing amplitude. Here, unlike the mesons
with only either scalars or fermions, we have to consider explicitly
the symmetrization by both pairs of flavor indices. We refer to the
case $i'=i,j'=j$ as the unpermuted diagrams. We also keep the shorthand notation
$A=\alpha\dot{\alpha},B=\beta\dot{\beta},A'=\beta\dot{\alpha},B'=\alpha\dot{\beta}$. 
The additional diagrams are shown in Figure \ref{fig:double mixed meson diagrams} - diagrams I-IV for the unpermuted case, and diagram V for the fermion-unpermuted cases (diagram V is uncolored because the scalar permutation is kept generic).

For the (unpermuted) tree-level, multiplying two copies of (\ref{eq:mixed meson tree level}) gives the amplitude:
\begin{equation}
    \label{eq:tree-level mixed 2-meson}
    \frac{\cancel{p_2}_{\alpha \dot{\alpha}} \cancel{p_4}_{\beta \dot{\beta}}}{p_1^2 p_2^2 p_3^2 p_4^2} \delta_{ab}\delta_{cd}.
\end{equation}

\subsubsection{Unpermuted diagrams - I-IV}

We omit color factors here and deal with them separately. Diagrams I-IV are 
respectively equal to:
\begin{align}
 I & =II = \frac{g^{2}}{\left(4\pi\right)^{2}\epsilon}\left(p_{E}^{2}\right)^{-\epsilon}\frac{\cancel{p_{2}}_{A}\cancel{p_{4}}_{B}}{p_{1}^{2}p_{2}^{2}p_{3}^{2}p_{4}^{2}}  \\
 III & = -\frac{g^{2}}{4\left(4\pi\right)^{2}\epsilon}\frac{\left[\gamma^{\nu}\gamma^{\mu}\cancel{p_{2}}\right]_{A}\left[\gamma_{\nu}\gamma_{\mu}\cancel{p_{4}}\right]_{B}}{p_{1}^{2}p_{2}^{2}p_{3}^{2}p_{4}^{2}}\left(p_{E}^{2}\right)^{-\epsilon} \\
 IV & = -\frac{g^{2}}{\left(4\pi\right)^{2}\epsilon}\frac{\cancel{p_{2}}_{A}\cancel{p_{4}}_{B}}{p_{1}^{2}p_{2}^{2}p_{3}^{2}p_{4}^{2}}\left(p_{E}^{2}\right)^{-\epsilon}.
\end{align}

The color factor for diagrams I-IV is $\frac{1}{2} \delta_{bc}\delta_{da}$, similarly to the previous cases.

\subsubsection{Diagram V}

Diagram V needs a more delicate consideration of the scalar flavor
factors, due to the structure of the $\phi^{4}$ couplings, so we
keep them explicit and deal with the color factors at the same time.
The index part is:
\begin{align}
 & \delta_{bb'}\delta_{dd'}\left[\tilde{f}\left(\delta_{ii'}\delta_{aa'}\delta_{kk'}\delta_{cc'}+\delta_{ik'}\delta_{ca'}\delta_{ki'}\delta_{ac'}\right)+\tilde{h}\left(\delta_{ii'}\delta_{ac'}\delta_{kk'}\delta_{ca'}+\delta_{ik'}\delta_{cc'}\delta_{ki'}\delta_{aa'}\right)\right]\delta_{a'b'}\delta_{c'd'}= \nonumber\\
 & =\left[\tilde{f}\left(\delta_{ii'}\delta_{ab}\delta_{kk'}\delta_{cd}+\delta_{ik'}\delta_{cb}\delta_{ki'}\delta_{ad}\right)+\tilde{h}\left(\delta_{ii'}\delta_{ad}\delta_{kk'}\delta_{cb}+\delta_{ik'}\delta_{cd}\delta_{ki'}\delta_{ab}\right)\right].
\end{align}
This is a contribution of the same order as in the case of scalar mesons. As we argue later, the $\tilde{f}$ terms are suppressed by $O\left(\frac{1}{N}\right)$ compared to the $\tilde{h}$
terms and can be neglected going forward.

The other factor (coming from the momentum loop, including a symmetry factor) is:

\begin{equation}
 -\frac{1}{2\left(4\pi\right)^{2}\epsilon}\frac{\cancel{p_{2}}_{A}\cancel{p_{4}}_{B}}{p_{1}^{2}p_{2}^{2}p_{3}^{2}p_{4}^{2}}\left(p_{E}^{2}\right)^{-\epsilon}\label{eq:mixed diagram V}.
\end{equation}
Taking permutations into account as well, this term gives $\tilde{h}\delta_{ad}\delta_{cb}$
in the cases without scalar permutation, and $\tilde{h}\delta_{ab}\delta_{cd}$
in the cases with it (times the other factor (\ref{eq:mixed diagram V})). 

\subsubsection{Permuted diagrams, Fermion operator signs and spin structure}

The fermion permutation $j\swap l$ changes the spinor and color structures as:
\begin{align}
\cancel{p_{2}}_{A}\cancel{p_{4}}_{B} & \swap\cancel{p_{2}}_{A'}\cancel{p_{4}}_{B'}\\
\delta_{ab}\delta_{cd} & \swap\delta_{ad}\delta_{cb}
\end{align}
while the scalar permutation only changes the color structures. Making both
permutations then returns the color structure to its unpermuted form,
while the spinor structure remains permuted.

Similarly to the fermionic meson case, when contracting the fermionic operators, we see that a change in sign only occurs
if the external fermion legs are permuted, i.e. in the cases of a fermion
permutation with or without a scalar permutation. 

Here the spin structure is simpler than that of the fermionic double
meson case. Diagrams I,II,IV,V
simply have $\cancel{p_{2}}_{A}\cancel{p_{4}}_{B}$ (in the unpermuted
case), and so do all the intra-meson diagrams.
The only exception is diagram III, which has the structure:
\begin{equation}
 \left[\gamma^{\nu}\gamma^{\mu}\cancel{p_{2}}\right]_{A}\left[\gamma_{\nu}\gamma_{\mu}\cancel{p_{4}}\right]_{B}=
 4\cancel{p_{2}}_{A}\cancel{p_{4}}_{B}+\left[\gamma^{\mu\nu}\cancel{p_{2}}\right]_{A}\left[\gamma_{\mu\nu}\cancel{p_{4}}\right]_{B}.
\end{equation}

\subsubsection{Diagram resummation}

We know that the anomalous dimension of the single meson is uniform (indeed, one can show that there is only one spinor degree of freedom for this operator), so we can subtract it in advance and only
consider the inter-meson contributions. The sum of the unpermuted diagrams I-IV is

\begin{equation}
    -\frac{g^{2}}{8\left(4\pi\right)^{2}\epsilon}\frac{\left(p_{E}^{2}\right)^{-\epsilon}}{p_{1}^{2}p_{2}^{2}p_{3}^{2}p_{4}^{2}}\left[\gamma^{\mu\nu}\cancel{p_{2}}\right]_{A}\left[\gamma_{\mu\nu}\cancel{p_{4}}\right]_{B} \delta_{bc}\delta_{da}.
\end{equation}

The permutations affect the color and spinor structure; summing over them gives:

\begin{align}
 & \left[\gamma^{\mu\nu}\cancel{p_{2}}\right]_{A}\left[\gamma_{\mu\nu}\cancel{p_{4}}\right]_{B}\left(\delta_{bc}\delta_{da}-\frac{1}{N}\delta_{ba}\delta_{dc}\right)\goto\nonumber \\
 & \goto\left[\gamma^{\mu\nu}\cancel{p_{2}}\right]_{A}\left[\gamma_{\mu\nu}\cancel{p_{4}}\right]_{B} \delta_{bc}\delta_{da} +\left[\gamma^{\mu\nu}\cancel{p_{2}}\right]_{A}\left[\gamma_{\mu\nu}\cancel{p_{4}}\right]_{B} \delta_{ba}\delta_{dc} \nonumber \\
 & -\left[\gamma^{\mu\nu}\cancel{p_{2}}\right]_{A'}\left[\gamma_{\mu\nu}\cancel{p_{4}}\right]_{B'} \delta_{ba}\delta_{dc}- \left[\gamma^{\mu\nu}\cancel{p_{2}}\right]_{A'}\left[\gamma_{\mu\nu}\cancel{p_{4}}\right]_{B'} \delta_{bc}\delta_{da} = \nonumber \\
 & =\left(\delta_{bc}\delta_{da}+\delta_{ba}\delta_{dc}\right)\left(\left[\gamma^{\mu\nu}\cancel{p_{2}}\right]_{A}\left[\gamma_{\mu\nu}\cancel{p_{4}}\right]_{B}-\left[\gamma^{\mu\nu}\cancel{p_{2}}\right]_{A'}\left[\gamma_{\mu\nu}\cancel{p_{4}}\right]_{B'}\right).
\end{align}

We can use the Fierz identity of equation (33) of \cite{paper15} (modified to our notation of $\gamma^{\mu \nu}$) to convert this result to the spinor structure of the tree level.
This is most conveniently done if we retroactively insert a left/right projection $\frac{1}{2}\left(1\pm\gamma^{5}\right)$
to each meson. The results are equivalent between the projections, so we can consider
left-handed fermions.
\begin{equation}
 \left[P_{L}\gamma^{\mu\nu}\cancel{p_{2}}\right]_{A}\left[P_{L}\gamma_{\mu\nu}\cancel{p_{4}}\right]_{B}=
 -\left\{ \frac{1}{2}\left[P_{L}\gamma^{\mu\nu}\cancel{p_{2}}\right]_{A'}\left[P_{L}\gamma_{\mu\nu}\cancel{p_{4}}\right]_{B'}+6\left[P_{L}\cancel{p_{2}}\right]_{A'}\left[P_{L}\cancel{p_{4}}\right]_{B'}\right\}.
\end{equation}
and similarly with the fermion-permuted contribution:
\begin{align}
 & \left[P_{L}\gamma^{\mu\nu}\cancel{p_{2}}\right]_{A'}\left[P_{L}\gamma_{\mu\nu}\cancel{p_{4}}\right]_{B'}=\frac{1}{2}\left[P_{L}\gamma^{\mu\nu}\cancel{p_{2}}\right]_{A}\left[P_{L}\gamma_{\mu\nu}\cancel{p_{4}}\right]_{B}-6\left[P_{L}\cancel{p_{2}}\right]_{A}\left[P_{L}\cancel{p_{4}}\right]_{B}.
\end{align}
Subtracting them, as they appear in the correlation function, gives::
\begin{align}
 & \left[P_{L}\gamma^{\mu\nu}\cancel{p_{2}}\right]_{A}\left[P_{L}\gamma_{\mu\nu}\cancel{p_{4}}\right]_{B}-\left[P_{L}\gamma^{\mu\nu}\cancel{p_{2}}\right]_{A'}\left[P_{L}\gamma_{\mu\nu}\cancel{p_{4}}\right]_{B'}= \nonumber\\
 & =-\left\{ \frac{1}{2}\left[P_{L}\gamma^{\mu\nu}\cancel{p_{2}}\right]_{A'}\left[P_{L}\gamma_{\mu\nu}\cancel{p_{4}}\right]_{B'}+6\left[P_{L}\cancel{p_{2}}\right]_{A'}\left[P_{L}\cancel{p_{4}}\right]_{B'}\right\} 
 +\left\{ \frac{1}{2}\left[P_{L}\gamma^{\mu\nu}\cancel{p_{2}}\right]_{A}\left[P_{L}\gamma_{\mu\nu}\cancel{p_{4}}\right]_{B}+6\left[P_{L}\cancel{p_{2}}\right]_{A}\left[P_{L}\cancel{p_{4}}\right]_{B} \right\} \nonumber\\
 & \ergo\left[P_{L}\gamma^{\mu\nu}\cancel{p_{2}}\right]_{A}\left[P_{L}\gamma_{\mu\nu}\cancel{p_{4}}\right]_{B}-\left[P_{L}\gamma^{\mu\nu}\cancel{p_{2}}\right]_{A'}\left[P_{L}\gamma_{\mu\nu}\cancel{p_{4}}\right]_{B'}= 
 12\left\{ \left[P_{L}\cancel{p_{2}}\right]_{A}\left[P_{L}\cancel{p_{4}}\right]_{B}-\left[P_{L}\cancel{p_{2}}\right]_{A'}\left[P_{L}\cancel{p_{4}}\right]_{B'}\right\}.
\end{align}

We see that the antisymmetrized spin structure of diagram III gives
a relative factor of $12$.
The same applies to the right-handed projections.

The permuted versions of diagram V give:
\begin{align}
    & -\frac{\tilde{h}}{2\left(4\pi\right)^{2}\epsilon}\frac{\left(p_{E}^{2}\right)^{-\epsilon}}{p_{1}^{2}p_{2}^{2}p_{3}^{2}p_{4}^{2}}\delta_{ad}\delta_{cb}\cancel{p_{2}}_{A}\cancel{p_{4}}_{B} \goto \nonumber\\
    & \goto -\frac{\tilde{h}}{2\left(4\pi\right)^{2}\epsilon}\frac{\left(p_{E}^{2}\right)^{-\epsilon}}{p_{1}^{2}p_{2}^{2}p_{3}^{2}p_{4}^{2}}\left(\delta_{ad}\delta_{cb}\cancel{p_{2}}_{A}\cancel{p_{4}}_{B}+\delta_{cd}\delta_{ab}\cancel{p_{2}}_{A}\cancel{p_{4}}_{B}-\delta_{ad}\delta_{cb}\cancel{p_{2}}_{A'}\cancel{p_{4}}_{B'}-\delta_{cd}\delta_{ab}\cancel{p_{2}}_{A'}\cancel{p_{4}}_{B'}\right) =\nonumber \\
    & = -\frac{\tilde{h}}{2\left(4\pi\right)^{2}\epsilon}\frac{\left(p_{E}^{2}\right)^{-\epsilon}}{p_{1}^{2}p_{2}^{2}p_{3}^{2}p_{4}^{2}}\left(\delta_{ad}\delta_{cb}+\delta_{cd}\delta_{ab}\right) \left(\cancel{p_{2}}_{A}\cancel{p_{4}}_{B}-\cancel{p_{2}}_{A'}\cancel{p_{4}}_{B'}\right).
\end{align}

A similar result holds for the permutation of the tree-level diagram (\ref{eq:tree-level mixed 2-meson}), giving:
\begin{equation}
        \frac{1 }{p_1^2 p_2^2 p_3^2 p_4^2} \left( \delta_{ab}\delta_{cd} + \delta_{cb} \delta_{ad} \right) \left( \cancel{p_2}_A \cancel{p_4}_B - \cancel{p_2}_{A'} \cancel{p_4}_{B'} \right).
\end{equation}
In total, the correlation function correction coming from the inter-meson
interactions is 
\begin{equation}
    -\frac{\tilde{h}+3g^{2}}{2\left(4\pi\right)^{2}\epsilon}\frac{\left(p_{E}^{2}\right)^{-\epsilon}}{p_{1}^{2}p_{2}^{2}p_{3}^{2}p_{4}^{2}}\left(\delta_{bc}\delta_{da}+\delta_{ba}\delta_{dc}\right)\left\{ \left[P_{L}\cancel{p_{2}}\right]_{A}\left[P_{L}\cancel{p_{4}}\right]_{B}-\left[P_{L}\cancel{p_{2}}\right]_{A'}\left[P_{L}\cancel{p_{4}}\right]_{B'}\right\} 
\end{equation}
The correction to the correlation function coming from the inter-meson
diagrams is thus (at scale $M$):
\begin{equation}
\delta'Z_{\left(\phi^{*}\psi\right)^{2}}
=\delta Z_{\left(\phi^{*}\psi\right)^{2}} - \delta\left[ Z^2_{\phi^{*}\psi} \right]
=-\frac{\tilde{h}+3g^{2}}{2\left(4\pi\right)^{2}\epsilon}M^{-2\epsilon}
\end{equation}

Seeing as the coefficient here is always negative for $g^{2},\tilde{h}>0$, we have an agreement with the CCC.

\subsubsection*{Notes about the order of limits}

To consider the Caswell-Banks-Zaks fixed point, we work in the order of limits:
\begin{equation}
    \frac{1}{N} \ll \lambda,h,f \ll 1.
\end{equation}

We saw for scalar mesons that the contribution of gluon exchange diagrams is at most of the order $O \left( 
\frac{\lambda^2}{N} \right)$.
The $\phi^4$ diagrams give a leading contribution of $(\Tilde{h} + \Tilde{f})$, without any additional factors of $N$.
Then we compare these to the 't Hooft couplings: $\Tilde{h}\propto \frac{h}{N}, \Tilde{f}\propto \frac{f}{N_s N}\sim \frac{f}{N^2}$. From this analysis it is clear that the $\Tilde{h}$ term in fact dominates the contribution to the anomalous dimension relevant to the CCC, and satisfies it. The contribution to the 2-fermion operator found in section \ref{section 4.2} is of the same order of magnitude as that of the $\tilde{h}$ diagram.
As for the mixed mesons, the gluon exchange diagrams I-IV give an $O(\frac{\lambda}{N})$ correction, and the $\phi^4$ diagram V gives an $O(\Tilde{h}+\Tilde{f})$ correction, as in the scalar double meson case. This justifies neglecting the $\Tilde{f}$ terms.

When we evaluate the correction to the anomalous dimension of the double meson operator, relative to the part attributed to the single meson operator, we consider inter-meson diagrams that are suppressed by $\frac{1}{N}$ relative to the intra-meson diagrams, but remain at one-loop order in the couplings. This is allowed because the correction is known to also be suppressed by $\frac{1}{N}$ -- and thus so is any contribution to it involving intra-meson interactions. However, contributions including non-trivial intra-meson interactions will also be subleading in perturbation theory, and therefore will be overall suppressed.

\section*{Acknowledgements}

We would like to thank Eran Palti, Adar Sharon, Tomer Solberg and Masataka Watanabe for useful discussions. This work was supported in part
by an Israel Science Foundation (ISF) center for excellence grant (grant number 2289/18), by ISF
grant no. 2159/22, by Simons Foundation grant 994296 (Simons Collaboration on Confinement and
QCD Strings), by grant no. 2018068 from the United States-Israel Binational Science Foundation
(BSF), by the Minerva foundation with funding from the Federal German Ministry for Education and
Research, by the German Research Foundation through a German-Israeli Project Cooperation (DIP)
grant “Holography and the Swampland”, and by a research grant from Martin Eisenstein. OA is the
Samuel Sebba Professorial Chair of Pure and Applied Physics.

\end{document}